\documentclass{svjour3}                     
\smartqed  
\usepackage[
pdftitle={Digital Ecosystems}, 
pdfauthor={Gerard Briscoe},
pdfsubject={Ecosystem-Oriented Architectures},
colorlinks=true, linkcolor=black, citecolor=black, filecolor=black, urlcolor=black,
hyperfootnotes=false]{hyperref} 

\usepackage[printonlyused]{acronym}

\usepackage{setspace, ifthen, substr, color}

\usepackage[numbers]{natbib}

\newboolean{final}\setboolean{final}{true}

\ifthenelse{\boolean{final}}{}{\usepackage{showlabels}\usepackage{citeext}}

\newcommand{\bold}{\textbf}
\newcommand{\italics}{\textit}

\newcommand{\Circ}[1]{\textcircled{\scriptsize #1}}

\newcommand{\added}[1]{#1}
\newcommand{\changed}[1]{#1}

\newcommand{\setCap}[2]{#1\immediate\write18{./mkcaption.sh #2}}
\newcommand{\getCap}[1]{\acl*{#1}}

\newcommand{\tfigure}[9]
	{
	\IfSubStringInString{!}{#7}{\begin{figure}[#7]}{\begin{figure}[!t]}
	\IfSubStringInString{mm}{#8}{\vspace{#8}}{}
	\centering
	
	\IfSubStringInString{pdf}{#3}
		{
		{\includegraphics[#1]{images/#2}}
		}
		{\IfSubStringInString{graph}{#3}
			{
				{\includegraphics[#1]{images/#2}}
			}
			{
				{\includegraphics[#1]{images/#2-crop.pdf}}
			}
		}
		
	\vspace{#6}
	\caption[#4]
		{
		\label{#2}
		\tcaption{#4}{#5}
		}
	\IfSubStringInString{mm}{#9}{\vspace{#9}}{}
	\end{figure}
	}

\newcommand{\mfigure}[7]
	{
	\vspace{#5}
		\IfSubStringInString{!}{#7}
			{\begin{figure}[#7]}{\begin{figure}[!t]}
		#1
		\vspace{#6}
		\caption[#2]{\italics{#2: #3}}
		\label{#4} 
	\end{figure}
	}

\newcommand{\tcaption}[2]
	{
	\IfSubStringInString{:}{#2}{\italics{#1 #2}}{\italics{#1: #2}}
	}

\definecolor{tred}{RGB}{255,0,0} 
\definecolor{t2blue}{RGB}{0,0,255} 
\definecolor{tblue}{RGB}{102,204,255} 
\definecolor{red}{RGB}{128,0,0} 
\definecolor{blue}{RGB}{0,0,128} 
\definecolor{green}{RGB}{0,128,0} 
\definecolor{yellow}{RGB}{128,128,0} 
\definecolor{purple}{RGB}{128,0,128} 
\definecolor{turquoise}{RGB}{0,128,128} 
\definecolor{grey}{RGB}{76,76,76}
\definecolor{brown}{RGB}{128,64,0}

\def\tred#1{#1}

\newcommand{\red}[1]{\color{red}#1\normalcolor}
\newcommand{\blue}[1]{\color{blue}#1\normalcolor}
\newcommand{\green}[1]{\color{green}#1\normalcolor}
\newcommand{\yellow}[1]{\color{yellow}#1\normalcolor}
\newcommand{\purple}[1]{\color{purple}#1\normalcolor}
\newcommand{\turquoise}[1]{\color{turquoise}#1\normalcolor}
\newcommand{\grey}[1]{\color{grey}#1\normalcolor}
\newcommand{\brown}[1]{\color{brown}#1\normalcolor}

\usepackage{graphicx}
\begin{document}

\acrodef{bidpCap}{starts with identifying some useful behaviour in a biological system. Next, the system is studied to isolate the mechanisms by which it performs the useful behaviour. Last, the mechanisms are mimicked in a computational system and their performance evaluated \cite{bidpSpiral}.}
\acrodef{introEcoCap}{made up of one or more communities of organisms, consisting of species in their habitats, with their populations existing in their respective micro-habitats \cite{begon96}. A community is a naturally occurring group of populations from different species that live together in the same habitat. A habitat is a distinct part of the environment \cite{begon96}}
\acrodef{im1}{are different probabilities of going from island \Circ{1} to island \Circ{2}, as there is of going from island \Circ{2} to island \Circ{1}.}
\acrodef{im2}{mirrors the naturally inspired quality that although two populations have the same physical separation, it may be easier to migrate in one direction than the other, i.e. fish migration is easier downstream than upstream.}
\acrodef{DBEdescription}{A wealthy ecosystem sees a balance between co-operation and competition in a dynamic free market.}
\acrodef{serviceCap}{lightweight entity consisting primarily of a pointer to the \ac{SWS} it represents,}
\acrodef{service2cap}{executable component and semantic description. A software service can be a software only service, e.g. data encryption, or provide a front-end to a real-world service, e.g. selling books}
\acrodef{structureCap}{The executable component of a \ac{SWS}, that an Agent represents \added{via a pointer to the \ac{SWS}}, is equivalent to an organism's DNA and is the gene (functional unit) in the evolutionary process \cite{lawrence1989hsd}. So, the Agents should be aggregated as a sequence, like the sequencing of genes in DNA \cite{lawrence1989hsd}\added{, such that an aggregation of Agents is a sequence of Agents}.}
\acrodef{structure2}{an unordered set, or, based on service orchestration, into a tree or workflow}
\acrodef{habnet}{the {agent stations} from mobile agent systems \cite{agentStation} (to provide a distributed environment in which Agent migration can occur), with evolutionary computing \cite{eiben2003iec} for the Agent interaction (instead of traditional agent interaction mechanisms \cite{wooldridge}), and the island-model of \acl*{DEC} \cite{lin1994cgp} for the connectivity between Habitats}
\acrodef{picUser}{will formulate queries to the Digital Ecosystem by creating a request as a {semantic description}, like those being used and developed in \acp{SOA} \cite{SOAsemantic}, specifying an application they desire and submitting it to their Habitat}
\acrodef{picUserReq}{A Population is then instantiated in the user's Habitat in response to the user's request, seeded from the Agents available at their Habitat}
\acrodef{digEco}{with the Agents, the Populations, the Agent migration for \acl*{DEC}, and the environmental selection pressures provided by the user base, then the union of the Habitats creates the Digital Ecosystem}
\acrodef{archComTop}{many strongly connected clusters (communities), called sub-networks (quasi-complete graphs), with a few connections between these clusters (communities) \cite{swn1}. Graphs with this topology have a very high clustering coefficient and small characteristic path lengths \cite{swn1}.}
\acrodef{bizEcoCap}{As the connections between Habitats are reconfigured depending on the connectivity of the user base, the Habitat clustering will therefore be parallel to the business sector communities}
\acrodef{similarCap}{requests are evaluated on separate {islands} (Populations), with their evolution accelerated by the sharing of solutions between the evolving Populations (islands), because they are working to solve similar requests (problems).}
\acrodef{similar2}{yellow lines connecting the evolving Populations indicate similarity in the requests being managed.}
\acrodef{lifeCycleCap}{with deployment to its owner's Habitat for distribution within the Habitat network.}
\acrodef{lifeCycle2}{used in evolving the optimal Agent-sequence in response to a user request. The optimal Agent-sequence is then registered at the Habitat}
\acrodef{lifeCycle3}{If an Agent-sequence solution is then executed, an attempt is made to migrate (copy) it to every other connected Habitat, success depending on the probability associated with the connection.}
\acrodef{as3}{with an abstract representation consisting of a set of}
\acrodef{agentSemantic2}{tuple representing an {attribute} of the {semantic description}, one integer for the {attribute identifier} and one for the {attribute value}, with both ranging between one and a hundred.}
\acrodef{as4}{Each simulated Agent had a semantic description}
\acrodef{semanticRequest}{A simulated user request consisted of an abstract {semantic description}, as a list of sets of numeric tuples to represent the properties of a desired business application}
\acrodef{bmlcap1}{the numerical semantic descriptions, of the simulated services (Agents) and user requests, in a human readable form.}
\acrodef{capbml2}{translated numerical semantic descriptions for one community within the user base, showing it in the context of the travel industry}
\acrodef{capbml3}{The simulation still operated on the numerical representation for operational efficiency, but the semantic filter essentially assigned meaning to the numbers.}
\acrodef{succession}{So, it becomes increasingly more complex through this process of succession, driven by the evolution of the populations within the ecosystem \cite{connell111msn}.}
\acrodef{succession2}{The formation of a mature ecosystem}
\acrodef{succession3}{is the slow, predictable, and orderly changes in the composition and structure of an ecological community, for which there are defined stages in the increasing complexity \cite{begon96}, as shown}
\acrodef{DigEcoSuc2Cap}{the end of the simulation run, the Agent-sequences had evolved and migrated over an average of only ten user requests per Habitat, and collectively had already reached near 70\% effectiveness for the user base.}
\acrodef{DigEcoSucCap}{The formation of a mature biological ecosystem, ecological succession, is a relatively slow process \cite{begon96}, and the simulated Digital Ecosystem acted similarly in reaching a mature state.}
\acrodef{speciesAbundance}{is a measure of the proportion of all organisms in a community belonging to a particular species \cite{Bell}. A relative abundance distribution provides the inequalities in population size within an ecosystem and therefore an indicator of biodiversity, with the distribution of most biological ecosystems taking a log-normal form \cite{Bell}.}
\acrodef{specAbund2}{the Digital Ecosystem did not conform to the expected log-normal}
\acrodef{speciesArea}{In ecology the {species-area} relationship measures diversity relative to the spatial scale, showing the number of species found in a defined area of a particular habitat or habitats of different areas \cite{sizling2004pls}, and is commonly found to follow a power law in biological ecosystems}
\acrodef{ecoCapClass}{then the Digital Ecosystem and biological ecosystem classes would both inherit from the abstract ecosystem class, but implement its attributes differently}
\acrodef{ecoCap2Class}{So, we would argue that the apparent compromises in mimicking biological ecosystems are actually features unique to Digital Ecosystems.}
\acrodef{bidpCap}{starts with identifying some useful behaviour in a biological system. Next, the system is studied to isolate the mechanisms by which it performs the useful behaviour. Last, the mechanisms are mimicked in a computational system and their performance evaluated \cite{bidpSpiral}.}
\acrodef{introEcoCap}{made up of one or more communities of organisms, consisting of species in their habitats, with their populations existing in their respective micro-habitats \cite{begon96}. A community is a naturally occurring group of populations from different species that live together in the same habitat. A habitat is a distinct part of the environment \cite{begon96}}
\acrodef{im1}{are different probabilities of going from island \Circ{1} to island \Circ{2}, as there is of going from island \Circ{2} to island \Circ{1}.}
\acrodef{im2}{mirrors the naturally inspired quality that although two populations have the same physical separation, it may be easier to migrate in one direction than the other, i.e. fish migration is easier downstream than upstream.}
\acrodef{DBEdescription}{A wealthy ecosystem sees a balance between co-operation and competition in a dynamic free market.}
\acrodef{serviceCap}{lightweight entity consisting primarily of a pointer to the \ac{SWS} it represents,}
\acrodef{service2cap}{executable component and semantic description. A software service can be a software only service, e.g. data encryption, or provide a front-end to a real-world service, e.g. selling books}
\acrodef{structureCap}{The executable component of a \ac{SWS}, that an Agent represents \added{via a pointer to the \ac{SWS}}, is equivalent to an organism's DNA and is the gene (functional unit) in the evolutionary process \cite{lawrence1989hsd}. So, the Agents should be aggregated as a sequence, like the sequencing of genes in DNA \cite{lawrence1989hsd}\added{, such that an aggregation of Agents is a sequence of Agents}.}
\acrodef{structure2}{an unordered set, or, based on service orchestration, into a tree or workflow}
\acrodef{habnet}{the {agent stations} from mobile agent systems \cite{agentStation} (to provide a distributed environment in which Agent migration can occur), with evolutionary computing \cite{eiben2003iec} for the Agent interaction (instead of traditional agent interaction mechanisms \cite{wooldridge}), and the island-model of \acl*{DEC} \cite{lin1994cgp} for the connectivity between Habitats}
\acrodef{picUser}{will formulate queries to the Digital Ecosystem by creating a request as a {semantic description}, like those being used and developed in \acp{SOA} \cite{SOAsemantic}, specifying an application they desire and submitting it to their Habitat}
\acrodef{picUserReq}{A Population is then instantiated in the user's Habitat in response to the user's request, seeded from the Agents available at their Habitat}
\acrodef{digEco}{with the Agents, the Populations, the Agent migration for \acl*{DEC}, and the environmental selection pressures provided by the user base, then the union of the Habitats creates the Digital Ecosystem}
\acrodef{archComTop}{many strongly connected clusters (communities), called sub-networks (quasi-complete graphs), with a few connections between these clusters (communities) \cite{swn1}. Graphs with this topology have a very high clustering coefficient and small characteristic path lengths \cite{swn1}.}
\acrodef{bizEcoCap}{As the connections between Habitats are reconfigured depending on the connectivity of the user base, the Habitat clustering will therefore be parallel to the business sector communities}
\acrodef{similarCap}{requests are evaluated on separate {islands} (Populations), with their evolution accelerated by the sharing of solutions between the evolving Populations (islands), because they are working to solve similar requests (problems).}
\acrodef{similar2}{yellow lines connecting the evolving Populations indicate similarity in the requests being managed.}
\acrodef{lifeCycleCap}{with deployment to its owner's Habitat for distribution within the Habitat network.}
\acrodef{lifeCycle2}{used in evolving the optimal Agent-sequence in response to a user request. The optimal Agent-sequence is then registered at the Habitat}
\acrodef{lifeCycle3}{If an Agent-sequence solution is then executed, an attempt is made to migrate (copy) it to every other connected Habitat, success depending on the probability associated with the connection.}
\acrodef{as3}{with an abstract representation consisting of a set of}
\acrodef{agentSemantic2}{tuple representing an {attribute} of the {semantic description}, one integer for the {attribute identifier} and one for the {attribute value}, with both ranging between one and a hundred.}
\acrodef{as4}{Each simulated Agent had a semantic description}
\acrodef{semanticRequest}{A simulated user request consisted of an abstract {semantic description}, as a list of sets of numeric tuples to represent the properties of a desired business application}
\acrodef{bmlcap1}{the numerical semantic descriptions, of the simulated services (Agents) and user requests, in a human readable form.}
\acrodef{capbml2}{translated numerical semantic descriptions for one community within the user base, showing it in the context of the travel industry}
\acrodef{capbml3}{The simulation still operated on the numerical representation for operational efficiency, but the semantic filter essentially assigned meaning to the numbers.}
\acrodef{succession}{So, it becomes increasingly more complex through this process of succession, driven by the evolution of the populations within the ecosystem \cite{connell111msn}.}
\acrodef{succession2}{The formation of a mature ecosystem}
\acrodef{succession3}{is the slow, predictable, and orderly changes in the composition and structure of an ecological community, for which there are defined stages in the increasing complexity \cite{begon96}, as shown}
\acrodef{DigEcoSuc2Cap}{the end of the simulation run, the Agent-sequences had evolved and migrated over an average of only ten user requests per Habitat, and collectively had already reached near 70\% effectiveness for the user base.}
\acrodef{DigEcoSucCap}{The formation of a mature biological ecosystem, ecological succession, is a relatively slow process \cite{begon96}, and the simulated Digital Ecosystem acted similarly in reaching a mature state.}
\acrodef{speciesAbundance}{is a measure of the proportion of all organisms in a community belonging to a particular species \cite{Bell}. A relative abundance distribution provides the inequalities in population size within an ecosystem and therefore an indicator of biodiversity, with the distribution of most biological ecosystems taking a log-normal form \cite{Bell}.}
\acrodef{specAbund2}{the Digital Ecosystem did not conform to the expected log-normal}
\acrodef{speciesArea}{In ecology the {species-area} relationship measures diversity relative to the spatial scale, showing the number of species found in a defined area of a particular habitat or habitats of different areas \cite{sizling2004pls}, and is commonly found to follow a power law in biological ecosystems}
\acrodef{ecoCapClass}{then the Digital Ecosystem and biological ecosystem classes would both inherit from the abstract ecosystem class, but implement its attributes differently}
\acrodef{ecoCap2Class}{So, we would argue that the apparent compromises in mimicking biological ecosystems are actually features unique to Digital Ecosystems.}
\acrodef{bidpCap}{starts with identifying some useful behaviour in a biological system. Next, the system is studied to isolate the mechanisms by which it performs the useful behaviour. Last, the mechanisms are mimicked in a computational system and their performance evaluated \cite{bidpSpiral}.}
\acrodef{introEcoCap}{made up of one or more communities of organisms, consisting of species in their habitats, with their populations existing in their respective micro-habitats \cite{begon96}. A community is a naturally occurring group of populations from different species that live together in the same habitat. A habitat is a distinct part of the environment \cite{begon96}}
\acrodef{im1}{are different probabilities of going from island \Circ{1} to island \Circ{2}, as there is of going from island \Circ{2} to island \Circ{1}.}
\acrodef{im2}{mirrors the naturally inspired quality that although two populations have the same physical separation, it may be easier to migrate in one direction than the other, i.e. fish migration is easier downstream than upstream.}
\acrodef{DBEdescription}{A wealthy ecosystem sees a balance between co-operation and competition in a dynamic free market.}
\acrodef{serviceCap}{lightweight entity consisting primarily of a pointer to the \ac{SWS} it represents,}
\acrodef{service2cap}{executable component and semantic description. A software service can be a software only service, e.g. data encryption, or provide a front-end to a real-world service, e.g. selling books}
\acrodef{structureCap}{The executable component of a \ac{SWS}, that an Agent represents \added{via a pointer to the \ac{SWS}}, is equivalent to an organism's DNA and is the gene (functional unit) in the evolutionary process \cite{lawrence1989hsd}. So, the Agents should be aggregated as a sequence, like the sequencing of genes in DNA \cite{lawrence1989hsd}\added{, such that an aggregation of Agents is a sequence of Agents}.}
\acrodef{structure2}{an unordered set, or, based on service orchestration, into a tree or workflow}
\acrodef{habnet}{the {agent stations} from mobile agent systems \cite{agentStation} (to provide a distributed environment in which Agent migration can occur), with evolutionary computing \cite{eiben2003iec} for the Agent interaction (instead of traditional agent interaction mechanisms \cite{wooldridge}), and the island-model of \acl*{DEC} \cite{lin1994cgp} for the connectivity between Habitats}
\acrodef{picUser}{will formulate queries to the Digital Ecosystem by creating a request as a {semantic description}, like those being used and developed in \acp{SOA} \cite{SOAsemantic}, specifying an application they desire and submitting it to their Habitat}
\acrodef{picUserReq}{A Population is then instantiated in the user's Habitat in response to the user's request, seeded from the Agents available at their Habitat}
\acrodef{digEco}{with the Agents, the Populations, the Agent migration for \acl*{DEC}, and the environmental selection pressures provided by the user base, then the union of the Habitats creates the Digital Ecosystem}
\acrodef{archComTop}{many strongly connected clusters (communities), called sub-networks (quasi-complete graphs), with a few connections between these clusters (communities) \cite{swn1}. Graphs with this topology have a very high clustering coefficient and small characteristic path lengths \cite{swn1}.}
\acrodef{bizEcoCap}{As the connections between Habitats are reconfigured depending on the connectivity of the user base, the Habitat clustering will therefore be parallel to the business sector communities}
\acrodef{similarCap}{requests are evaluated on separate {islands} (Populations), with their evolution accelerated by the sharing of solutions between the evolving Populations (islands), because they are working to solve similar requests (problems).}
\acrodef{similar2}{yellow lines connecting the evolving Populations indicate similarity in the requests being managed.}
\acrodef{lifeCycleCap}{with deployment to its owner's Habitat for distribution within the Habitat network.}
\acrodef{lifeCycle2}{used in evolving the optimal Agent-sequence in response to a user request. The optimal Agent-sequence is then registered at the Habitat}
\acrodef{lifeCycle3}{If an Agent-sequence solution is then executed, an attempt is made to migrate (copy) it to every other connected Habitat, success depending on the probability associated with the connection.}
\acrodef{as3}{with an abstract representation consisting of a set of}
\acrodef{agentSemantic2}{tuple representing an {attribute} of the {semantic description}, one integer for the {attribute identifier} and one for the {attribute value}, with both ranging between one and a hundred.}
\acrodef{as4}{Each simulated Agent had a semantic description}
\acrodef{semanticRequest}{A simulated user request consisted of an abstract {semantic description}, as a list of sets of numeric tuples to represent the properties of a desired business application}
\acrodef{bmlcap1}{the numerical semantic descriptions, of the simulated services (Agents) and user requests, in a human readable form.}
\acrodef{capbml2}{translated numerical semantic descriptions for one community within the user base, showing it in the context of the travel industry}
\acrodef{capbml3}{The simulation still operated on the numerical representation for operational efficiency, but the semantic filter essentially assigned meaning to the numbers.}
\acrodef{succession}{So, it becomes increasingly more complex through this process of succession, driven by the evolution of the populations within the ecosystem \cite{connell111msn}.}
\acrodef{succession2}{The formation of a mature ecosystem}
\acrodef{succession3}{is the slow, predictable, and orderly changes in the composition and structure of an ecological community, for which there are defined stages in the increasing complexity \cite{begon96}, as shown}
\acrodef{DigEcoSuc2Cap}{the end of the simulation run, the Agent-sequences had evolved and migrated over an average of only ten user requests per Habitat, and collectively had already reached near 70\% effectiveness for the user base.}
\acrodef{DigEcoSucCap}{The formation of a mature biological ecosystem, ecological succession, is a relatively slow process \cite{begon96}, and the simulated Digital Ecosystem acted similarly in reaching a mature state.}
\acrodef{speciesAbundance}{is a measure of the proportion of all organisms in a community belonging to a particular species \cite{Bell}. A relative abundance distribution provides the inequalities in population size within an ecosystem and therefore an indicator of biodiversity, with the distribution of most biological ecosystems taking a log-normal form \cite{Bell}.}
\acrodef{specAbund2}{the Digital Ecosystem did not conform to the expected log-normal}
\acrodef{speciesArea}{In ecology the {species-area} relationship measures diversity relative to the spatial scale, showing the number of species found in a defined area of a particular habitat or habitats of different areas \cite{sizling2004pls}, and is commonly found to follow a power law in biological ecosystems}
\acrodef{ecoCapClass}{then the Digital Ecosystem and biological ecosystem classes would both inherit from the abstract ecosystem class, but implement its attributes differently}
\acrodef{ecoCap2Class}{So, we would argue that the apparent compromises in mimicking biological ecosystems are actually features unique to Digital Ecosystems.}
\acrodef{bidpCap}{starts with identifying some useful behaviour in a biological system. Next, the system is studied to isolate the mechanisms by which it performs the useful behaviour. Last, the mechanisms are mimicked in a computational system and their performance evaluated \cite{bidpSpiral}.}
\acrodef{introEcoCap}{made up of one or more communities of organisms, consisting of species in their habitats, with their populations existing in their respective micro-habitats \cite{begon96}. A community is a naturally occurring group of populations from different species that live together in the same habitat. A habitat is a distinct part of the environment \cite{begon96}}
\acrodef{im1}{are different probabilities of going from island \Circ{1} to island \Circ{2}, as there is of going from island \Circ{2} to island \Circ{1}.}
\acrodef{im2}{mirrors the naturally inspired quality that although two populations have the same physical separation, it may be easier to migrate in one direction than the other, i.e. fish migration is easier downstream than upstream.}
\acrodef{DBEdescription}{A wealthy ecosystem sees a balance between co-operation and competition in a dynamic free market.}
\acrodef{serviceCap}{lightweight entity consisting primarily of a pointer to the \ac{SWS} it represents,}
\acrodef{service2cap}{executable component and semantic description. A software service can be a software only service, e.g. data encryption, or provide a front-end to a real-world service, e.g. selling books}
\acrodef{structureCap}{The executable component of a \ac{SWS}, that an Agent represents \added{via a pointer to the \ac{SWS}}, is equivalent to an organism's DNA and is the gene (functional unit) in the evolutionary process \cite{lawrence1989hsd}. So, the Agents should be aggregated as a sequence, like the sequencing of genes in DNA \cite{lawrence1989hsd}\added{, such that an aggregation of Agents is a sequence of Agents}.}
\acrodef{structure2}{an unordered set, or, based on service orchestration, into a tree or workflow}
\acrodef{habnet}{the {agent stations} from mobile agent systems \cite{agentStation} (to provide a distributed environment in which Agent migration can occur), with evolutionary computing \cite{eiben2003iec} for the Agent interaction (instead of traditional agent interaction mechanisms \cite{wooldridge}), and the island-model of \acl*{DEC} \cite{lin1994cgp} for the connectivity between Habitats}
\acrodef{picUser}{will formulate queries to the Digital Ecosystem by creating a request as a {semantic description}, like those being used and developed in \acp{SOA} \cite{SOAsemantic}, specifying an application they desire and submitting it to their Habitat}
\acrodef{picUserReq}{A Population is then instantiated in the user's Habitat in response to the user's request, seeded from the Agents available at their Habitat}
\acrodef{digEco}{with the Agents, the Populations, the Agent migration for \acl*{DEC}, and the environmental selection pressures provided by the user base, then the union of the Habitats creates the Digital Ecosystem}
\acrodef{archComTop}{many strongly connected clusters (communities), called sub-networks (quasi-complete graphs), with a few connections between these clusters (communities) \cite{swn1}. Graphs with this topology have a very high clustering coefficient and small characteristic path lengths \cite{swn1}.}
\acrodef{bizEcoCap}{As the connections between Habitats are reconfigured depending on the connectivity of the user base, the Habitat clustering will therefore be parallel to the business sector communities}
\acrodef{similarCap}{requests are evaluated on separate {islands} (Populations), with their evolution accelerated by the sharing of solutions between the evolving Populations (islands), because they are working to solve similar requests (problems).}
\acrodef{similar2}{yellow lines connecting the evolving Populations indicate similarity in the requests being managed.}
\acrodef{lifeCycleCap}{with deployment to its owner's Habitat for distribution within the Habitat network.}
\acrodef{lifeCycle2}{used in evolving the optimal Agent-sequence in response to a user request. The optimal Agent-sequence is then registered at the Habitat}
\acrodef{lifeCycle3}{If an Agent-sequence solution is then executed, an attempt is made to migrate (copy) it to every other connected Habitat, success depending on the probability associated with the connection.}
\acrodef{as3}{with an abstract representation consisting of a set of}
\acrodef{agentSemantic2}{tuple representing an {attribute} of the {semantic description}, one integer for the {attribute identifier} and one for the {attribute value}, with both ranging between one and a hundred.}
\acrodef{as4}{Each simulated Agent had a semantic description}
\acrodef{semanticRequest}{A simulated user request consisted of an abstract {semantic description}, as a list of sets of numeric tuples to represent the properties of a desired business application}
\acrodef{bmlcap1}{the numerical semantic descriptions, of the simulated services (Agents) and user requests, in a human readable form.}
\acrodef{capbml2}{translated numerical semantic descriptions for one community within the user base, showing it in the context of the travel industry}
\acrodef{capbml3}{The simulation still operated on the numerical representation for operational efficiency, but the semantic filter essentially assigned meaning to the numbers.}
\acrodef{succession}{So, it becomes increasingly more complex through this process of succession, driven by the evolution of the populations within the ecosystem \cite{connell111msn}.}
\acrodef{succession2}{The formation of a mature ecosystem}
\acrodef{succession3}{is the slow, predictable, and orderly changes in the composition and structure of an ecological community, for which there are defined stages in the increasing complexity \cite{begon96}, as shown}
\acrodef{DigEcoSuc2Cap}{the end of the simulation run, the Agent-sequences had evolved and migrated over an average of only ten user requests per Habitat, and collectively had already reached near 70\% effectiveness for the user base.}
\acrodef{DigEcoSucCap}{The formation of a mature biological ecosystem, ecological succession, is a relatively slow process \cite{begon96}, and the simulated Digital Ecosystem acted similarly in reaching a mature state.}
\acrodef{speciesAbundance}{is a measure of the proportion of all organisms in a community belonging to a particular species \cite{Bell}. A relative abundance distribution provides the inequalities in population size within an ecosystem and therefore an indicator of biodiversity, with the distribution of most biological ecosystems taking a log-normal form \cite{Bell}.}
\acrodef{specAbund2}{the Digital Ecosystem did not conform to the expected log-normal}
\acrodef{speciesArea}{In ecology the {species-area} relationship measures diversity relative to the spatial scale, showing the number of species found in a defined area of a particular habitat or habitats of different areas \cite{sizling2004pls}, and is commonly found to follow a power law in biological ecosystems}
\acrodef{ecoCapClass}{then the Digital Ecosystem and biological ecosystem classes would both inherit from the abstract ecosystem class, but implement its attributes differently}
\acrodef{ecoCap2Class}{So, we would argue that the apparent compromises in mimicking biological ecosystems are actually features unique to Digital Ecosystems.}
\acrodef{bidpCap}{starts with identifying some useful behaviour in a biological system. Next, the system is studied to isolate the mechanisms by which it performs the useful behaviour. Last, the mechanisms are mimicked in a computational system and their performance evaluated \cite{bidpSpiral}.}
\acrodef{introEcoCap}{made up of one or more communities of organisms, consisting of species in their habitats, with their populations existing in their respective micro-habitats \cite{begon96}. A community is a naturally occurring group of populations from different species that live together in the same habitat. A habitat is a distinct part of the environment \cite{begon96}}
\acrodef{im1}{are different probabilities of going from island \Circ{1} to island \Circ{2}, as there is of going from island \Circ{2} to island \Circ{1}.}
\acrodef{im2}{mirrors the naturally inspired quality that although two populations have the same physical separation, it may be easier to migrate in one direction than the other, i.e. fish migration is easier downstream than upstream.}
\acrodef{DBEdescription}{A wealthy ecosystem sees a balance between co-operation and competition in a dynamic free market.}
\acrodef{serviceCap}{lightweight entity consisting primarily of a pointer to the \ac{SWS} it represents,}
\acrodef{service2cap}{executable component and semantic description. A software service can be a software only service, e.g. data encryption, or provide a front-end to a real-world service, e.g. selling books}
\acrodef{structureCap}{The executable component of a \ac{SWS}, that an Agent represents \added{via a pointer to the \ac{SWS}}, is equivalent to an organism's DNA and is the gene (functional unit) in the evolutionary process \cite{lawrence1989hsd}. So, the Agents should be aggregated as a sequence, like the sequencing of genes in DNA \cite{lawrence1989hsd}\added{, such that an aggregation of Agents is a sequence of Agents}.}
\acrodef{structure2}{an unordered set, or, based on service orchestration, into a tree or workflow}
\acrodef{habnet}{the {agent stations} from mobile agent systems \cite{agentStation} (to provide a distributed environment in which Agent migration can occur), with evolutionary computing \cite{eiben2003iec} for the Agent interaction (instead of traditional agent interaction mechanisms \cite{wooldridge}), and the island-model of \acl*{DEC} \cite{lin1994cgp} for the connectivity between Habitats}
\acrodef{picUser}{will formulate queries to the Digital Ecosystem by creating a request as a {semantic description}, like those being used and developed in \acp{SOA} \cite{SOAsemantic}, specifying an application they desire and submitting it to their Habitat}
\acrodef{picUserReq}{A Population is then instantiated in the user's Habitat in response to the user's request, seeded from the Agents available at their Habitat}
\acrodef{digEco}{with the Agents, the Populations, the Agent migration for \acl*{DEC}, and the environmental selection pressures provided by the user base, then the union of the Habitats creates the Digital Ecosystem}
\acrodef{archComTop}{many strongly connected clusters (communities), called sub-networks (quasi-complete graphs), with a few connections between these clusters (communities) \cite{swn1}. Graphs with this topology have a very high clustering coefficient and small characteristic path lengths \cite{swn1}.}
\acrodef{bizEcoCap}{As the connections between Habitats are reconfigured depending on the connectivity of the user base, the Habitat clustering will therefore be parallel to the business sector communities}
\acrodef{similarCap}{requests are evaluated on separate {islands} (Populations), with their evolution accelerated by the sharing of solutions between the evolving Populations (islands), because they are working to solve similar requests (problems).}
\acrodef{similar2}{yellow lines connecting the evolving Populations indicate similarity in the requests being managed.}
\acrodef{lifeCycleCap}{with deployment to its owner's Habitat for distribution within the Habitat network.}
\acrodef{lifeCycle2}{used in evolving the optimal Agent-sequence in response to a user request. The optimal Agent-sequence is then registered at the Habitat}
\acrodef{lifeCycle3}{If an Agent-sequence solution is then executed, an attempt is made to migrate (copy) it to every other connected Habitat, success depending on the probability associated with the connection.}
\acrodef{as3}{with an abstract representation consisting of a set of}
\acrodef{agentSemantic2}{tuple representing an {attribute} of the {semantic description}, one integer for the {attribute identifier} and one for the {attribute value}, with both ranging between one and a hundred.}
\acrodef{as4}{Each simulated Agent had a semantic description}
\acrodef{semanticRequest}{A simulated user request consisted of an abstract {semantic description}, as a list of sets of numeric tuples to represent the properties of a desired business application}
\acrodef{bmlcap1}{the numerical semantic descriptions, of the simulated services (Agents) and user requests, in a human readable form.}
\acrodef{capbml2}{translated numerical semantic descriptions for one community within the user base, showing it in the context of the travel industry}
\acrodef{capbml3}{The simulation still operated on the numerical representation for operational efficiency, but the semantic filter essentially assigned meaning to the numbers.}
\acrodef{succession}{So, it becomes increasingly more complex through this process of succession, driven by the evolution of the populations within the ecosystem \cite{connell111msn}.}
\acrodef{succession2}{The formation of a mature ecosystem}
\acrodef{succession3}{is the slow, predictable, and orderly changes in the composition and structure of an ecological community, for which there are defined stages in the increasing complexity \cite{begon96}, as shown}
\acrodef{DigEcoSuc2Cap}{the end of the simulation run, the Agent-sequences had evolved and migrated over an average of only ten user requests per Habitat, and collectively had already reached near 70\% effectiveness for the user base.}
\acrodef{DigEcoSucCap}{The formation of a mature biological ecosystem, ecological succession, is a relatively slow process \cite{begon96}, and the simulated Digital Ecosystem acted similarly in reaching a mature state.}
\acrodef{speciesAbundance}{is a measure of the proportion of all organisms in a community belonging to a particular species \cite{Bell}. A relative abundance distribution provides the inequalities in population size within an ecosystem and therefore an indicator of biodiversity, with the distribution of most biological ecosystems taking a log-normal form \cite{Bell}.}
\acrodef{specAbund2}{the Digital Ecosystem did not conform to the expected log-normal}
\acrodef{speciesArea}{In ecology the {species-area} relationship measures diversity relative to the spatial scale, showing the number of species found in a defined area of a particular habitat or habitats of different areas \cite{sizling2004pls}, and is commonly found to follow a power law in biological ecosystems}
\acrodef{ecoCapClass}{then the Digital Ecosystem and biological ecosystem classes would both inherit from the abstract ecosystem class, but implement its attributes differently}
\acrodef{ecoCap2Class}{So, we would argue that the apparent compromises in mimicking biological ecosystems are actually features unique to Digital Ecosystems.}
\acrodef{bidpCap}{starts with identifying some useful behaviour in a biological system. Next, the system is studied to isolate the mechanisms by which it performs the useful behaviour. Last, the mechanisms are mimicked in a computational system and their performance evaluated \cite{bidpSpiral}.}
\acrodef{introEcoCap}{made up of one or more communities of organisms, consisting of species in their habitats, with their populations existing in their respective micro-habitats \cite{begon96}. A community is a naturally occurring group of populations from different species that live together in the same habitat. A habitat is a distinct part of the environment \cite{begon96}}
\acrodef{im1}{are different probabilities of going from island \Circ{1} to island \Circ{2}, as there is of going from island \Circ{2} to island \Circ{1}.}
\acrodef{im2}{mirrors the naturally inspired quality that although two populations have the same physical separation, it may be easier to migrate in one direction than the other, i.e. fish migration is easier downstream than upstream.}
\acrodef{DBEdescription}{A wealthy ecosystem sees a balance between co-operation and competition in a dynamic free market.}
\acrodef{serviceCap}{lightweight entity consisting primarily of a pointer to the \ac{SWS} it represents,}
\acrodef{service2cap}{executable component and semantic description. A software service can be a software only service, e.g. data encryption, or provide a front-end to a real-world service, e.g. selling books}
\acrodef{structureCap}{The executable component of a \ac{SWS}, that an Agent represents \added{via a pointer to the \ac{SWS}}, is equivalent to an organism's DNA and is the gene (functional unit) in the evolutionary process \cite{lawrence1989hsd}. So, the Agents should be aggregated as a sequence, like the sequencing of genes in DNA \cite{lawrence1989hsd}\added{, such that an aggregation of Agents is a sequence of Agents}.}
\acrodef{structure2}{an unordered set, or, based on service orchestration, into a tree or workflow}
\acrodef{habnet}{the {agent stations} from mobile agent systems \cite{agentStation} (to provide a distributed environment in which Agent migration can occur), with evolutionary computing \cite{eiben2003iec} for the Agent interaction (instead of traditional agent interaction mechanisms \cite{wooldridge}), and the island-model of \acl*{DEC} \cite{lin1994cgp} for the connectivity between Habitats}
\acrodef{picUser}{will formulate queries to the Digital Ecosystem by creating a request as a {semantic description}, like those being used and developed in \acp{SOA} \cite{SOAsemantic}, specifying an application they desire and submitting it to their Habitat}
\acrodef{picUserReq}{A Population is then instantiated in the user's Habitat in response to the user's request, seeded from the Agents available at their Habitat}
\acrodef{digEco}{with the Agents, the Populations, the Agent migration for \acl*{DEC}, and the environmental selection pressures provided by the user base, then the union of the Habitats creates the Digital Ecosystem}
\acrodef{archComTop}{many strongly connected clusters (communities), called sub-networks (quasi-complete graphs), with a few connections between these clusters (communities) \cite{swn1}. Graphs with this topology have a very high clustering coefficient and small characteristic path lengths \cite{swn1}.}
\acrodef{bizEcoCap}{As the connections between Habitats are reconfigured depending on the connectivity of the user base, the Habitat clustering will therefore be parallel to the business sector communities}
\acrodef{similarCap}{requests are evaluated on separate {islands} (Populations), with their evolution accelerated by the sharing of solutions between the evolving Populations (islands), because they are working to solve similar requests (problems).}
\acrodef{similar2}{yellow lines connecting the evolving Populations indicate similarity in the requests being managed.}
\acrodef{lifeCycleCap}{with deployment to its owner's Habitat for distribution within the Habitat network.}
\acrodef{lifeCycle2}{used in evolving the optimal Agent-sequence in response to a user request. The optimal Agent-sequence is then registered at the Habitat}
\acrodef{lifeCycle3}{If an Agent-sequence solution is then executed, an attempt is made to migrate (copy) it to every other connected Habitat, success depending on the probability associated with the connection.}
\acrodef{as3}{with an abstract representation consisting of a set of}
\acrodef{agentSemantic2}{tuple representing an {attribute} of the {semantic description}, one integer for the {attribute identifier} and one for the {attribute value}, with both ranging between one and a hundred.}
\acrodef{as4}{Each simulated Agent had a semantic description}
\acrodef{semanticRequest}{A simulated user request consisted of an abstract {semantic description}, as a list of sets of numeric tuples to represent the properties of a desired business application}
\acrodef{bmlcap1}{the numerical semantic descriptions, of the simulated services (Agents) and user requests, in a human readable form.}
\acrodef{capbml2}{translated numerical semantic descriptions for one community within the user base, showing it in the context of the travel industry}
\acrodef{capbml3}{The simulation still operated on the numerical representation for operational efficiency, but the semantic filter essentially assigned meaning to the numbers.}
\acrodef{succession}{So, it becomes increasingly more complex through this process of succession, driven by the evolution of the populations within the ecosystem \cite{connell111msn}.}
\acrodef{succession2}{The formation of a mature ecosystem}
\acrodef{succession3}{is the slow, predictable, and orderly changes in the composition and structure of an ecological community, for which there are defined stages in the increasing complexity \cite{begon96}, as shown}
\acrodef{DigEcoSuc2Cap}{the end of the simulation run, the Agent-sequences had evolved and migrated over an average of only ten user requests per Habitat, and collectively had already reached near 70\% effectiveness for the user base.}
\acrodef{DigEcoSucCap}{The formation of a mature biological ecosystem, ecological succession, is a relatively slow process \cite{begon96}, and the simulated Digital Ecosystem acted similarly in reaching a mature state.}
\acrodef{speciesAbundance}{is a measure of the proportion of all organisms in a community belonging to a particular species \cite{Bell}. A relative abundance distribution provides the inequalities in population size within an ecosystem and therefore an indicator of biodiversity, with the distribution of most biological ecosystems taking a log-normal form \cite{Bell}.}
\acrodef{specAbund2}{the Digital Ecosystem did not conform to the expected log-normal}
\acrodef{speciesArea}{In ecology the {species-area} relationship measures diversity relative to the spatial scale, showing the number of species found in a defined area of a particular habitat or habitats of different areas \cite{sizling2004pls}, and is commonly found to follow a power law in biological ecosystems}
\acrodef{ecoCapClass}{then the Digital Ecosystem and biological ecosystem classes would both inherit from the abstract ecosystem class, but implement its attributes differently}
\acrodef{ecoCap2Class}{So, we would argue that the apparent compromises in mimicking biological ecosystems are actually features unique to Digital Ecosystems.}
\acrodef{bidpCap}{starts with identifying some useful behaviour in a biological system. Next, the system is studied to isolate the mechanisms by which it performs the useful behaviour. Last, the mechanisms are mimicked in a computational system and their performance evaluated \cite{bidpSpiral}.}
\acrodef{introEcoCap}{made up of one or more communities of organisms, consisting of species in their habitats, with their populations existing in their respective micro-habitats \cite{begon96}. A community is a naturally occurring group of populations from different species that live together in the same habitat. A habitat is a distinct part of the environment \cite{begon96}}
\acrodef{im1}{are different probabilities of going from island \Circ{1} to island \Circ{2}, as there is of going from island \Circ{2} to island \Circ{1}.}
\acrodef{im2}{mirrors the naturally inspired quality that although two populations have the same physical separation, it may be easier to migrate in one direction than the other, i.e. fish migration is easier downstream than upstream.}
\acrodef{DBEdescription}{A wealthy ecosystem sees a balance between co-operation and competition in a dynamic free market.}
\acrodef{serviceCap}{lightweight entity consisting primarily of a pointer to the \ac{SWS} it represents,}
\acrodef{service2cap}{executable component and semantic description. A software service can be a software only service, e.g. data encryption, or provide a front-end to a real-world service, e.g. selling books}
\acrodef{structureCap}{The executable component of a \ac{SWS}, that an Agent represents \added{via a pointer to the \ac{SWS}}, is equivalent to an organism's DNA and is the gene (functional unit) in the evolutionary process \cite{lawrence1989hsd}. So, the Agents should be aggregated as a sequence, like the sequencing of genes in DNA \cite{lawrence1989hsd}\added{, such that an aggregation of Agents is a sequence of Agents}.}
\acrodef{structure2}{an unordered set, or, based on service orchestration, into a tree or workflow}
\acrodef{habnet}{the {agent stations} from mobile agent systems \cite{agentStation} (to provide a distributed environment in which Agent migration can occur), with evolutionary computing \cite{eiben2003iec} for the Agent interaction (instead of traditional agent interaction mechanisms \cite{wooldridge}), and the island-model of \acl*{DEC} \cite{lin1994cgp} for the connectivity between Habitats}
\acrodef{picUser}{will formulate queries to the Digital Ecosystem by creating a request as a {semantic description}, like those being used and developed in \acp{SOA} \cite{SOAsemantic}, specifying an application they desire and submitting it to their Habitat}
\acrodef{picUserReq}{A Population is then instantiated in the user's Habitat in response to the user's request, seeded from the Agents available at their Habitat}
\acrodef{digEco}{with the Agents, the Populations, the Agent migration for \acl*{DEC}, and the environmental selection pressures provided by the user base, then the union of the Habitats creates the Digital Ecosystem}
\acrodef{archComTop}{many strongly connected clusters (communities), called sub-networks (quasi-complete graphs), with a few connections between these clusters (communities) \cite{swn1}. Graphs with this topology have a very high clustering coefficient and small characteristic path lengths \cite{swn1}.}
\acrodef{bizEcoCap}{As the connections between Habitats are reconfigured depending on the connectivity of the user base, the Habitat clustering will therefore be parallel to the business sector communities}
\acrodef{similarCap}{requests are evaluated on separate {islands} (Populations), with their evolution accelerated by the sharing of solutions between the evolving Populations (islands), because they are working to solve similar requests (problems).}
\acrodef{similar2}{yellow lines connecting the evolving Populations indicate similarity in the requests being managed.}
\acrodef{lifeCycleCap}{with deployment to its owner's Habitat for distribution within the Habitat network.}
\acrodef{lifeCycle2}{used in evolving the optimal Agent-sequence in response to a user request. The optimal Agent-sequence is then registered at the Habitat}
\acrodef{lifeCycle3}{If an Agent-sequence solution is then executed, an attempt is made to migrate (copy) it to every other connected Habitat, success depending on the probability associated with the connection.}
\acrodef{as3}{with an abstract representation consisting of a set of}
\acrodef{agentSemantic2}{tuple representing an {attribute} of the {semantic description}, one integer for the {attribute identifier} and one for the {attribute value}, with both ranging between one and a hundred.}
\acrodef{as4}{Each simulated Agent had a semantic description}
\acrodef{semanticRequest}{A simulated user request consisted of an abstract {semantic description}, as a list of sets of numeric tuples to represent the properties of a desired business application}
\acrodef{bmlcap1}{the numerical semantic descriptions, of the simulated services (Agents) and user requests, in a human readable form.}
\acrodef{capbml2}{translated numerical semantic descriptions for one community within the user base, showing it in the context of the travel industry}
\acrodef{capbml3}{The simulation still operated on the numerical representation for operational efficiency, but the semantic filter essentially assigned meaning to the numbers.}
\acrodef{succession}{So, it becomes increasingly more complex through this process of succession, driven by the evolution of the populations within the ecosystem \cite{connell111msn}.}
\acrodef{succession2}{The formation of a mature ecosystem}
\acrodef{succession3}{is the slow, predictable, and orderly changes in the composition and structure of an ecological community, for which there are defined stages in the increasing complexity \cite{begon96}, as shown}
\acrodef{DigEcoSuc2Cap}{the end of the simulation run, the Agent-sequences had evolved and migrated over an average of only ten user requests per Habitat, and collectively had already reached near 70\% effectiveness for the user base.}
\acrodef{DigEcoSucCap}{The formation of a mature biological ecosystem, ecological succession, is a relatively slow process \cite{begon96}, and the simulated Digital Ecosystem acted similarly in reaching a mature state.}
\acrodef{speciesAbundance}{is a measure of the proportion of all organisms in a community belonging to a particular species \cite{Bell}. A relative abundance distribution provides the inequalities in population size within an ecosystem and therefore an indicator of biodiversity, with the distribution of most biological ecosystems taking a log-normal form \cite{Bell}.}
\acrodef{specAbund2}{the Digital Ecosystem did not conform to the expected log-normal}
\acrodef{speciesArea}{In ecology the {species-area} relationship measures diversity relative to the spatial scale, showing the number of species found in a defined area of a particular habitat or habitats of different areas \cite{sizling2004pls}, and is commonly found to follow a power law in biological ecosystems}
\acrodef{ecoCapClass}{then the Digital Ecosystem and biological ecosystem classes would both inherit from the abstract ecosystem class, but implement its attributes differently}
\acrodef{ecoCap2Class}{So, we would argue that the apparent compromises in mimicking biological ecosystems are actually features unique to Digital Ecosystems.}
\acrodef{bidpCap}{starts with identifying some useful behaviour in a biological system. Next, the system is studied to isolate the mechanisms by which it performs the useful behaviour. Last, the mechanisms are mimicked in a computational system and their performance evaluated \cite{bidpSpiral}.}
\acrodef{introEcoCap}{made up of one or more communities of organisms, consisting of species in their habitats, with their populations existing in their respective micro-habitats \cite{begon96}. A community is a naturally occurring group of populations from different species that live together in the same habitat. A habitat is a distinct part of the environment \cite{begon96}}
\acrodef{im1}{are different probabilities of going from island \Circ{1} to island \Circ{2}, as there is of going from island \Circ{2} to island \Circ{1}.}
\acrodef{im2}{mirrors the naturally inspired quality that although two populations have the same physical separation, it may be easier to migrate in one direction than the other, i.e. fish migration is easier downstream than upstream.}
\acrodef{DBEdescription}{A wealthy ecosystem sees a balance between co-operation and competition in a dynamic free market.}
\acrodef{serviceCap}{lightweight entity consisting primarily of a pointer to the \ac{SWS} it represents,}
\acrodef{service2cap}{executable component and semantic description. A software service can be a software only service, e.g. data encryption, or provide a front-end to a real-world service, e.g. selling books}
\acrodef{structureCap}{The executable component of a \ac{SWS}, that an Agent represents \added{via a pointer to the \ac{SWS}}, is equivalent to an organism's DNA and is the gene (functional unit) in the evolutionary process \cite{lawrence1989hsd}. So, the Agents should be aggregated as a sequence, like the sequencing of genes in DNA \cite{lawrence1989hsd}\added{, such that an aggregation of Agents is a sequence of Agents}.}
\acrodef{structure2}{an unordered set, or, based on service orchestration, into a tree or workflow}
\acrodef{habnet}{the {agent stations} from mobile agent systems \cite{agentStation} (to provide a distributed environment in which Agent migration can occur), with evolutionary computing \cite{eiben2003iec} for the Agent interaction (instead of traditional agent interaction mechanisms \cite{wooldridge}), and the island-model of \acl*{DEC} \cite{lin1994cgp} for the connectivity between Habitats}
\acrodef{picUser}{will formulate queries to the Digital Ecosystem by creating a request as a {semantic description}, like those being used and developed in \acp{SOA} \cite{SOAsemantic}, specifying an application they desire and submitting it to their Habitat}
\acrodef{picUserReq}{A Population is then instantiated in the user's Habitat in response to the user's request, seeded from the Agents available at their Habitat}
\acrodef{digEco}{with the Agents, the Populations, the Agent migration for \acl*{DEC}, and the environmental selection pressures provided by the user base, then the union of the Habitats creates the Digital Ecosystem}
\acrodef{archComTop}{many strongly connected clusters (communities), called sub-networks (quasi-complete graphs), with a few connections between these clusters (communities) \cite{swn1}. Graphs with this topology have a very high clustering coefficient and small characteristic path lengths \cite{swn1}.}
\acrodef{bizEcoCap}{As the connections between Habitats are reconfigured depending on the connectivity of the user base, the Habitat clustering will therefore be parallel to the business sector communities}
\acrodef{similarCap}{requests are evaluated on separate {islands} (Populations), with their evolution accelerated by the sharing of solutions between the evolving Populations (islands), because they are working to solve similar requests (problems).}
\acrodef{similar2}{yellow lines connecting the evolving Populations indicate similarity in the requests being managed.}
\acrodef{lifeCycleCap}{with deployment to its owner's Habitat for distribution within the Habitat network.}
\acrodef{lifeCycle2}{used in evolving the optimal Agent-sequence in response to a user request. The optimal Agent-sequence is then registered at the Habitat}
\acrodef{lifeCycle3}{If an Agent-sequence solution is then executed, an attempt is made to migrate (copy) it to every other connected Habitat, success depending on the probability associated with the connection.}
\acrodef{as3}{with an abstract representation consisting of a set of}
\acrodef{agentSemantic2}{tuple representing an {attribute} of the {semantic description}, one integer for the {attribute identifier} and one for the {attribute value}, with both ranging between one and a hundred.}
\acrodef{as4}{Each simulated Agent had a semantic description}
\acrodef{semanticRequest}{A simulated user request consisted of an abstract {semantic description}, as a list of sets of numeric tuples to represent the properties of a desired business application}
\acrodef{bmlcap1}{the numerical semantic descriptions, of the simulated services (Agents) and user requests, in a human readable form.}
\acrodef{capbml2}{translated numerical semantic descriptions for one community within the user base, showing it in the context of the travel industry}
\acrodef{capbml3}{The simulation still operated on the numerical representation for operational efficiency, but the semantic filter essentially assigned meaning to the numbers.}
\acrodef{succession}{So, it becomes increasingly more complex through this process of succession, driven by the evolution of the populations within the ecosystem \cite{connell111msn}.}
\acrodef{succession2}{The formation of a mature ecosystem}
\acrodef{succession3}{is the slow, predictable, and orderly changes in the composition and structure of an ecological community, for which there are defined stages in the increasing complexity \cite{begon96}, as shown}
\acrodef{DigEcoSuc2Cap}{the end of the simulation run, the Agent-sequences had evolved and migrated over an average of only ten user requests per Habitat, and collectively had already reached near 70\% effectiveness for the user base.}
\acrodef{DigEcoSucCap}{The formation of a mature biological ecosystem, ecological succession, is a relatively slow process \cite{begon96}, and the simulated Digital Ecosystem acted similarly in reaching a mature state.}
\acrodef{speciesAbundance}{is a measure of the proportion of all organisms in a community belonging to a particular species \cite{Bell}. A relative abundance distribution provides the inequalities in population size within an ecosystem and therefore an indicator of biodiversity, with the distribution of most biological ecosystems taking a log-normal form \cite{Bell}.}
\acrodef{specAbund2}{the Digital Ecosystem did not conform to the expected log-normal}
\acrodef{speciesArea}{In ecology the {species-area} relationship measures diversity relative to the spatial scale, showing the number of species found in a defined area of a particular habitat or habitats of different areas \cite{sizling2004pls}, and is commonly found to follow a power law in biological ecosystems}
\acrodef{ecoCapClass}{then the Digital Ecosystem and biological ecosystem classes would both inherit from the abstract ecosystem class, but implement its attributes differently}
\acrodef{ecoCap2Class}{So, we would argue that the apparent compromises in mimicking biological ecosystems are actually features unique to Digital Ecosystems.}
\acrodef{bidpCap}{starts with identifying some useful behaviour in a biological system. Next, the system is studied to isolate the mechanisms by which it performs the useful behaviour. Last, the mechanisms are mimicked in a computational system and their performance evaluated \cite{bidpSpiral}.}
\acrodef{introEcoCap}{made up of one or more communities of organisms, consisting of species in their habitats, with their populations existing in their respective micro-habitats \cite{begon96}. A community is a naturally occurring group of populations from different species that live together in the same habitat. A habitat is a distinct part of the environment \cite{begon96}}
\acrodef{im1}{are different probabilities of going from island \Circ{1} to island \Circ{2}, as there is of going from island \Circ{2} to island \Circ{1}.}
\acrodef{im2}{mirrors the naturally inspired quality that although two populations have the same physical separation, it may be easier to migrate in one direction than the other, i.e. fish migration is easier downstream than upstream.}
\acrodef{DBEdescription}{A wealthy ecosystem sees a balance between co-operation and competition in a dynamic free market.}
\acrodef{serviceCap}{lightweight entity consisting primarily of a pointer to the \ac{SWS} it represents,}
\acrodef{service2cap}{executable component and semantic description. A software service can be a software only service, e.g. data encryption, or provide a front-end to a real-world service, e.g. selling books}
\acrodef{structureCap}{The executable component of a \ac{SWS}, that an Agent represents \added{via a pointer to the \ac{SWS}}, is equivalent to an organism's DNA and is the gene (functional unit) in the evolutionary process \cite{lawrence1989hsd}. So, the Agents should be aggregated as a sequence, like the sequencing of genes in DNA \cite{lawrence1989hsd}\added{, such that an aggregation of Agents is a sequence of Agents}.}
\acrodef{structure2}{an unordered set, or, based on service orchestration, into a tree or workflow}
\acrodef{habnet}{the {agent stations} from mobile agent systems \cite{agentStation} (to provide a distributed environment in which Agent migration can occur), with evolutionary computing \cite{eiben2003iec} for the Agent interaction (instead of traditional agent interaction mechanisms \cite{wooldridge}), and the island-model of \acl*{DEC} \cite{lin1994cgp} for the connectivity between Habitats}
\acrodef{picUser}{will formulate queries to the Digital Ecosystem by creating a request as a {semantic description}, like those being used and developed in \acp{SOA} \cite{SOAsemantic}, specifying an application they desire and submitting it to their Habitat}
\acrodef{picUserReq}{A Population is then instantiated in the user's Habitat in response to the user's request, seeded from the Agents available at their Habitat}
\acrodef{digEco}{with the Agents, the Populations, the Agent migration for \acl*{DEC}, and the environmental selection pressures provided by the user base, then the union of the Habitats creates the Digital Ecosystem}
\acrodef{archComTop}{many strongly connected clusters (communities), called sub-networks (quasi-complete graphs), with a few connections between these clusters (communities) \cite{swn1}. Graphs with this topology have a very high clustering coefficient and small characteristic path lengths \cite{swn1}.}
\acrodef{bizEcoCap}{As the connections between Habitats are reconfigured depending on the connectivity of the user base, the Habitat clustering will therefore be parallel to the business sector communities}
\acrodef{similarCap}{requests are evaluated on separate {islands} (Populations), with their evolution accelerated by the sharing of solutions between the evolving Populations (islands), because they are working to solve similar requests (problems).}
\acrodef{similar2}{yellow lines connecting the evolving Populations indicate similarity in the requests being managed.}
\acrodef{lifeCycleCap}{with deployment to its owner's Habitat for distribution within the Habitat network.}
\acrodef{lifeCycle2}{used in evolving the optimal Agent-sequence in response to a user request. The optimal Agent-sequence is then registered at the Habitat}
\acrodef{lifeCycle3}{If an Agent-sequence solution is then executed, an attempt is made to migrate (copy) it to every other connected Habitat, success depending on the probability associated with the connection.}
\acrodef{as3}{with an abstract representation consisting of a set of}
\acrodef{agentSemantic2}{tuple representing an {attribute} of the {semantic description}, one integer for the {attribute identifier} and one for the {attribute value}, with both ranging between one and a hundred.}
\acrodef{as4}{Each simulated Agent had a semantic description}
\acrodef{semanticRequest}{A simulated user request consisted of an abstract {semantic description}, as a list of sets of numeric tuples to represent the properties of a desired business application}
\acrodef{bmlcap1}{the numerical semantic descriptions, of the simulated services (Agents) and user requests, in a human readable form.}
\acrodef{capbml2}{translated numerical semantic descriptions for one community within the user base, showing it in the context of the travel industry}
\acrodef{capbml3}{The simulation still operated on the numerical representation for operational efficiency, but the semantic filter essentially assigned meaning to the numbers.}
\acrodef{succession}{So, it becomes increasingly more complex through this process of succession, driven by the evolution of the populations within the ecosystem \cite{connell111msn}.}
\acrodef{succession2}{The formation of a mature ecosystem}
\acrodef{succession3}{is the slow, predictable, and orderly changes in the composition and structure of an ecological community, for which there are defined stages in the increasing complexity \cite{begon96}, as shown}
\acrodef{DigEcoSuc2Cap}{the end of the simulation run, the Agent-sequences had evolved and migrated over an average of only ten user requests per Habitat, and collectively had already reached near 70\% effectiveness for the user base.}
\acrodef{DigEcoSucCap}{The formation of a mature biological ecosystem, ecological succession, is a relatively slow process \cite{begon96}, and the simulated Digital Ecosystem acted similarly in reaching a mature state.}
\acrodef{speciesAbundance}{is a measure of the proportion of all organisms in a community belonging to a particular species \cite{Bell}. A relative abundance distribution provides the inequalities in population size within an ecosystem and therefore an indicator of biodiversity, with the distribution of most biological ecosystems taking a log-normal form \cite{Bell}.}
\acrodef{specAbund2}{the Digital Ecosystem did not conform to the expected log-normal}
\acrodef{speciesArea}{In ecology the {species-area} relationship measures diversity relative to the spatial scale, showing the number of species found in a defined area of a particular habitat or habitats of different areas \cite{sizling2004pls}, and is commonly found to follow a power law in biological ecosystems}
\acrodef{ecoCapClass}{then the Digital Ecosystem and biological ecosystem classes would both inherit from the abstract ecosystem class, but implement its attributes differently}
\acrodef{ecoCap2Class}{So, we would argue that the apparent compromises in mimicking biological ecosystems are actually features unique to Digital Ecosystems.}
\acrodef{bidpCap}{starts with identifying some useful behaviour in a biological system. Next, the system is studied to isolate the mechanisms by which it performs the useful behaviour. Last, the mechanisms are mimicked in a computational system and their performance evaluated \cite{bidpSpiral}.}
\acrodef{introEcoCap}{made up of one or more communities of organisms, consisting of species in their habitats, with their populations existing in their respective micro-habitats \cite{begon96}. A community is a naturally occurring group of populations from different species that live together in the same habitat. A habitat is a distinct part of the environment \cite{begon96}}
\acrodef{im1}{are different probabilities of going from island \Circ{1} to island \Circ{2}, as there is of going from island \Circ{2} to island \Circ{1}.}
\acrodef{im2}{mirrors the naturally inspired quality that although two populations have the same physical separation, it may be easier to migrate in one direction than the other, i.e. fish migration is easier downstream than upstream.}
\acrodef{DBEdescription}{A wealthy ecosystem sees a balance between co-operation and competition in a dynamic free market.}
\acrodef{serviceCap}{lightweight entity consisting primarily of a pointer to the \ac{SWS} it represents,}
\acrodef{service2cap}{executable component and semantic description. A software service can be a software only service, e.g. data encryption, or provide a front-end to a real-world service, e.g. selling books}
\acrodef{structureCap}{The executable component of a \ac{SWS}, that an Agent represents \added{via a pointer to the \ac{SWS}}, is equivalent to an organism's DNA and is the gene (functional unit) in the evolutionary process \cite{lawrence1989hsd}. So, the Agents should be aggregated as a sequence, like the sequencing of genes in DNA \cite{lawrence1989hsd}\added{, such that an aggregation of Agents is a sequence of Agents}.}
\acrodef{structure2}{an unordered set, or, based on service orchestration, into a tree or workflow}
\acrodef{habnet}{the {agent stations} from mobile agent systems \cite{agentStation} (to provide a distributed environment in which Agent migration can occur), with evolutionary computing \cite{eiben2003iec} for the Agent interaction (instead of traditional agent interaction mechanisms \cite{wooldridge}), and the island-model of \acl*{DEC} \cite{lin1994cgp} for the connectivity between Habitats}
\acrodef{picUser}{will formulate queries to the Digital Ecosystem by creating a request as a {semantic description}, like those being used and developed in \acp{SOA} \cite{SOAsemantic}, specifying an application they desire and submitting it to their Habitat}
\acrodef{picUserReq}{A Population is then instantiated in the user's Habitat in response to the user's request, seeded from the Agents available at their Habitat}
\acrodef{digEco}{with the Agents, the Populations, the Agent migration for \acl*{DEC}, and the environmental selection pressures provided by the user base, then the union of the Habitats creates the Digital Ecosystem}
\acrodef{archComTop}{many strongly connected clusters (communities), called sub-networks (quasi-complete graphs), with a few connections between these clusters (communities) \cite{swn1}. Graphs with this topology have a very high clustering coefficient and small characteristic path lengths \cite{swn1}.}
\acrodef{bizEcoCap}{As the connections between Habitats are reconfigured depending on the connectivity of the user base, the Habitat clustering will therefore be parallel to the business sector communities}
\acrodef{similarCap}{requests are evaluated on separate {islands} (Populations), with their evolution accelerated by the sharing of solutions between the evolving Populations (islands), because they are working to solve similar requests (problems).}
\acrodef{similar2}{yellow lines connecting the evolving Populations indicate similarity in the requests being managed.}
\acrodef{lifeCycleCap}{with deployment to its owner's Habitat for distribution within the Habitat network.}
\acrodef{lifeCycle2}{used in evolving the optimal Agent-sequence in response to a user request. The optimal Agent-sequence is then registered at the Habitat}
\acrodef{lifeCycle3}{If an Agent-sequence solution is then executed, an attempt is made to migrate (copy) it to every other connected Habitat, success depending on the probability associated with the connection.}
\acrodef{as3}{with an abstract representation consisting of a set of}
\acrodef{agentSemantic2}{tuple representing an {attribute} of the {semantic description}, one integer for the {attribute identifier} and one for the {attribute value}, with both ranging between one and a hundred.}
\acrodef{as4}{Each simulated Agent had a semantic description}
\acrodef{semanticRequest}{A simulated user request consisted of an abstract {semantic description}, as a list of sets of numeric tuples to represent the properties of a desired business application}
\acrodef{bmlcap1}{the numerical semantic descriptions, of the simulated services (Agents) and user requests, in a human readable form.}
\acrodef{capbml2}{translated numerical semantic descriptions for one community within the user base, showing it in the context of the travel industry}
\acrodef{capbml3}{The simulation still operated on the numerical representation for operational efficiency, but the semantic filter essentially assigned meaning to the numbers.}
\acrodef{succession}{So, it becomes increasingly more complex through this process of succession, driven by the evolution of the populations within the ecosystem \cite{connell111msn}.}
\acrodef{succession2}{The formation of a mature ecosystem}
\acrodef{succession3}{is the slow, predictable, and orderly changes in the composition and structure of an ecological community, for which there are defined stages in the increasing complexity \cite{begon96}, as shown}
\acrodef{DigEcoSuc2Cap}{the end of the simulation run, the Agent-sequences had evolved and migrated over an average of only ten user requests per Habitat, and collectively had already reached near 70\% effectiveness for the user base.}
\acrodef{DigEcoSucCap}{The formation of a mature biological ecosystem, ecological succession, is a relatively slow process \cite{begon96}, and the simulated Digital Ecosystem acted similarly in reaching a mature state.}
\acrodef{speciesAbundance}{is a measure of the proportion of all organisms in a community belonging to a particular species \cite{Bell}. A relative abundance distribution provides the inequalities in population size within an ecosystem and therefore an indicator of biodiversity, with the distribution of most biological ecosystems taking a log-normal form \cite{Bell}.}
\acrodef{specAbund2}{the Digital Ecosystem did not conform to the expected log-normal}
\acrodef{speciesArea}{In ecology the {species-area} relationship measures diversity relative to the spatial scale, showing the number of species found in a defined area of a particular habitat or habitats of different areas \cite{sizling2004pls}, and is commonly found to follow a power law in biological ecosystems}
\acrodef{ecoCapClass}{then the Digital Ecosystem and biological ecosystem classes would both inherit from the abstract ecosystem class, but implement its attributes differently}
\acrodef{ecoCap2Class}{So, we would argue that the apparent compromises in mimicking biological ecosystems are actually features unique to Digital Ecosystems.}
\acrodef{bidpCap}{starts with identifying some useful behaviour in a biological system. Next, the system is studied to isolate the mechanisms by which it performs the useful behaviour. Last, the mechanisms are mimicked in a computational system and their performance evaluated \cite{bidpSpiral}.}
\acrodef{introEcoCap}{made up of one or more communities of organisms, consisting of species in their habitats, with their populations existing in their respective micro-habitats \cite{begon96}. A community is a naturally occurring group of populations from different species that live together in the same habitat. A habitat is a distinct part of the environment \cite{begon96}}
\acrodef{im1}{are different probabilities of going from island \Circ{1} to island \Circ{2}, as there is of going from island \Circ{2} to island \Circ{1}.}
\acrodef{im2}{mirrors the naturally inspired quality that although two populations have the same physical separation, it may be easier to migrate in one direction than the other, i.e. fish migration is easier downstream than upstream.}
\acrodef{DBEdescription}{A wealthy ecosystem sees a balance between co-operation and competition in a dynamic free market.}
\acrodef{serviceCap}{lightweight entity consisting primarily of a pointer to the \ac{SWS} it represents,}
\acrodef{service2cap}{executable component and semantic description. A software service can be a software only service, e.g. data encryption, or provide a front-end to a real-world service, e.g. selling books}
\acrodef{structureCap}{The executable component of a \ac{SWS}, that an Agent represents \added{via a pointer to the \ac{SWS}}, is equivalent to an organism's DNA and is the gene (functional unit) in the evolutionary process \cite{lawrence1989hsd}. So, the Agents should be aggregated as a sequence, like the sequencing of genes in DNA \cite{lawrence1989hsd}\added{, such that an aggregation of Agents is a sequence of Agents}.}
\acrodef{structure2}{an unordered set, or, based on service orchestration, into a tree or workflow}
\acrodef{habnet}{the {agent stations} from mobile agent systems \cite{agentStation} (to provide a distributed environment in which Agent migration can occur), with evolutionary computing \cite{eiben2003iec} for the Agent interaction (instead of traditional agent interaction mechanisms \cite{wooldridge}), and the island-model of \acl*{DEC} \cite{lin1994cgp} for the connectivity between Habitats}
\acrodef{picUser}{will formulate queries to the Digital Ecosystem by creating a request as a {semantic description}, like those being used and developed in \acp{SOA} \cite{SOAsemantic}, specifying an application they desire and submitting it to their Habitat}
\acrodef{picUserReq}{A Population is then instantiated in the user's Habitat in response to the user's request, seeded from the Agents available at their Habitat}
\acrodef{digEco}{with the Agents, the Populations, the Agent migration for \acl*{DEC}, and the environmental selection pressures provided by the user base, then the union of the Habitats creates the Digital Ecosystem}
\acrodef{archComTop}{many strongly connected clusters (communities), called sub-networks (quasi-complete graphs), with a few connections between these clusters (communities) \cite{swn1}. Graphs with this topology have a very high clustering coefficient and small characteristic path lengths \cite{swn1}.}
\acrodef{bizEcoCap}{As the connections between Habitats are reconfigured depending on the connectivity of the user base, the Habitat clustering will therefore be parallel to the business sector communities}
\acrodef{similarCap}{requests are evaluated on separate {islands} (Populations), with their evolution accelerated by the sharing of solutions between the evolving Populations (islands), because they are working to solve similar requests (problems).}
\acrodef{similar2}{yellow lines connecting the evolving Populations indicate similarity in the requests being managed.}
\acrodef{lifeCycleCap}{with deployment to its owner's Habitat for distribution within the Habitat network.}
\acrodef{lifeCycle2}{used in evolving the optimal Agent-sequence in response to a user request. The optimal Agent-sequence is then registered at the Habitat}
\acrodef{lifeCycle3}{If an Agent-sequence solution is then executed, an attempt is made to migrate (copy) it to every other connected Habitat, success depending on the probability associated with the connection.}
\acrodef{as3}{with an abstract representation consisting of a set of}
\acrodef{agentSemantic2}{tuple representing an {attribute} of the {semantic description}, one integer for the {attribute identifier} and one for the {attribute value}, with both ranging between one and a hundred.}
\acrodef{as4}{Each simulated Agent had a semantic description}
\acrodef{semanticRequest}{A simulated user request consisted of an abstract {semantic description}, as a list of sets of numeric tuples to represent the properties of a desired business application}
\acrodef{bmlcap1}{the numerical semantic descriptions, of the simulated services (Agents) and user requests, in a human readable form.}
\acrodef{capbml2}{translated numerical semantic descriptions for one community within the user base, showing it in the context of the travel industry}
\acrodef{capbml3}{The simulation still operated on the numerical representation for operational efficiency, but the semantic filter essentially assigned meaning to the numbers.}
\acrodef{succession}{So, it becomes increasingly more complex through this process of succession, driven by the evolution of the populations within the ecosystem \cite{connell111msn}.}
\acrodef{succession2}{The formation of a mature ecosystem}
\acrodef{succession3}{is the slow, predictable, and orderly changes in the composition and structure of an ecological community, for which there are defined stages in the increasing complexity \cite{begon96}, as shown}
\acrodef{DigEcoSuc2Cap}{the end of the simulation run, the Agent-sequences had evolved and migrated over an average of only ten user requests per Habitat, and collectively had already reached near 70\% effectiveness for the user base.}
\acrodef{DigEcoSucCap}{The formation of a mature biological ecosystem, ecological succession, is a relatively slow process \cite{begon96}, and the simulated Digital Ecosystem acted similarly in reaching a mature state.}
\acrodef{speciesAbundance}{is a measure of the proportion of all organisms in a community belonging to a particular species \cite{Bell}. A relative abundance distribution provides the inequalities in population size within an ecosystem and therefore an indicator of biodiversity, with the distribution of most biological ecosystems taking a log-normal form \cite{Bell}.}
\acrodef{specAbund2}{the Digital Ecosystem did not conform to the expected log-normal}
\acrodef{speciesArea}{In ecology the {species-area} relationship measures diversity relative to the spatial scale, showing the number of species found in a defined area of a particular habitat or habitats of different areas \cite{sizling2004pls}, and is commonly found to follow a power law in biological ecosystems}
\acrodef{ecoCapClass}{then the Digital Ecosystem and biological ecosystem classes would both inherit from the abstract ecosystem class, but implement its attributes differently}
\acrodef{ecoCap2Class}{So, we would argue that the apparent compromises in mimicking biological ecosystems are actually features unique to Digital Ecosystems.}
\acrodef{bidpCap}{starts with identifying some useful behaviour in a biological system. Next, the system is studied to isolate the mechanisms by which it performs the useful behaviour. Last, the mechanisms are mimicked in a computational system and their performance evaluated \cite{bidpSpiral}.}
\acrodef{introEcoCap}{made up of one or more communities of organisms, consisting of species in their habitats, with their populations existing in their respective micro-habitats \cite{begon96}. A community is a naturally occurring group of populations from different species that live together in the same habitat. A habitat is a distinct part of the environment \cite{begon96}}
\acrodef{im1}{are different probabilities of going from island \Circ{1} to island \Circ{2}, as there is of going from island \Circ{2} to island \Circ{1}.}
\acrodef{im2}{mirrors the naturally inspired quality that although two populations have the same physical separation, it may be easier to migrate in one direction than the other, i.e. fish migration is easier downstream than upstream.}
\acrodef{DBEdescription}{A wealthy ecosystem sees a balance between co-operation and competition in a dynamic free market.}
\acrodef{serviceCap}{lightweight entity consisting primarily of a pointer to the \ac{SWS} it represents,}
\acrodef{service2cap}{executable component and semantic description. A software service can be a software only service, e.g. data encryption, or provide a front-end to a real-world service, e.g. selling books}
\acrodef{structureCap}{The executable component of a \ac{SWS}, that an Agent represents \added{via a pointer to the \ac{SWS}}, is equivalent to an organism's DNA and is the gene (functional unit) in the evolutionary process \cite{lawrence1989hsd}. So, the Agents should be aggregated as a sequence, like the sequencing of genes in DNA \cite{lawrence1989hsd}\added{, such that an aggregation of Agents is a sequence of Agents}.}
\acrodef{structure2}{an unordered set, or, based on service orchestration, into a tree or workflow}
\acrodef{habnet}{the {agent stations} from mobile agent systems \cite{agentStation} (to provide a distributed environment in which Agent migration can occur), with evolutionary computing \cite{eiben2003iec} for the Agent interaction (instead of traditional agent interaction mechanisms \cite{wooldridge}), and the island-model of \acl*{DEC} \cite{lin1994cgp} for the connectivity between Habitats}
\acrodef{picUser}{will formulate queries to the Digital Ecosystem by creating a request as a {semantic description}, like those being used and developed in \acp{SOA} \cite{SOAsemantic}, specifying an application they desire and submitting it to their Habitat}
\acrodef{picUserReq}{A Population is then instantiated in the user's Habitat in response to the user's request, seeded from the Agents available at their Habitat}
\acrodef{digEco}{with the Agents, the Populations, the Agent migration for \acl*{DEC}, and the environmental selection pressures provided by the user base, then the union of the Habitats creates the Digital Ecosystem}
\acrodef{archComTop}{many strongly connected clusters (communities), called sub-networks (quasi-complete graphs), with a few connections between these clusters (communities) \cite{swn1}. Graphs with this topology have a very high clustering coefficient and small characteristic path lengths \cite{swn1}.}
\acrodef{bizEcoCap}{As the connections between Habitats are reconfigured depending on the connectivity of the user base, the Habitat clustering will therefore be parallel to the business sector communities}
\acrodef{similarCap}{requests are evaluated on separate {islands} (Populations), with their evolution accelerated by the sharing of solutions between the evolving Populations (islands), because they are working to solve similar requests (problems).}
\acrodef{similar2}{yellow lines connecting the evolving Populations indicate similarity in the requests being managed.}
\acrodef{lifeCycleCap}{with deployment to its owner's Habitat for distribution within the Habitat network.}
\acrodef{lifeCycle2}{used in evolving the optimal Agent-sequence in response to a user request. The optimal Agent-sequence is then registered at the Habitat}
\acrodef{lifeCycle3}{If an Agent-sequence solution is then executed, an attempt is made to migrate (copy) it to every other connected Habitat, success depending on the probability associated with the connection.}
\acrodef{as3}{with an abstract representation consisting of a set of}
\acrodef{agentSemantic2}{tuple representing an {attribute} of the {semantic description}, one integer for the {attribute identifier} and one for the {attribute value}, with both ranging between one and a hundred.}
\acrodef{as4}{Each simulated Agent had a semantic description}
\acrodef{semanticRequest}{A simulated user request consisted of an abstract {semantic description}, as a list of sets of numeric tuples to represent the properties of a desired business application}
\acrodef{bmlcap1}{the numerical semantic descriptions, of the simulated services (Agents) and user requests, in a human readable form.}
\acrodef{capbml2}{translated numerical semantic descriptions for one community within the user base, showing it in the context of the travel industry}
\acrodef{capbml3}{The simulation still operated on the numerical representation for operational efficiency, but the semantic filter essentially assigned meaning to the numbers.}
\acrodef{succession}{So, it becomes increasingly more complex through this process of succession, driven by the evolution of the populations within the ecosystem \cite{connell111msn}.}
\acrodef{succession2}{The formation of a mature ecosystem}
\acrodef{succession3}{is the slow, predictable, and orderly changes in the composition and structure of an ecological community, for which there are defined stages in the increasing complexity \cite{begon96}, as shown}
\acrodef{DigEcoSuc2Cap}{the end of the simulation run, the Agent-sequences had evolved and migrated over an average of only ten user requests per Habitat, and collectively had already reached near 70\% effectiveness for the user base.}
\acrodef{DigEcoSucCap}{The formation of a mature biological ecosystem, ecological succession, is a relatively slow process \cite{begon96}, and the simulated Digital Ecosystem acted similarly in reaching a mature state.}
\acrodef{speciesAbundance}{is a measure of the proportion of all organisms in a community belonging to a particular species \cite{Bell}. A relative abundance distribution provides the inequalities in population size within an ecosystem and therefore an indicator of biodiversity, with the distribution of most biological ecosystems taking a log-normal form \cite{Bell}.}
\acrodef{specAbund2}{the Digital Ecosystem did not conform to the expected log-normal}
\acrodef{speciesArea}{In ecology the {species-area} relationship measures diversity relative to the spatial scale, showing the number of species found in a defined area of a particular habitat or habitats of different areas \cite{sizling2004pls}, and is commonly found to follow a power law in biological ecosystems}
\acrodef{ecoCapClass}{then the Digital Ecosystem and biological ecosystem classes would both inherit from the abstract ecosystem class, but implement its attributes differently}
\acrodef{ecoCap2Class}{So, we would argue that the apparent compromises in mimicking biological ecosystems are actually features unique to Digital Ecosystems.}
\acrodef{bidpCap}{starts with identifying some useful behaviour in a biological system. Next, the system is studied to isolate the mechanisms by which it performs the useful behaviour. Last, the mechanisms are mimicked in a computational system and their performance evaluated \cite{bidpSpiral}.}
\acrodef{introEcoCap}{made up of one or more communities of organisms, consisting of species in their habitats, with their populations existing in their respective micro-habitats \cite{begon96}. A community is a naturally occurring group of populations from different species that live together in the same habitat. A habitat is a distinct part of the environment \cite{begon96}}
\acrodef{im1}{are different probabilities of going from island \Circ{1} to island \Circ{2}, as there is of going from island \Circ{2} to island \Circ{1}.}
\acrodef{im2}{mirrors the naturally inspired quality that although two populations have the same physical separation, it may be easier to migrate in one direction than the other, i.e. fish migration is easier downstream than upstream.}
\acrodef{DBEdescription}{A wealthy ecosystem sees a balance between co-operation and competition in a dynamic free market.}
\acrodef{serviceCap}{lightweight entity consisting primarily of a pointer to the \ac{SWS} it represents,}
\acrodef{service2cap}{executable component and semantic description. A software service can be a software only service, e.g. data encryption, or provide a front-end to a real-world service, e.g. selling books}
\acrodef{structureCap}{The executable component of a \ac{SWS}, that an Agent represents \added{via a pointer to the \ac{SWS}}, is equivalent to an organism's DNA and is the gene (functional unit) in the evolutionary process \cite{lawrence1989hsd}. So, the Agents should be aggregated as a sequence, like the sequencing of genes in DNA \cite{lawrence1989hsd}\added{, such that an aggregation of Agents is a sequence of Agents}.}
\acrodef{structure2}{an unordered set, or, based on service orchestration, into a tree or workflow}
\acrodef{habnet}{the {agent stations} from mobile agent systems \cite{agentStation} (to provide a distributed environment in which Agent migration can occur), with evolutionary computing \cite{eiben2003iec} for the Agent interaction (instead of traditional agent interaction mechanisms \cite{wooldridge}), and the island-model of \acl*{DEC} \cite{lin1994cgp} for the connectivity between Habitats}
\acrodef{picUser}{will formulate queries to the Digital Ecosystem by creating a request as a {semantic description}, like those being used and developed in \acp{SOA} \cite{SOAsemantic}, specifying an application they desire and submitting it to their Habitat}
\acrodef{picUserReq}{A Population is then instantiated in the user's Habitat in response to the user's request, seeded from the Agents available at their Habitat}
\acrodef{digEco}{with the Agents, the Populations, the Agent migration for \acl*{DEC}, and the environmental selection pressures provided by the user base, then the union of the Habitats creates the Digital Ecosystem}
\acrodef{archComTop}{many strongly connected clusters (communities), called sub-networks (quasi-complete graphs), with a few connections between these clusters (communities) \cite{swn1}. Graphs with this topology have a very high clustering coefficient and small characteristic path lengths \cite{swn1}.}
\acrodef{bizEcoCap}{As the connections between Habitats are reconfigured depending on the connectivity of the user base, the Habitat clustering will therefore be parallel to the business sector communities}
\acrodef{similarCap}{requests are evaluated on separate {islands} (Populations), with their evolution accelerated by the sharing of solutions between the evolving Populations (islands), because they are working to solve similar requests (problems).}
\acrodef{similar2}{yellow lines connecting the evolving Populations indicate similarity in the requests being managed.}
\acrodef{lifeCycleCap}{with deployment to its owner's Habitat for distribution within the Habitat network.}
\acrodef{lifeCycle2}{used in evolving the optimal Agent-sequence in response to a user request. The optimal Agent-sequence is then registered at the Habitat}
\acrodef{lifeCycle3}{If an Agent-sequence solution is then executed, an attempt is made to migrate (copy) it to every other connected Habitat, success depending on the probability associated with the connection.}
\acrodef{as3}{with an abstract representation consisting of a set of}
\acrodef{agentSemantic2}{tuple representing an {attribute} of the {semantic description}, one integer for the {attribute identifier} and one for the {attribute value}, with both ranging between one and a hundred.}
\acrodef{as4}{Each simulated Agent had a semantic description}
\acrodef{semanticRequest}{A simulated user request consisted of an abstract {semantic description}, as a list of sets of numeric tuples to represent the properties of a desired business application}
\acrodef{bmlcap1}{the numerical semantic descriptions, of the simulated services (Agents) and user requests, in a human readable form.}
\acrodef{capbml2}{translated numerical semantic descriptions for one community within the user base, showing it in the context of the travel industry}
\acrodef{capbml3}{The simulation still operated on the numerical representation for operational efficiency, but the semantic filter essentially assigned meaning to the numbers.}
\acrodef{succession}{So, it becomes increasingly more complex through this process of succession, driven by the evolution of the populations within the ecosystem \cite{connell111msn}.}
\acrodef{succession2}{The formation of a mature ecosystem}
\acrodef{succession3}{is the slow, predictable, and orderly changes in the composition and structure of an ecological community, for which there are defined stages in the increasing complexity \cite{begon96}, as shown}
\acrodef{DigEcoSuc2Cap}{the end of the simulation run, the Agent-sequences had evolved and migrated over an average of only ten user requests per Habitat, and collectively had already reached near 70\% effectiveness for the user base.}
\acrodef{DigEcoSucCap}{The formation of a mature biological ecosystem, ecological succession, is a relatively slow process \cite{begon96}, and the simulated Digital Ecosystem acted similarly in reaching a mature state.}
\acrodef{speciesAbundance}{is a measure of the proportion of all organisms in a community belonging to a particular species \cite{Bell}. A relative abundance distribution provides the inequalities in population size within an ecosystem and therefore an indicator of biodiversity, with the distribution of most biological ecosystems taking a log-normal form \cite{Bell}.}
\acrodef{specAbund2}{the Digital Ecosystem did not conform to the expected log-normal}
\acrodef{speciesArea}{In ecology the {species-area} relationship measures diversity relative to the spatial scale, showing the number of species found in a defined area of a particular habitat or habitats of different areas \cite{sizling2004pls}, and is commonly found to follow a power law in biological ecosystems}
\acrodef{ecoCapClass}{then the Digital Ecosystem and biological ecosystem classes would both inherit from the abstract ecosystem class, but implement its attributes differently}
\acrodef{ecoCap2Class}{So, we would argue that the apparent compromises in mimicking biological ecosystems are actually features unique to Digital Ecosystems.}
\acrodef{bidpCap}{starts with identifying some useful behaviour in a biological system. Next, the system is studied to isolate the mechanisms by which it performs the useful behaviour. Last, the mechanisms are mimicked in a computational system and their performance evaluated \cite{bidpSpiral}.}
\acrodef{introEcoCap}{made up of one or more communities of organisms, consisting of species in their habitats, with their populations existing in their respective micro-habitats \cite{begon96}. A community is a naturally occurring group of populations from different species that live together in the same habitat. A habitat is a distinct part of the environment \cite{begon96}}
\acrodef{im1}{are different probabilities of going from island \Circ{1} to island \Circ{2}, as there is of going from island \Circ{2} to island \Circ{1}.}
\acrodef{im2}{mirrors the naturally inspired quality that although two populations have the same physical separation, it may be easier to migrate in one direction than the other, i.e. fish migration is easier downstream than upstream.}
\acrodef{DBEdescription}{A wealthy ecosystem sees a balance between co-operation and competition in a dynamic free market.}
\acrodef{serviceCap}{lightweight entity consisting primarily of a pointer to the \ac{SWS} it represents,}
\acrodef{service2cap}{executable component and semantic description. A software service can be a software only service, e.g. data encryption, or provide a front-end to a real-world service, e.g. selling books}
\acrodef{structureCap}{The executable component of a \ac{SWS}, that an Agent represents \added{via a pointer to the \ac{SWS}}, is equivalent to an organism's DNA and is the gene (functional unit) in the evolutionary process \cite{lawrence1989hsd}. So, the Agents should be aggregated as a sequence, like the sequencing of genes in DNA \cite{lawrence1989hsd}\added{, such that an aggregation of Agents is a sequence of Agents}.}
\acrodef{structure2}{an unordered set, or, based on service orchestration, into a tree or workflow}
\acrodef{habnet}{the {agent stations} from mobile agent systems \cite{agentStation} (to provide a distributed environment in which Agent migration can occur), with evolutionary computing \cite{eiben2003iec} for the Agent interaction (instead of traditional agent interaction mechanisms \cite{wooldridge}), and the island-model of \acl*{DEC} \cite{lin1994cgp} for the connectivity between Habitats}
\acrodef{picUser}{will formulate queries to the Digital Ecosystem by creating a request as a {semantic description}, like those being used and developed in \acp{SOA} \cite{SOAsemantic}, specifying an application they desire and submitting it to their Habitat}
\acrodef{picUserReq}{A Population is then instantiated in the user's Habitat in response to the user's request, seeded from the Agents available at their Habitat}
\acrodef{digEco}{with the Agents, the Populations, the Agent migration for \acl*{DEC}, and the environmental selection pressures provided by the user base, then the union of the Habitats creates the Digital Ecosystem}
\acrodef{archComTop}{many strongly connected clusters (communities), called sub-networks (quasi-complete graphs), with a few connections between these clusters (communities) \cite{swn1}. Graphs with this topology have a very high clustering coefficient and small characteristic path lengths \cite{swn1}.}
\acrodef{bizEcoCap}{As the connections between Habitats are reconfigured depending on the connectivity of the user base, the Habitat clustering will therefore be parallel to the business sector communities}
\acrodef{similarCap}{requests are evaluated on separate {islands} (Populations), with their evolution accelerated by the sharing of solutions between the evolving Populations (islands), because they are working to solve similar requests (problems).}
\acrodef{similar2}{yellow lines connecting the evolving Populations indicate similarity in the requests being managed.}
\acrodef{lifeCycleCap}{with deployment to its owner's Habitat for distribution within the Habitat network.}
\acrodef{lifeCycle2}{used in evolving the optimal Agent-sequence in response to a user request. The optimal Agent-sequence is then registered at the Habitat}
\acrodef{lifeCycle3}{If an Agent-sequence solution is then executed, an attempt is made to migrate (copy) it to every other connected Habitat, success depending on the probability associated with the connection.}
\acrodef{as3}{with an abstract representation consisting of a set of}
\acrodef{agentSemantic2}{tuple representing an {attribute} of the {semantic description}, one integer for the {attribute identifier} and one for the {attribute value}, with both ranging between one and a hundred.}
\acrodef{as4}{Each simulated Agent had a semantic description}
\acrodef{semanticRequest}{A simulated user request consisted of an abstract {semantic description}, as a list of sets of numeric tuples to represent the properties of a desired business application}
\acrodef{bmlcap1}{the numerical semantic descriptions, of the simulated services (Agents) and user requests, in a human readable form.}
\acrodef{capbml2}{translated numerical semantic descriptions for one community within the user base, showing it in the context of the travel industry}
\acrodef{capbml3}{The simulation still operated on the numerical representation for operational efficiency, but the semantic filter essentially assigned meaning to the numbers.}
\acrodef{succession}{So, it becomes increasingly more complex through this process of succession, driven by the evolution of the populations within the ecosystem \cite{connell111msn}.}
\acrodef{succession2}{The formation of a mature ecosystem}
\acrodef{succession3}{is the slow, predictable, and orderly changes in the composition and structure of an ecological community, for which there are defined stages in the increasing complexity \cite{begon96}, as shown}
\acrodef{DigEcoSuc2Cap}{the end of the simulation run, the Agent-sequences had evolved and migrated over an average of only ten user requests per Habitat, and collectively had already reached near 70\% effectiveness for the user base.}
\acrodef{DigEcoSucCap}{The formation of a mature biological ecosystem, ecological succession, is a relatively slow process \cite{begon96}, and the simulated Digital Ecosystem acted similarly in reaching a mature state.}
\acrodef{speciesAbundance}{is a measure of the proportion of all organisms in a community belonging to a particular species \cite{Bell}. A relative abundance distribution provides the inequalities in population size within an ecosystem and therefore an indicator of biodiversity, with the distribution of most biological ecosystems taking a log-normal form \cite{Bell}.}
\acrodef{specAbund2}{the Digital Ecosystem did not conform to the expected log-normal}
\acrodef{speciesArea}{In ecology the {species-area} relationship measures diversity relative to the spatial scale, showing the number of species found in a defined area of a particular habitat or habitats of different areas \cite{sizling2004pls}, and is commonly found to follow a power law in biological ecosystems}
\acrodef{ecoCapClass}{then the Digital Ecosystem and biological ecosystem classes would both inherit from the abstract ecosystem class, but implement its attributes differently}
\acrodef{ecoCap2Class}{So, we would argue that the apparent compromises in mimicking biological ecosystems are actually features unique to Digital Ecosystems.}
\acrodef{bidpCap}{starts with identifying some useful behaviour in a biological system. Next, the system is studied to isolate the mechanisms by which it performs the useful behaviour. Last, the mechanisms are mimicked in a computational system and their performance evaluated \cite{bidpSpiral}.}
\acrodef{introEcoCap}{made up of one or more communities of organisms, consisting of species in their habitats, with their populations existing in their respective micro-habitats \cite{begon96}. A community is a naturally occurring group of populations from different species that live together in the same habitat. A habitat is a distinct part of the environment \cite{begon96}}
\acrodef{im1}{are different probabilities of going from island \Circ{1} to island \Circ{2}, as there is of going from island \Circ{2} to island \Circ{1}.}
\acrodef{im2}{mirrors the naturally inspired quality that although two populations have the same physical separation, it may be easier to migrate in one direction than the other, i.e. fish migration is easier downstream than upstream.}
\acrodef{DBEdescription}{A wealthy ecosystem sees a balance between co-operation and competition in a dynamic free market.}
\acrodef{serviceCap}{lightweight entity consisting primarily of a pointer to the \ac{SWS} it represents,}
\acrodef{service2cap}{executable component and semantic description. A software service can be a software only service, e.g. data encryption, or provide a front-end to a real-world service, e.g. selling books}
\acrodef{structureCap}{The executable component of a \ac{SWS}, that an Agent represents \added{via a pointer to the \ac{SWS}}, is equivalent to an organism's DNA and is the gene (functional unit) in the evolutionary process \cite{lawrence1989hsd}. So, the Agents should be aggregated as a sequence, like the sequencing of genes in DNA \cite{lawrence1989hsd}\added{, such that an aggregation of Agents is a sequence of Agents}.}
\acrodef{structure2}{an unordered set, or, based on service orchestration, into a tree or workflow}
\acrodef{habnet}{the {agent stations} from mobile agent systems \cite{agentStation} (to provide a distributed environment in which Agent migration can occur), with evolutionary computing \cite{eiben2003iec} for the Agent interaction (instead of traditional agent interaction mechanisms \cite{wooldridge}), and the island-model of \acl*{DEC} \cite{lin1994cgp} for the connectivity between Habitats}
\acrodef{picUser}{will formulate queries to the Digital Ecosystem by creating a request as a {semantic description}, like those being used and developed in \acp{SOA} \cite{SOAsemantic}, specifying an application they desire and submitting it to their Habitat}
\acrodef{picUserReq}{A Population is then instantiated in the user's Habitat in response to the user's request, seeded from the Agents available at their Habitat}
\acrodef{digEco}{with the Agents, the Populations, the Agent migration for \acl*{DEC}, and the environmental selection pressures provided by the user base, then the union of the Habitats creates the Digital Ecosystem}
\acrodef{archComTop}{many strongly connected clusters (communities), called sub-networks (quasi-complete graphs), with a few connections between these clusters (communities) \cite{swn1}. Graphs with this topology have a very high clustering coefficient and small characteristic path lengths \cite{swn1}.}
\acrodef{bizEcoCap}{As the connections between Habitats are reconfigured depending on the connectivity of the user base, the Habitat clustering will therefore be parallel to the business sector communities}
\acrodef{similarCap}{requests are evaluated on separate {islands} (Populations), with their evolution accelerated by the sharing of solutions between the evolving Populations (islands), because they are working to solve similar requests (problems).}
\acrodef{similar2}{yellow lines connecting the evolving Populations indicate similarity in the requests being managed.}
\acrodef{lifeCycleCap}{with deployment to its owner's Habitat for distribution within the Habitat network.}
\acrodef{lifeCycle2}{used in evolving the optimal Agent-sequence in response to a user request. The optimal Agent-sequence is then registered at the Habitat}
\acrodef{lifeCycle3}{If an Agent-sequence solution is then executed, an attempt is made to migrate (copy) it to every other connected Habitat, success depending on the probability associated with the connection.}
\acrodef{as3}{with an abstract representation consisting of a set of}
\acrodef{agentSemantic2}{tuple representing an {attribute} of the {semantic description}, one integer for the {attribute identifier} and one for the {attribute value}, with both ranging between one and a hundred.}
\acrodef{as4}{Each simulated Agent had a semantic description}
\acrodef{semanticRequest}{A simulated user request consisted of an abstract {semantic description}, as a list of sets of numeric tuples to represent the properties of a desired business application}
\acrodef{bmlcap1}{the numerical semantic descriptions, of the simulated services (Agents) and user requests, in a human readable form.}
\acrodef{capbml2}{translated numerical semantic descriptions for one community within the user base, showing it in the context of the travel industry}
\acrodef{capbml3}{The simulation still operated on the numerical representation for operational efficiency, but the semantic filter essentially assigned meaning to the numbers.}
\acrodef{succession}{So, it becomes increasingly more complex through this process of succession, driven by the evolution of the populations within the ecosystem \cite{connell111msn}.}
\acrodef{succession2}{The formation of a mature ecosystem}
\acrodef{succession3}{is the slow, predictable, and orderly changes in the composition and structure of an ecological community, for which there are defined stages in the increasing complexity \cite{begon96}, as shown}
\acrodef{DigEcoSuc2Cap}{the end of the simulation run, the Agent-sequences had evolved and migrated over an average of only ten user requests per Habitat, and collectively had already reached near 70\% effectiveness for the user base.}
\acrodef{DigEcoSucCap}{The formation of a mature biological ecosystem, ecological succession, is a relatively slow process \cite{begon96}, and the simulated Digital Ecosystem acted similarly in reaching a mature state.}
\acrodef{speciesAbundance}{is a measure of the proportion of all organisms in a community belonging to a particular species \cite{Bell}. A relative abundance distribution provides the inequalities in population size within an ecosystem and therefore an indicator of biodiversity, with the distribution of most biological ecosystems taking a log-normal form \cite{Bell}.}
\acrodef{specAbund2}{the Digital Ecosystem did not conform to the expected log-normal}
\acrodef{speciesArea}{In ecology the {species-area} relationship measures diversity relative to the spatial scale, showing the number of species found in a defined area of a particular habitat or habitats of different areas \cite{sizling2004pls}, and is commonly found to follow a power law in biological ecosystems}
\acrodef{ecoCapClass}{then the Digital Ecosystem and biological ecosystem classes would both inherit from the abstract ecosystem class, but implement its attributes differently}
\acrodef{ecoCap2Class}{So, we would argue that the apparent compromises in mimicking biological ecosystems are actually features unique to Digital Ecosystems.}
\acrodef{bidpCap}{starts with identifying some useful behaviour in a biological system. Next, the system is studied to isolate the mechanisms by which it performs the useful behaviour. Last, the mechanisms are mimicked in a computational system and their performance evaluated \cite{bidpSpiral}.}
\acrodef{introEcoCap}{made up of one or more communities of organisms, consisting of species in their habitats, with their populations existing in their respective micro-habitats \cite{begon96}. A community is a naturally occurring group of populations from different species that live together in the same habitat. A habitat is a distinct part of the environment \cite{begon96}}
\acrodef{im1}{are different probabilities of going from island \Circ{1} to island \Circ{2}, as there is of going from island \Circ{2} to island \Circ{1}.}
\acrodef{im2}{mirrors the naturally inspired quality that although two populations have the same physical separation, it may be easier to migrate in one direction than the other, i.e. fish migration is easier downstream than upstream.}
\acrodef{DBEdescription}{A wealthy ecosystem sees a balance between co-operation and competition in a dynamic free market.}
\acrodef{serviceCap}{lightweight entity consisting primarily of a pointer to the \ac{SWS} it represents,}
\acrodef{service2cap}{executable component and semantic description. A software service can be a software only service, e.g. data encryption, or provide a front-end to a real-world service, e.g. selling books}
\acrodef{structureCap}{The executable component of a \ac{SWS}, that an Agent represents \added{via a pointer to the \ac{SWS}}, is equivalent to an organism's DNA and is the gene (functional unit) in the evolutionary process \cite{lawrence1989hsd}. So, the Agents should be aggregated as a sequence, like the sequencing of genes in DNA \cite{lawrence1989hsd}\added{, such that an aggregation of Agents is a sequence of Agents}.}
\acrodef{structure2}{an unordered set, or, based on service orchestration, into a tree or workflow}
\acrodef{habnet}{the {agent stations} from mobile agent systems \cite{agentStation} (to provide a distributed environment in which Agent migration can occur), with evolutionary computing \cite{eiben2003iec} for the Agent interaction (instead of traditional agent interaction mechanisms \cite{wooldridge}), and the island-model of \acl*{DEC} \cite{lin1994cgp} for the connectivity between Habitats}
\acrodef{picUser}{will formulate queries to the Digital Ecosystem by creating a request as a {semantic description}, like those being used and developed in \acp{SOA} \cite{SOAsemantic}, specifying an application they desire and submitting it to their Habitat}
\acrodef{picUserReq}{A Population is then instantiated in the user's Habitat in response to the user's request, seeded from the Agents available at their Habitat}
\acrodef{digEco}{with the Agents, the Populations, the Agent migration for \acl*{DEC}, and the environmental selection pressures provided by the user base, then the union of the Habitats creates the Digital Ecosystem}
\acrodef{archComTop}{many strongly connected clusters (communities), called sub-networks (quasi-complete graphs), with a few connections between these clusters (communities) \cite{swn1}. Graphs with this topology have a very high clustering coefficient and small characteristic path lengths \cite{swn1}.}
\acrodef{bizEcoCap}{As the connections between Habitats are reconfigured depending on the connectivity of the user base, the Habitat clustering will therefore be parallel to the business sector communities}
\acrodef{similarCap}{requests are evaluated on separate {islands} (Populations), with their evolution accelerated by the sharing of solutions between the evolving Populations (islands), because they are working to solve similar requests (problems).}
\acrodef{similar2}{yellow lines connecting the evolving Populations indicate similarity in the requests being managed.}
\acrodef{lifeCycleCap}{with deployment to its owner's Habitat for distribution within the Habitat network.}
\acrodef{lifeCycle2}{used in evolving the optimal Agent-sequence in response to a user request. The optimal Agent-sequence is then registered at the Habitat}
\acrodef{lifeCycle3}{If an Agent-sequence solution is then executed, an attempt is made to migrate (copy) it to every other connected Habitat, success depending on the probability associated with the connection.}
\acrodef{as3}{with an abstract representation consisting of a set of}
\acrodef{agentSemantic2}{tuple representing an {attribute} of the {semantic description}, one integer for the {attribute identifier} and one for the {attribute value}, with both ranging between one and a hundred.}
\acrodef{as4}{Each simulated Agent had a semantic description}
\acrodef{semanticRequest}{A simulated user request consisted of an abstract {semantic description}, as a list of sets of numeric tuples to represent the properties of a desired business application}
\acrodef{bmlcap1}{the numerical semantic descriptions, of the simulated services (Agents) and user requests, in a human readable form.}
\acrodef{capbml2}{translated numerical semantic descriptions for one community within the user base, showing it in the context of the travel industry}
\acrodef{capbml3}{The simulation still operated on the numerical representation for operational efficiency, but the semantic filter essentially assigned meaning to the numbers.}
\acrodef{succession}{So, it becomes increasingly more complex through this process of succession, driven by the evolution of the populations within the ecosystem \cite{connell111msn}.}
\acrodef{succession2}{The formation of a mature ecosystem}
\acrodef{succession3}{is the slow, predictable, and orderly changes in the composition and structure of an ecological community, for which there are defined stages in the increasing complexity \cite{begon96}, as shown}
\acrodef{DigEcoSuc2Cap}{the end of the simulation run, the Agent-sequences had evolved and migrated over an average of only ten user requests per Habitat, and collectively had already reached near 70\% effectiveness for the user base.}
\acrodef{DigEcoSucCap}{The formation of a mature biological ecosystem, ecological succession, is a relatively slow process \cite{begon96}, and the simulated Digital Ecosystem acted similarly in reaching a mature state.}
\acrodef{speciesAbundance}{is a measure of the proportion of all organisms in a community belonging to a particular species \cite{Bell}. A relative abundance distribution provides the inequalities in population size within an ecosystem and therefore an indicator of biodiversity, with the distribution of most biological ecosystems taking a log-normal form \cite{Bell}.}
\acrodef{specAbund2}{the Digital Ecosystem did not conform to the expected log-normal}
\acrodef{speciesArea}{In ecology the {species-area} relationship measures diversity relative to the spatial scale, showing the number of species found in a defined area of a particular habitat or habitats of different areas \cite{sizling2004pls}, and is commonly found to follow a power law in biological ecosystems}
\acrodef{ecoCapClass}{then the Digital Ecosystem and biological ecosystem classes would both inherit from the abstract ecosystem class, but implement its attributes differently}
\acrodef{ecoCap2Class}{So, we would argue that the apparent compromises in mimicking biological ecosystems are actually features unique to Digital Ecosystems.}
\acrodef{bidpCap}{starts with identifying some useful behaviour in a biological system. Next, the system is studied to isolate the mechanisms by which it performs the useful behaviour. Last, the mechanisms are mimicked in a computational system and their performance evaluated \cite{bidpSpiral}.}
\acrodef{introEcoCap}{made up of one or more communities of organisms, consisting of species in their habitats, with their populations existing in their respective micro-habitats \cite{begon96}. A community is a naturally occurring group of populations from different species that live together in the same habitat. A habitat is a distinct part of the environment \cite{begon96}}
\acrodef{im1}{are different probabilities of going from island \Circ{1} to island \Circ{2}, as there is of going from island \Circ{2} to island \Circ{1}.}
\acrodef{im2}{mirrors the naturally inspired quality that although two populations have the same physical separation, it may be easier to migrate in one direction than the other, i.e. fish migration is easier downstream than upstream.}
\acrodef{DBEdescription}{A wealthy ecosystem sees a balance between co-operation and competition in a dynamic free market.}
\acrodef{serviceCap}{lightweight entity consisting primarily of a pointer to the \ac{SWS} it represents,}
\acrodef{service2cap}{executable component and semantic description. A software service can be a software only service, e.g. data encryption, or provide a front-end to a real-world service, e.g. selling books}
\acrodef{structureCap}{The executable component of a \ac{SWS}, that an Agent represents \added{via a pointer to the \ac{SWS}}, is equivalent to an organism's DNA and is the gene (functional unit) in the evolutionary process \cite{lawrence1989hsd}. So, the Agents should be aggregated as a sequence, like the sequencing of genes in DNA \cite{lawrence1989hsd}\added{, such that an aggregation of Agents is a sequence of Agents}.}
\acrodef{structure2}{an unordered set, or, based on service orchestration, into a tree or workflow}
\acrodef{habnet}{the {agent stations} from mobile agent systems \cite{agentStation} (to provide a distributed environment in which Agent migration can occur), with evolutionary computing \cite{eiben2003iec} for the Agent interaction (instead of traditional agent interaction mechanisms \cite{wooldridge}), and the island-model of \acl*{DEC} \cite{lin1994cgp} for the connectivity between Habitats}
\acrodef{picUser}{will formulate queries to the Digital Ecosystem by creating a request as a {semantic description}, like those being used and developed in \acp{SOA} \cite{SOAsemantic}, specifying an application they desire and submitting it to their Habitat}
\acrodef{picUserReq}{A Population is then instantiated in the user's Habitat in response to the user's request, seeded from the Agents available at their Habitat}
\acrodef{digEco}{with the Agents, the Populations, the Agent migration for \acl*{DEC}, and the environmental selection pressures provided by the user base, then the union of the Habitats creates the Digital Ecosystem}
\acrodef{archComTop}{many strongly connected clusters (communities), called sub-networks (quasi-complete graphs), with a few connections between these clusters (communities) \cite{swn1}. Graphs with this topology have a very high clustering coefficient and small characteristic path lengths \cite{swn1}.}
\acrodef{bizEcoCap}{As the connections between Habitats are reconfigured depending on the connectivity of the user base, the Habitat clustering will therefore be parallel to the business sector communities}
\acrodef{similarCap}{requests are evaluated on separate {islands} (Populations), with their evolution accelerated by the sharing of solutions between the evolving Populations (islands), because they are working to solve similar requests (problems).}
\acrodef{similar2}{yellow lines connecting the evolving Populations indicate similarity in the requests being managed.}
\acrodef{lifeCycleCap}{with deployment to its owner's Habitat for distribution within the Habitat network.}
\acrodef{lifeCycle2}{used in evolving the optimal Agent-sequence in response to a user request. The optimal Agent-sequence is then registered at the Habitat}
\acrodef{lifeCycle3}{If an Agent-sequence solution is then executed, an attempt is made to migrate (copy) it to every other connected Habitat, success depending on the probability associated with the connection.}
\acrodef{as3}{with an abstract representation consisting of a set of}
\acrodef{agentSemantic2}{tuple representing an {attribute} of the {semantic description}, one integer for the {attribute identifier} and one for the {attribute value}, with both ranging between one and a hundred.}
\acrodef{as4}{Each simulated Agent had a semantic description}
\acrodef{semanticRequest}{A simulated user request consisted of an abstract {semantic description}, as a list of sets of numeric tuples to represent the properties of a desired business application}
\acrodef{bmlcap1}{the numerical semantic descriptions, of the simulated services (Agents) and user requests, in a human readable form.}
\acrodef{capbml2}{translated numerical semantic descriptions for one community within the user base, showing it in the context of the travel industry}
\acrodef{capbml3}{The simulation still operated on the numerical representation for operational efficiency, but the semantic filter essentially assigned meaning to the numbers.}
\acrodef{succession}{So, it becomes increasingly more complex through this process of succession, driven by the evolution of the populations within the ecosystem \cite{connell111msn}.}
\acrodef{succession2}{The formation of a mature ecosystem}
\acrodef{succession3}{is the slow, predictable, and orderly changes in the composition and structure of an ecological community, for which there are defined stages in the increasing complexity \cite{begon96}, as shown}
\acrodef{DigEcoSuc2Cap}{the end of the simulation run, the Agent-sequences had evolved and migrated over an average of only ten user requests per Habitat, and collectively had already reached near 70\% effectiveness for the user base.}
\acrodef{DigEcoSucCap}{The formation of a mature biological ecosystem, ecological succession, is a relatively slow process \cite{begon96}, and the simulated Digital Ecosystem acted similarly in reaching a mature state.}
\acrodef{speciesAbundance}{is a measure of the proportion of all organisms in a community belonging to a particular species \cite{Bell}. A relative abundance distribution provides the inequalities in population size within an ecosystem and therefore an indicator of biodiversity, with the distribution of most biological ecosystems taking a log-normal form \cite{Bell}.}
\acrodef{specAbund2}{the Digital Ecosystem did not conform to the expected log-normal}
\acrodef{speciesArea}{In ecology the {species-area} relationship measures diversity relative to the spatial scale, showing the number of species found in a defined area of a particular habitat or habitats of different areas \cite{sizling2004pls}, and is commonly found to follow a power law in biological ecosystems}
\acrodef{ecoCapClass}{then the Digital Ecosystem and biological ecosystem classes would both inherit from the abstract ecosystem class, but implement its attributes differently}
\acrodef{ecoCap2Class}{So, we would argue that the apparent compromises in mimicking biological ecosystems are actually features unique to Digital Ecosystems.}
\acrodef{bidpCap}{starts with identifying some useful behaviour in a biological system. Next, the system is studied to isolate the mechanisms by which it performs the useful behaviour. Last, the mechanisms are mimicked in a computational system and their performance evaluated \cite{bidpSpiral}.}
\acrodef{introEcoCap}{made up of one or more communities of organisms, consisting of species in their habitats, with their populations existing in their respective micro-habitats \cite{begon96}. A community is a naturally occurring group of populations from different species that live together in the same habitat. A habitat is a distinct part of the environment \cite{begon96}}
\acrodef{im1}{are different probabilities of going from island \Circ{1} to island \Circ{2}, as there is of going from island \Circ{2} to island \Circ{1}.}
\acrodef{im2}{mirrors the naturally inspired quality that although two populations have the same physical separation, it may be easier to migrate in one direction than the other, i.e. fish migration is easier downstream than upstream.}
\acrodef{DBEdescription}{A wealthy ecosystem sees a balance between co-operation and competition in a dynamic free market.}
\acrodef{serviceCap}{lightweight entity consisting primarily of a pointer to the \ac{SWS} it represents,}
\acrodef{service2cap}{executable component and semantic description. A software service can be a software only service, e.g. data encryption, or provide a front-end to a real-world service, e.g. selling books}
\acrodef{structureCap}{The executable component of a \ac{SWS}, that an Agent represents \added{via a pointer to the \ac{SWS}}, is equivalent to an organism's DNA and is the gene (functional unit) in the evolutionary process \cite{lawrence1989hsd}. So, the Agents should be aggregated as a sequence, like the sequencing of genes in DNA \cite{lawrence1989hsd}\added{, such that an aggregation of Agents is a sequence of Agents}.}
\acrodef{structure2}{an unordered set, or, based on service orchestration, into a tree or workflow}
\acrodef{habnet}{the {agent stations} from mobile agent systems \cite{agentStation} (to provide a distributed environment in which Agent migration can occur), with evolutionary computing \cite{eiben2003iec} for the Agent interaction (instead of traditional agent interaction mechanisms \cite{wooldridge}), and the island-model of \acl*{DEC} \cite{lin1994cgp} for the connectivity between Habitats}
\acrodef{picUser}{will formulate queries to the Digital Ecosystem by creating a request as a {semantic description}, like those being used and developed in \acp{SOA} \cite{SOAsemantic}, specifying an application they desire and submitting it to their Habitat}
\acrodef{picUserReq}{A Population is then instantiated in the user's Habitat in response to the user's request, seeded from the Agents available at their Habitat}
\acrodef{digEco}{with the Agents, the Populations, the Agent migration for \acl*{DEC}, and the environmental selection pressures provided by the user base, then the union of the Habitats creates the Digital Ecosystem}
\acrodef{archComTop}{many strongly connected clusters (communities), called sub-networks (quasi-complete graphs), with a few connections between these clusters (communities) \cite{swn1}. Graphs with this topology have a very high clustering coefficient and small characteristic path lengths \cite{swn1}.}
\acrodef{bizEcoCap}{As the connections between Habitats are reconfigured depending on the connectivity of the user base, the Habitat clustering will therefore be parallel to the business sector communities}
\acrodef{similarCap}{requests are evaluated on separate {islands} (Populations), with their evolution accelerated by the sharing of solutions between the evolving Populations (islands), because they are working to solve similar requests (problems).}
\acrodef{similar2}{yellow lines connecting the evolving Populations indicate similarity in the requests being managed.}
\acrodef{lifeCycleCap}{with deployment to its owner's Habitat for distribution within the Habitat network.}
\acrodef{lifeCycle2}{used in evolving the optimal Agent-sequence in response to a user request. The optimal Agent-sequence is then registered at the Habitat}
\acrodef{lifeCycle3}{If an Agent-sequence solution is then executed, an attempt is made to migrate (copy) it to every other connected Habitat, success depending on the probability associated with the connection.}
\acrodef{as3}{with an abstract representation consisting of a set of}
\acrodef{agentSemantic2}{tuple representing an {attribute} of the {semantic description}, one integer for the {attribute identifier} and one for the {attribute value}, with both ranging between one and a hundred.}
\acrodef{as4}{Each simulated Agent had a semantic description}
\acrodef{semanticRequest}{A simulated user request consisted of an abstract {semantic description}, as a list of sets of numeric tuples to represent the properties of a desired business application}
\acrodef{bmlcap1}{the numerical semantic descriptions, of the simulated services (Agents) and user requests, in a human readable form.}
\acrodef{capbml2}{translated numerical semantic descriptions for one community within the user base, showing it in the context of the travel industry}
\acrodef{capbml3}{The simulation still operated on the numerical representation for operational efficiency, but the semantic filter essentially assigned meaning to the numbers.}
\acrodef{succession}{So, it becomes increasingly more complex through this process of succession, driven by the evolution of the populations within the ecosystem \cite{connell111msn}.}
\acrodef{succession2}{The formation of a mature ecosystem}
\acrodef{succession3}{is the slow, predictable, and orderly changes in the composition and structure of an ecological community, for which there are defined stages in the increasing complexity \cite{begon96}, as shown}
\acrodef{DigEcoSuc2Cap}{the end of the simulation run, the Agent-sequences had evolved and migrated over an average of only ten user requests per Habitat, and collectively had already reached near 70\% effectiveness for the user base.}
\acrodef{DigEcoSucCap}{The formation of a mature biological ecosystem, ecological succession, is a relatively slow process \cite{begon96}, and the simulated Digital Ecosystem acted similarly in reaching a mature state.}
\acrodef{speciesAbundance}{is a measure of the proportion of all organisms in a community belonging to a particular species \cite{Bell}. A relative abundance distribution provides the inequalities in population size within an ecosystem and therefore an indicator of biodiversity, with the distribution of most biological ecosystems taking a log-normal form \cite{Bell}.}
\acrodef{specAbund2}{the Digital Ecosystem did not conform to the expected log-normal}
\acrodef{speciesArea}{In ecology the {species-area} relationship measures diversity relative to the spatial scale, showing the number of species found in a defined area of a particular habitat or habitats of different areas \cite{sizling2004pls}, and is commonly found to follow a power law in biological ecosystems}
\acrodef{ecoCapClass}{then the Digital Ecosystem and biological ecosystem classes would both inherit from the abstract ecosystem class, but implement its attributes differently}
\acrodef{ecoCap2Class}{So, we would argue that the apparent compromises in mimicking biological ecosystems are actually features unique to Digital Ecosystems.}
\acrodef{bidpCap}{starts with identifying some useful behaviour in a biological system. Next, the system is studied to isolate the mechanisms by which it performs the useful behaviour. Last, the mechanisms are mimicked in a computational system and their performance evaluated \cite{bidpSpiral}.}
\acrodef{introEcoCap}{made up of one or more communities of organisms, consisting of species in their habitats, with their populations existing in their respective micro-habitats \cite{begon96}. A community is a naturally occurring group of populations from different species that live together in the same habitat. A habitat is a distinct part of the environment \cite{begon96}}
\acrodef{im1}{are different probabilities of going from island \Circ{1} to island \Circ{2}, as there is of going from island \Circ{2} to island \Circ{1}.}
\acrodef{im2}{mirrors the naturally inspired quality that although two populations have the same physical separation, it may be easier to migrate in one direction than the other, i.e. fish migration is easier downstream than upstream.}
\acrodef{DBEdescription}{A wealthy ecosystem sees a balance between co-operation and competition in a dynamic free market.}
\acrodef{serviceCap}{lightweight entity consisting primarily of a pointer to the \ac{SWS} it represents,}
\acrodef{service2cap}{executable component and semantic description. A software service can be a software only service, e.g. data encryption, or provide a front-end to a real-world service, e.g. selling books}
\acrodef{structureCap}{The executable component of a \ac{SWS}, that an Agent represents \added{via a pointer to the \ac{SWS}}, is equivalent to an organism's DNA and is the gene (functional unit) in the evolutionary process \cite{lawrence1989hsd}. So, the Agents should be aggregated as a sequence, like the sequencing of genes in DNA \cite{lawrence1989hsd}\added{, such that an aggregation of Agents is a sequence of Agents}.}
\acrodef{structure2}{an unordered set, or, based on service orchestration, into a tree or workflow}
\acrodef{habnet}{the {agent stations} from mobile agent systems \cite{agentStation} (to provide a distributed environment in which Agent migration can occur), with evolutionary computing \cite{eiben2003iec} for the Agent interaction (instead of traditional agent interaction mechanisms \cite{wooldridge}), and the island-model of \acl*{DEC} \cite{lin1994cgp} for the connectivity between Habitats}
\acrodef{picUser}{will formulate queries to the Digital Ecosystem by creating a request as a {semantic description}, like those being used and developed in \acp{SOA} \cite{SOAsemantic}, specifying an application they desire and submitting it to their Habitat}
\acrodef{picUserReq}{A Population is then instantiated in the user's Habitat in response to the user's request, seeded from the Agents available at their Habitat}
\acrodef{digEco}{with the Agents, the Populations, the Agent migration for \acl*{DEC}, and the environmental selection pressures provided by the user base, then the union of the Habitats creates the Digital Ecosystem}
\acrodef{archComTop}{many strongly connected clusters (communities), called sub-networks (quasi-complete graphs), with a few connections between these clusters (communities) \cite{swn1}. Graphs with this topology have a very high clustering coefficient and small characteristic path lengths \cite{swn1}.}
\acrodef{bizEcoCap}{As the connections between Habitats are reconfigured depending on the connectivity of the user base, the Habitat clustering will therefore be parallel to the business sector communities}
\acrodef{similarCap}{requests are evaluated on separate {islands} (Populations), with their evolution accelerated by the sharing of solutions between the evolving Populations (islands), because they are working to solve similar requests (problems).}
\acrodef{similar2}{yellow lines connecting the evolving Populations indicate similarity in the requests being managed.}
\acrodef{lifeCycleCap}{with deployment to its owner's Habitat for distribution within the Habitat network.}
\acrodef{lifeCycle2}{used in evolving the optimal Agent-sequence in response to a user request. The optimal Agent-sequence is then registered at the Habitat}
\acrodef{lifeCycle3}{If an Agent-sequence solution is then executed, an attempt is made to migrate (copy) it to every other connected Habitat, success depending on the probability associated with the connection.}
\acrodef{as3}{with an abstract representation consisting of a set of}
\acrodef{agentSemantic2}{tuple representing an {attribute} of the {semantic description}, one integer for the {attribute identifier} and one for the {attribute value}, with both ranging between one and a hundred.}
\acrodef{as4}{Each simulated Agent had a semantic description}
\acrodef{semanticRequest}{A simulated user request consisted of an abstract {semantic description}, as a list of sets of numeric tuples to represent the properties of a desired business application}
\acrodef{bmlcap1}{the numerical semantic descriptions, of the simulated services (Agents) and user requests, in a human readable form.}
\acrodef{capbml2}{translated numerical semantic descriptions for one community within the user base, showing it in the context of the travel industry}
\acrodef{capbml3}{The simulation still operated on the numerical representation for operational efficiency, but the semantic filter essentially assigned meaning to the numbers.}
\acrodef{succession}{So, it becomes increasingly more complex through this process of succession, driven by the evolution of the populations within the ecosystem \cite{connell111msn}.}
\acrodef{succession2}{The formation of a mature ecosystem}
\acrodef{succession3}{is the slow, predictable, and orderly changes in the composition and structure of an ecological community, for which there are defined stages in the increasing complexity \cite{begon96}, as shown}
\acrodef{DigEcoSuc2Cap}{the end of the simulation run, the Agent-sequences had evolved and migrated over an average of only ten user requests per Habitat, and collectively had already reached near 70\% effectiveness for the user base.}
\acrodef{DigEcoSucCap}{The formation of a mature biological ecosystem, ecological succession, is a relatively slow process \cite{begon96}, and the simulated Digital Ecosystem acted similarly in reaching a mature state.}
\acrodef{speciesAbundance}{is a measure of the proportion of all organisms in a community belonging to a particular species \cite{Bell}. A relative abundance distribution provides the inequalities in population size within an ecosystem and therefore an indicator of biodiversity, with the distribution of most biological ecosystems taking a log-normal form \cite{Bell}.}
\acrodef{specAbund2}{the Digital Ecosystem did not conform to the expected log-normal}
\acrodef{speciesArea}{In ecology the {species-area} relationship measures diversity relative to the spatial scale, showing the number of species found in a defined area of a particular habitat or habitats of different areas \cite{sizling2004pls}, and is commonly found to follow a power law in biological ecosystems}
\acrodef{ecoCapClass}{then the Digital Ecosystem and biological ecosystem classes would both inherit from the abstract ecosystem class, but implement its attributes differently}
\acrodef{ecoCap2Class}{So, we would argue that the apparent compromises in mimicking biological ecosystems are actually features unique to Digital Ecosystems.}
\acrodef{bidpCap}{starts with identifying some useful behaviour in a biological system. Next, the system is studied to isolate the mechanisms by which it performs the useful behaviour. Last, the mechanisms are mimicked in a computational system and their performance evaluated \cite{bidpSpiral}.}
\acrodef{introEcoCap}{made up of one or more communities of organisms, consisting of species in their habitats, with their populations existing in their respective micro-habitats \cite{begon96}. A community is a naturally occurring group of populations from different species that live together in the same habitat. A habitat is a distinct part of the environment \cite{begon96}}
\acrodef{im1}{are different probabilities of going from island \Circ{1} to island \Circ{2}, as there is of going from island \Circ{2} to island \Circ{1}.}
\acrodef{im2}{mirrors the naturally inspired quality that although two populations have the same physical separation, it may be easier to migrate in one direction than the other, i.e. fish migration is easier downstream than upstream.}
\acrodef{DBEdescription}{A wealthy ecosystem sees a balance between co-operation and competition in a dynamic free market.}
\acrodef{serviceCap}{lightweight entity consisting primarily of a pointer to the \ac{SWS} it represents,}
\acrodef{service2cap}{executable component and semantic description. A software service can be a software only service, e.g. data encryption, or provide a front-end to a real-world service, e.g. selling books}
\acrodef{structureCap}{The executable component of a \ac{SWS}, that an Agent represents \added{via a pointer to the \ac{SWS}}, is equivalent to an organism's DNA and is the gene (functional unit) in the evolutionary process \cite{lawrence1989hsd}. So, the Agents should be aggregated as a sequence, like the sequencing of genes in DNA \cite{lawrence1989hsd}\added{, such that an aggregation of Agents is a sequence of Agents}.}
\acrodef{structure2}{an unordered set, or, based on service orchestration, into a tree or workflow}
\acrodef{habnet}{the {agent stations} from mobile agent systems \cite{agentStation} (to provide a distributed environment in which Agent migration can occur), with evolutionary computing \cite{eiben2003iec} for the Agent interaction (instead of traditional agent interaction mechanisms \cite{wooldridge}), and the island-model of \acl*{DEC} \cite{lin1994cgp} for the connectivity between Habitats}
\acrodef{picUser}{will formulate queries to the Digital Ecosystem by creating a request as a {semantic description}, like those being used and developed in \acp{SOA} \cite{SOAsemantic}, specifying an application they desire and submitting it to their Habitat}
\acrodef{picUserReq}{A Population is then instantiated in the user's Habitat in response to the user's request, seeded from the Agents available at their Habitat}
\acrodef{digEco}{with the Agents, the Populations, the Agent migration for \acl*{DEC}, and the environmental selection pressures provided by the user base, then the union of the Habitats creates the Digital Ecosystem}
\acrodef{archComTop}{many strongly connected clusters (communities), called sub-networks (quasi-complete graphs), with a few connections between these clusters (communities) \cite{swn1}. Graphs with this topology have a very high clustering coefficient and small characteristic path lengths \cite{swn1}.}
\acrodef{bizEcoCap}{As the connections between Habitats are reconfigured depending on the connectivity of the user base, the Habitat clustering will therefore be parallel to the business sector communities}
\acrodef{similarCap}{requests are evaluated on separate {islands} (Populations), with their evolution accelerated by the sharing of solutions between the evolving Populations (islands), because they are working to solve similar requests (problems).}
\acrodef{similar2}{yellow lines connecting the evolving Populations indicate similarity in the requests being managed.}
\acrodef{lifeCycleCap}{with deployment to its owner's Habitat for distribution within the Habitat network.}
\acrodef{lifeCycle2}{used in evolving the optimal Agent-sequence in response to a user request. The optimal Agent-sequence is then registered at the Habitat}
\acrodef{lifeCycle3}{If an Agent-sequence solution is then executed, an attempt is made to migrate (copy) it to every other connected Habitat, success depending on the probability associated with the connection.}
\acrodef{as3}{with an abstract representation consisting of a set of}
\acrodef{agentSemantic2}{tuple representing an {attribute} of the {semantic description}, one integer for the {attribute identifier} and one for the {attribute value}, with both ranging between one and a hundred.}
\acrodef{as4}{Each simulated Agent had a semantic description}
\acrodef{semanticRequest}{A simulated user request consisted of an abstract {semantic description}, as a list of sets of numeric tuples to represent the properties of a desired business application}
\acrodef{bmlcap1}{the numerical semantic descriptions, of the simulated services (Agents) and user requests, in a human readable form.}
\acrodef{capbml2}{translated numerical semantic descriptions for one community within the user base, showing it in the context of the travel industry}
\acrodef{capbml3}{The simulation still operated on the numerical representation for operational efficiency, but the semantic filter essentially assigned meaning to the numbers.}
\acrodef{succession}{So, it becomes increasingly more complex through this process of succession, driven by the evolution of the populations within the ecosystem \cite{connell111msn}.}
\acrodef{succession2}{The formation of a mature ecosystem}
\acrodef{succession3}{is the slow, predictable, and orderly changes in the composition and structure of an ecological community, for which there are defined stages in the increasing complexity \cite{begon96}, as shown}
\acrodef{DigEcoSuc2Cap}{the end of the simulation run, the Agent-sequences had evolved and migrated over an average of only ten user requests per Habitat, and collectively had already reached near 70\% effectiveness for the user base.}
\acrodef{DigEcoSucCap}{The formation of a mature biological ecosystem, ecological succession, is a relatively slow process \cite{begon96}, and the simulated Digital Ecosystem acted similarly in reaching a mature state.}
\acrodef{speciesAbundance}{is a measure of the proportion of all organisms in a community belonging to a particular species \cite{Bell}. A relative abundance distribution provides the inequalities in population size within an ecosystem and therefore an indicator of biodiversity, with the distribution of most biological ecosystems taking a log-normal form \cite{Bell}.}
\acrodef{specAbund2}{the Digital Ecosystem did not conform to the expected log-normal}
\acrodef{speciesArea}{In ecology the {species-area} relationship measures diversity relative to the spatial scale, showing the number of species found in a defined area of a particular habitat or habitats of different areas \cite{sizling2004pls}, and is commonly found to follow a power law in biological ecosystems}
\acrodef{ecoCapClass}{then the Digital Ecosystem and biological ecosystem classes would both inherit from the abstract ecosystem class, but implement its attributes differently}
\acrodef{ecoCap2Class}{So, we would argue that the apparent compromises in mimicking biological ecosystems are actually features unique to Digital Ecosystems.}
\acrodef{bidpCap}{starts with identifying some useful behaviour in a biological system. Next, the system is studied to isolate the mechanisms by which it performs the useful behaviour. Last, the mechanisms are mimicked in a computational system and their performance evaluated \cite{bidpSpiral}.}
\acrodef{introEcoCap}{made up of one or more communities of organisms, consisting of species in their habitats, with their populations existing in their respective micro-habitats \cite{begon96}. A community is a naturally occurring group of populations from different species that live together in the same habitat. A habitat is a distinct part of the environment \cite{begon96}}
\acrodef{im1}{are different probabilities of going from island \Circ{1} to island \Circ{2}, as there is of going from island \Circ{2} to island \Circ{1}.}
\acrodef{im2}{mirrors the naturally inspired quality that although two populations have the same physical separation, it may be easier to migrate in one direction than the other, i.e. fish migration is easier downstream than upstream.}
\acrodef{DBEdescription}{A wealthy ecosystem sees a balance between co-operation and competition in a dynamic free market.}
\acrodef{serviceCap}{lightweight entity consisting primarily of a pointer to the \ac{SWS} it represents,}
\acrodef{service2cap}{executable component and semantic description. A software service can be a software only service, e.g. data encryption, or provide a front-end to a real-world service, e.g. selling books}
\acrodef{structureCap}{The executable component of a \ac{SWS}, that an Agent represents \added{via a pointer to the \ac{SWS}}, is equivalent to an organism's DNA and is the gene (functional unit) in the evolutionary process \cite{lawrence1989hsd}. So, the Agents should be aggregated as a sequence, like the sequencing of genes in DNA \cite{lawrence1989hsd}\added{, such that an aggregation of Agents is a sequence of Agents}.}
\acrodef{structure2}{an unordered set, or, based on service orchestration, into a tree or workflow}
\acrodef{habnet}{the {agent stations} from mobile agent systems \cite{agentStation} (to provide a distributed environment in which Agent migration can occur), with evolutionary computing \cite{eiben2003iec} for the Agent interaction (instead of traditional agent interaction mechanisms \cite{wooldridge}), and the island-model of \acl*{DEC} \cite{lin1994cgp} for the connectivity between Habitats}
\acrodef{picUser}{will formulate queries to the Digital Ecosystem by creating a request as a {semantic description}, like those being used and developed in \acp{SOA} \cite{SOAsemantic}, specifying an application they desire and submitting it to their Habitat}
\acrodef{picUserReq}{A Population is then instantiated in the user's Habitat in response to the user's request, seeded from the Agents available at their Habitat}
\acrodef{digEco}{with the Agents, the Populations, the Agent migration for \acl*{DEC}, and the environmental selection pressures provided by the user base, then the union of the Habitats creates the Digital Ecosystem}
\acrodef{archComTop}{many strongly connected clusters (communities), called sub-networks (quasi-complete graphs), with a few connections between these clusters (communities) \cite{swn1}. Graphs with this topology have a very high clustering coefficient and small characteristic path lengths \cite{swn1}.}
\acrodef{bizEcoCap}{As the connections between Habitats are reconfigured depending on the connectivity of the user base, the Habitat clustering will therefore be parallel to the business sector communities}
\acrodef{similarCap}{requests are evaluated on separate {islands} (Populations), with their evolution accelerated by the sharing of solutions between the evolving Populations (islands), because they are working to solve similar requests (problems).}
\acrodef{similar2}{yellow lines connecting the evolving Populations indicate similarity in the requests being managed.}
\acrodef{lifeCycleCap}{with deployment to its owner's Habitat for distribution within the Habitat network.}
\acrodef{lifeCycle2}{used in evolving the optimal Agent-sequence in response to a user request. The optimal Agent-sequence is then registered at the Habitat}
\acrodef{lifeCycle3}{If an Agent-sequence solution is then executed, an attempt is made to migrate (copy) it to every other connected Habitat, success depending on the probability associated with the connection.}
\acrodef{as3}{with an abstract representation consisting of a set of}
\acrodef{agentSemantic2}{tuple representing an {attribute} of the {semantic description}, one integer for the {attribute identifier} and one for the {attribute value}, with both ranging between one and a hundred.}
\acrodef{as4}{Each simulated Agent had a semantic description}
\acrodef{semanticRequest}{A simulated user request consisted of an abstract {semantic description}, as a list of sets of numeric tuples to represent the properties of a desired business application}
\acrodef{bmlcap1}{the numerical semantic descriptions, of the simulated services (Agents) and user requests, in a human readable form.}
\acrodef{capbml2}{translated numerical semantic descriptions for one community within the user base, showing it in the context of the travel industry}
\acrodef{capbml3}{The simulation still operated on the numerical representation for operational efficiency, but the semantic filter essentially assigned meaning to the numbers.}
\acrodef{succession}{So, it becomes increasingly more complex through this process of succession, driven by the evolution of the populations within the ecosystem \cite{connell111msn}.}
\acrodef{succession2}{The formation of a mature ecosystem}
\acrodef{succession3}{is the slow, predictable, and orderly changes in the composition and structure of an ecological community, for which there are defined stages in the increasing complexity \cite{begon96}, as shown}
\acrodef{DigEcoSuc2Cap}{the end of the simulation run, the Agent-sequences had evolved and migrated over an average of only ten user requests per Habitat, and collectively had already reached near 70\% effectiveness for the user base.}
\acrodef{DigEcoSucCap}{The formation of a mature biological ecosystem, ecological succession, is a relatively slow process \cite{begon96}, and the simulated Digital Ecosystem acted similarly in reaching a mature state.}
\acrodef{speciesAbundance}{is a measure of the proportion of all organisms in a community belonging to a particular species \cite{Bell}. A relative abundance distribution provides the inequalities in population size within an ecosystem and therefore an indicator of biodiversity, with the distribution of most biological ecosystems taking a log-normal form \cite{Bell}.}
\acrodef{specAbund2}{the Digital Ecosystem did not conform to the expected log-normal}
\acrodef{speciesArea}{In ecology the {species-area} relationship measures diversity relative to the spatial scale, showing the number of species found in a defined area of a particular habitat or habitats of different areas \cite{sizling2004pls}, and is commonly found to follow a power law in biological ecosystems}
\acrodef{ecoCapClass}{then the Digital Ecosystem and biological ecosystem classes would both inherit from the abstract ecosystem class, but implement its attributes differently}
\acrodef{ecoCap2Class}{So, we would argue that the apparent compromises in mimicking biological ecosystems are actually features unique to Digital Ecosystems.}

\title{Digital Ecosystems: Ecosystem-Oriented Architectures}

\author{Gerard Briscoe         \and
Suzanne Sadedin
\and
        Philippe De Wilde 
}

\institute{G Briscoe \at
              Systems Research Group\\
              Computer Laboratory\\
              University of Cambridge\\
              \email{gerard.briscoe@cl.cam.ac.uk}           
           \and
S Sadedin \at
Department of Organismic and Evolutionary Biology\\
Harvard University\\
\email{sadedin@fas.harvard.edu}
           \and
           P De Wilde \at
       Intelligent Systems Lab\\
Department of Computer Science\\
Heriot-Watt University\\
\email{p.de\_wilde@hw.ac.uk}
}

\maketitle

\begin{abstract}
We view Digital Ecosystems to be the digital counterparts of biological ecosystems. \changed{Here, we are concerned with the creation of these Digital Ecosystems, exploiting the self-organising properties of biological ecosystems} \added{to evolve high-level software applications}. Therefore, we created the Digital Ecosystem, a novel optimisation technique inspired by biological ecosystems, where the optimisation works at two levels: a first optimisation, migration of agents which are distributed in a decentralised peer-to-peer network, operating continuously in time; this process feeds a second optimisation based on evolutionary computing that operates locally on single peers and is aimed at finding solutions to satisfy locally relevant constraints. \changed{The Digital Ecosystem was then measured experimentally through simulations, with measures originating from theoretical ecology, evaluating its likeness to biological ecosystems.} This included its responsiveness to requests for applications from the user base, as a measure of the ecological succession (ecosystem maturity). Overall, we have advanced the understanding of Digital Ecosystems, creating \aclp*{EOA} where the word ecosystem is more than just a metaphor.


\end{abstract}

\vspace{-6mm}

\section*{Abbreviations}
\vspace{-4mm}
\begin{acronym}[HBCIBCIBC] 
\acro{PCG}{Projected Conjugate Gradient} 
\acro{QP}{quadratic programming}
\acro{RBF}{Radial-Basis Function}
\acro{ABM}{Agent-Based Modelling}
\acro{AI}{Artificial Intelligence}
\acro{DAI}{Distributed Artificial Intelligence}
\acro{API}{Application Programming Interface}
\acro{ARF}{p14ARF human tumor-suppressor gene}
\acro{B2B}{business-to-business}
\acro{BDP}{Biological Design Pattern}
\acro{BGS}{Best Guess Solution}
\acro{BIC}{Biologically-Inspired Computing}
\acro{BML}{Business Modelling Language}
\acro{BPEL}{Business Process Execution Language}
\acro{BPMN}{Business Process Modelling Notation}
\acro{CAS}{Complex Adaptive Systems}
\acro{COBOL}{COmmon Business-Oriented Language}
\acro{DBE}{Digital Business Ecosystem}
\acro{DE}{Digital Ecosystem}
\acro{DEC}{distributed evolutionary computing}
\acro{DGA}{Distributed genetic algorithms}
\acro{DIS}{Distributed Intelligence System}
\acro{DNA}{Deoxyribose Nucleic Acid}
\acro{DOP}{DBE Open Protocol}
\acro{DSS}{Distributed Storage System}
\acro{EAP}{Evolving Agent Population}
\acro{ebXML}{e-business eXtensible Markup Language}
\acro{EC}{Evolutionary Computing}
\acro{ECJ}{Evolutionary Computing in Java}
\acro{EE}{Evolutionary Environment}
\acro{EFL}{Evolutionary Framework for Language}
\acro{FLE}{Framework for Language Ecosystems}
\acro{EOA}{Ecosystem-Oriented Architecture}
\acro{ESS}{evolutionary stable strategy}
\acro{EvE}{Evolutionary Environment}
\acro{ExE}{Execution Environment}
\acro{FCB}{Framework for Computational Biomimicry}
\acro{FFF}{Fitness Function Framework}
\acro{FL}{Fitness Landscape}
\acro{HWU}{Heriot-Watt University}
\acro{ICL}{Imperial College London}
\acro{ICT}{Information and Communications Technology}
\acro{INTEL}{Intel Ireland}
\acro{IPA}{International Phonetic Alphabet}
\acro{ISUFI}{Istituto Superiore Universitario di Formazione Interdisciplinare}
\acro{JDJ}{Java Developer's Journal}
\acro{KC}{Kolmogorov-Chaitin}
\acro{LAN}{local area network}
\acro{LSE}{London School of Economics and Political Science}
\acro{MAS}{Multi-Agent System}
\acro{MDL}{Minimum Description Length}
\acro{MDM2}{murine double minute 2}
\acro{MFT}{Mean Field Theory}
\acro{MoAS}{Mobile Agent System}
\acro{MOF}{Meta Object Facility}
\acro{MUH}{migration and usage history}
\acro{NIC}{Nature Inspired Computing}
\acro{NN}{Neural Network}
\acro{NoE}{Network of Excellence}
\acro{OMG}{Open Mac Grid}
\acro{OPAALS}{Open Philosophies for Associative Autopoietic Digital Ecosystems}
\acro{P2P}{peer-to-peer}
\acro{P53}{protein 53}
\acro{PDA}{Personal Digital Assistant}
\acro{QoS}{quality of service}
\acro{REST}{REpresentational State Transfer}
\acro{RNA}{Deoxyribose Nucleic Acid}
\acro{SAE}{Software Agent Ecosystem}
\acro{SBML}{Systems Biology Modelling Language}
\acro{SBVR}{Semantics of Business Vocabulary and Business Rules}
\acro{SDL}{Service Description Language}
\acro{SF}{Service Factory}
\acro{SIM}{Social Interaction Mechanism}
\acro{SM}{Service Manifest}
\acro{SME}{Small and Medium sized Enterprise}
\acro{SML}{Service Modelling Language}
\acro{SMO}{Sequential Minimal Optimisation}
\acro{SOA}{Service-Oriented Architecture}
\acro{SOAP}{Simple Object Access Protocol}
\acro{SOC}{Self-Organised Criticality}
\acro{SOLUTA}{SOLUTA.NET}
\acro{SOM}{Self-Organising Map}
\acro{SSL}{Semantic Service Language}
\acro{STU}{Salzburg Technical University}
\acro{SUN}{Sun Microsystems}
\acro{SVM}{Support Vector Machine}
\acro{TM}{Turing Machine}
\acro{UBHAM}{University of Birmingham}
\acro{UDDI}{Universal Description Discovery and Integration}
\acro{UML}{Unified Modelling Language}
\acro{URI}{Uniform Resource Identifier}
\acro{UTM}{Universal Turing Machine}
\acro{VLP}{variable length population}
\acro{VLS}{variable length sequences}
\acro{vls}{variable length sequence}
\acro{WP}{Work-Package}
\acro{WSDL}{Web Services Definition Language}
\acro{XMI}{XML Metadata Interchange}
\acro{XML}{eXtensible Markup Language}
\acro{MD5}{Message-Digest algorithm 5}
\acro{GA}{genetic algorithm}
\acro{GP}{genetic programming}
\acro{MASON}{Multi-Agent Simulator Of Neighbourhoods}
\acro{Repast}{Recursive Porous Agent Simulation Toolkit}
\acro{JCLEC}{Java Computing Library for Evolutionary Computing}
\acro{OWL-S}{Web Ontology Language - Service}
\acro{EGT}{Evolutionary Game Theory}
\acro{RBF}{Radial Basis Functions}
\acro{SWS}{Semantic Web Services}
\acro{HDD}{Hard Disk Drive}
\acro{SSD}{Solid-State Drive}
\end{acronym}

\section{Introduction}
\label{intro}
Is mimicking ecosystems the future of information systems? A key challenge in modern computing is to develop systems that address complex, dynamic problems in a scalable and efficient way, because the increasing complexity of software makes designing and maintaining efficient and flexible systems a growing challenge \cite{newsArticle1, slashdot, newsArticle3}. With the ever expanding number of services being offered online from \acp{API} being made public, there is an ever growing number of computational units available to be combined in the creation of applications. \changed{We propose that these computational units could be assembled without the intervention of a programmer.} \added{Furthermore, we would argue that manual programming is becoming too complex to manage, which could be overcome by automating the search for new algorithms through automated service composition.} \changed{There are existing efforts aimed at achieving automated service composition \cite{reef3, reef5, reef1, reef4}, the most prevalent of which is \acp{SOA} and its associated standards and technologies \cite{curbera2002uws, SOAstandards}.}

Alternatively, nature has been in the research business for 3.8 billion years and in that time has accumulated close to 30 million well-adjusted solutions to a plethora of design challenges that humankind struggles to address with mixed results \cite{biomimicry}. Biomimicry is a discipline that seeks solutions by emulating nature's designs and processes, and there is considerable opportunity to learn elegant solutions for human-made problems \cite{biomimicry}. Biological ecosystems are thought to be robust, scalable architectures that can automatically solve complex, dynamic problems, possessing several properties that may be useful in automated systems. These properties include self-organisation, self-management, scalability, the ability to provide complex solutions, and the automated composition of these complex solutions \cite{Levin}.

\changed{So, our approach to the aforementioned challenge is to develop Digital Ecosystems, artificial systems that harness the dynamics that underlie the complex and diverse adaptations of living organisms in biological ecosystems.} While evolutionary theory may be widely appreciated in computer science under the auspices of evolutionary computing \cite{eiben2003iec}, ecological theory is not. Possible connections between Digital Ecosystems and their biological counterparts are yet to be closely examined, so potential exists to create an \acl*{EOA} with the essential elements of biological ecosystems. We propose that an approach inspired by ecosystems \added{to evolving high-level software applications} may be \changed{feasible}, because it can build upon the scalable and self-organising properties of biological ecosystems \cite{Levin}. \added{We must first introduce the relevant theoretical biology, before considering the complementary computer science, in order to justify creating the digital counterparts of biological ecosystems.}

\section{Background Theory}
\label{background}
Relevant theory for digital ecosystems spans several academic domains. Here, we briefly introduce the core concepts of \ac{NIC}, followed by related work in theoretical biology and computer science \cite{bionetics, thesis}. Within theoretical biology, we consider how properties of biological ecosystems influence functions that are relevant to developing Digital Ecosystems. This leads us to suggest ways in which concepts from ecology can be used to create Digital Ecosystems that solve practical problems.

\subsection{Existing Digital Ecosystems}

Our focus is on creating the digital counterpart of biological ecosystems. However, the term \emph{digital ecosystem} has described a variety of concepts, ranging from the existing networking infrastructure of the Internet \cite{debook2, ballmer, fiorina, XIMBIOTIX}, to digital ecosystem services which enable customers to use existing e-business solutions \cite{accenture, syntel, xewow}. The term is also increasingly linked to future development of \ac{ICT} adoption for e-business, to support \emph{business ecosystems} \cite{moore1996}. However, perhaps the most frequent references to digital ecosystems arise in Artificial Life research, where they are created primarily to investigate biological and other complex systems \cite{sorakugun1995eas, grand1998ces, deAI1}.

The extent to which these disparate systems resemble biological ecosystems varies, and frequently the word ecosystem is merely used for branding purposes without any inherent ecological properties. We consider Digital Ecosystems to be software systems that exploit the properties of biological ecosystems, and suggest that several key features of biological ecosystems have not been fully explored in existing digital ecosystems. Mimicking these features may create Digital Ecosystems which are robust, scalable, and self-organising.

\subsection{Nature-Inspired Computing}

Biomimicry (bios, meaning life, and mimesis, meaning to imitate) is the science that studies natural systems and processes, and takes creative inspiration from them to design engineered systems \cite{biomimicry}. This concept is far from new, with humans having long been inspired by the animals and plants of the natural world; Leonardo Da Vinci once said, \emph{Those who are inspired by a model other than Nature, a mistress above all masters, are labouring in vain} \cite{bramly1994laa}. Albeit overstating the point, it reminds us that the transfer of technology between life-forms and synthetic constructs is often desirable because evolutionary pressures often drive living organisms to become highly optimised and efficient at specific tasks. A classical example is the development of dirt and water repellent paint from the observation that the surface of the lotus flower plant is practically non-sticky for anything, commonly known as the \emph{lotus effect} \cite{barthlott1997psl}. However, biomimicry, when done well, is not slavish imitation; it is inspiration using the principles which nature has demonstrated to be successful design strategies. For example, in the early days of mechanised flight the best designs were not the ornithopters, which most completely imitated birds, but the fixed-wing craft that used the principle of aerofoil cross-section in their wings \cite{andersonFlight}. 

Biomimicry in computer science is called \ac{NIC} or Natural Computation, and the benefits of natural computation technologies often mimic those found in real natural systems, such as flexibility, adaptability, robustness, and decentralised control \cite{NICbook}. Their sources of inspiration come from many aspects of natural systems; evolution, ecology, development, cell and molecular phenomena, behaviour, cognition, and other areas \cite{ecpaper}. Such nature inspired techniques often lead to novel computing systems with applicability in many different areas \cite{ecpaper}. \ac{NIC} itself can be divided into three main branches \cite{NICbook}:

\begin{itemize}
\item \ac{BIC}: This makes use of nature as inspiration for the development of problem solving techniques. The main idea of this branch is to develop computational tools (algorithms) by taking inspiration from nature for the solution of complex problems.

\item The simulation and emulation of nature by computational means: This is the computational synthesis of patterns, forms, behaviours, and organisms that resemble \emph{life-as-we-know-it}. By mimicking various natural phenomena, computational simulation can improve our understanding of both nature and computer models.

\item Computing with natural materials: This corresponds to the use of natural materials to perform computation, to substitute or supplement the current silicon-based computers.
\end{itemize}

Many different natural systems inspire these approaches. For example, evolutionary algorithms use the concepts of mutation, recombination, and natural selection from biology; neural networks are inspired by the highly interconnected neural structures in the brain and the nervous system; molecular computing is based on paradigms from molecular biology; and quantum computing is based on how quantum physics exploits quantum parallelism \cite{NICbook}. There are also important methodological differences between various sub-areas. For example, evolutionary algorithms and algorithms based on neural networks are presently implemented on conventional computers. However, molecular computing aims to develop alternatives to silicon hardware by implementing algorithms in biological hardware using DNA molecules and enzymes, while quantum computing aims at non-traditional hardware that can make use of quantum effects \cite{NICbook}.

\tfigure{scale=0.65}{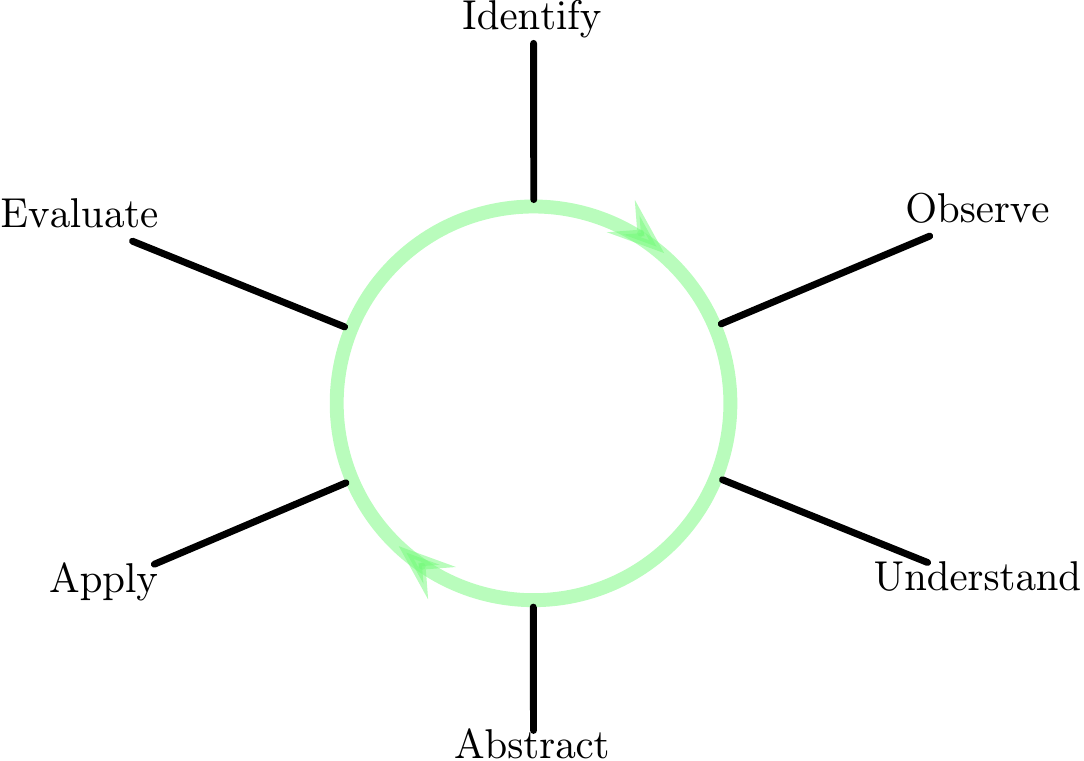}{graffle}{Biomimicry Design Spiral}{(modified from \cite{bidpSpiral}): The process of biomimicry \getCap{bidpCap}}{0mm}{!t}{}{-3mm}

We are concerned with \ac{BIC}, which relies heavily on the fields of biology, computer science, and mathematics. Briefly put, it is the study of nature to improve algorithms \cite{forbes2004ilb}, and should not to be confused with computational biology \cite{waterman1995icb}, which applies techniques from computer science, mathematics, and statistics to address biological problems. \ac{BIC} has produced \aclp*{NN}, swarm intelligence and evolutionary computing \cite{forbes2004ilb}. Introducing \ac{BIC}, one comes quickly to its applications, partly because this is the essence of the approach, and partly because biomimicry as a process tends to be ad hoc \cite{bidpSpiral}. It generally involves an engineer or scientist observing or being aware of an area of biological study, which seems applicable to a technology or research problem they are currently tackling, or which inspires the creation of a new technology \cite{NICbook}. However, there are some common steps in this process, which \setCap{starts with identifying some useful behaviour in a biological system. Next, the system is studied to isolate the mechanisms by which it performs the useful behaviour. Last, the mechanisms are mimicked in a computational system and their performance evaluated \cite{bidpSpiral}.}{bidpCap} This process is summarised in Figure \ref{bidp}.

\subsection{Biology of Digital Ecosystems}

\label{bioOfDE}

Natural science is the study of the universe via the rules or laws of natural order, and the term is also used to differentiate those fields using scientific method in the study of nature, in contrast with the social sciences which apply the scientific method to culture and human behaviour: economics, psychology, political economy, anthropology, etc \cite{hollis1994pss}. The fields of natural science are diverse, ranging from particle physics to astronomy \cite{salmon1999ips}, and while not all these fields of study will provide paradigms for Digital Ecosystems, the further one wishes to take the analogy of the word ecosystem, the more one has to consider the relevance of the fields of natural science, particularly the biological sciences.

A primary motivation for our research in Digital Ecosystems is the desire to exploit the self-organising properties of biological ecosystems. However, the biological processes that contribute to these properties have not been made explicit in Digital Ecosystems research. Here, we discuss how biological properties contribute to the self-organising features of biological ecosystems, including population dynamics, evolution, a complex dynamic environment, and spatial distributions for generating local interactions \cite{tilman1997ser}. The potential for exploiting these properties in artificial systems is then considered. We suggest that several key features of biological ecosystems have not been fully explored in existing digital ecosystems, and discuss how mimicking these features may assist in developing robust, scalable self-organising architectures \cite{de07oz, dbebkpub}.

Evolutionary computing uses natural selection to evolve solutions \cite{goldberg}; it starts with a set of possible solutions chosen arbitrarily, then selection, replication, recombination, and mutation are applied iteratively. Selection is based on conforming to a fitness function which is determined by a specific problem of interest, and so over time better solutions to the problem can thus evolve \cite{goldberg}. As Digital Ecosystems will likely solve problems by evolving solutions, they will probably incorporate some form of evolutionary computing. 

Notably, a fundamental difference between biological and digital ecosystems lies in the motivation and approach of their respective researchers. Biological ecosystems are ubiquitous natural phenomena whose maintenance is crucial to our survival \cite{balmford2002erc}, developing through the process of \emph{ecological succession} \cite{begon96}. In contrast, Digital Ecosystems will be defined here as a technology engineered to serve specific human purposes, developed to solve dynamic problems in parallel with high efficiency.

\subsubsection{Biological Ecosystems}

\tfigure{scale=0.8}{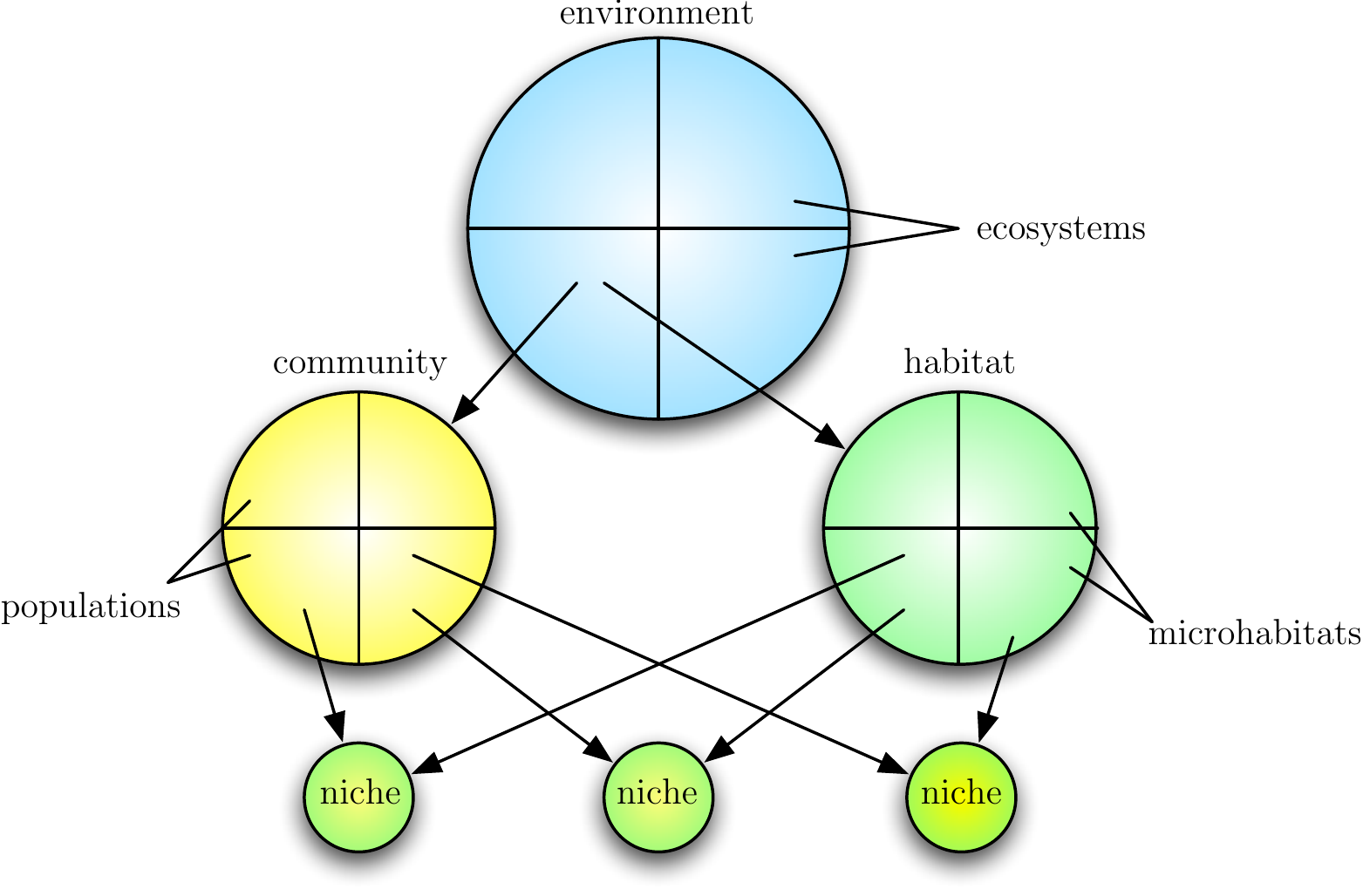}{graffle}{Ecosystem Structure}{(redrawn from \cite{longman}): A stable, self-perpetuating system \getCap{introEcoCap}.}{-2mm}{!b}{}{}

An ecosystem is a natural unit made up of living (biotic) and non-living (abiotic) components, from whose interactions emerge a self-perpetuating system. It is \setCap{made up of one or more communities of organisms, consisting of species in their habitats, with their populations existing in their respective micro-habitats \cite{begon96}. A community is a naturally occurring group of populations from different species that live together in the same habitat. A habitat is a distinct part of the environment \cite{begon96}}{introEcoCap}, for example, a stream. Individual organisms migrate through the ecosystem into different habitats competing with other organisms for limited resources, with a population being the aggregate of the individuals, of a particular species, inhabiting a specific location \cite{begon96}. A micro-habitat is a subdivision of a habitat that possesses its own unique properties, such as a micro-climate \cite{lawrence1989hsd}. Evolution occurs to all living components of an ecosystem, with the evolutionary pressures varying from one population to the next depending on the local environment. A population, in its micro-habitat, comes to occupy a niche, which is the functional relationship of a population to the environment that it occupies. A niche results in the highly specialised adaptation of a population to its micro-habitat \cite{lawrence1989hsd}. \tred{This is shown graphically in Figure \ref{abstractEcosystem}.}

\subsubsection{Fitness Landscapes and Agents}

\label{agents}
An ecosystem comprises both an environment and a set of interacting, reproducing entities (or agents) in that environment; with both the environment and other agents acting as a set of physical and chemical constraints on reproduction and survival \cite{begon96}. These constraints can be considered in abstract using the metaphor of a \emph{fitness landscape}, in which individuals are represented as solutions to the problem of survival and reproduction \cite{wright1932}. All possible solutions are distributed in a space whose dimensions are the possible properties of individuals. An additional dimension, height, indicates the relative fitness (in terms of survival and reproduction) of each solution. The fitness landscape is envisaged as a rugged, multidimensional landscape of hills, mountains, and valleys, because individuals with certain sets of properties are \emph{fitter} than others \cite{wright1932}, as visualised in Figure \ref{figure1new}. 

\tfigure{scale=0.6}{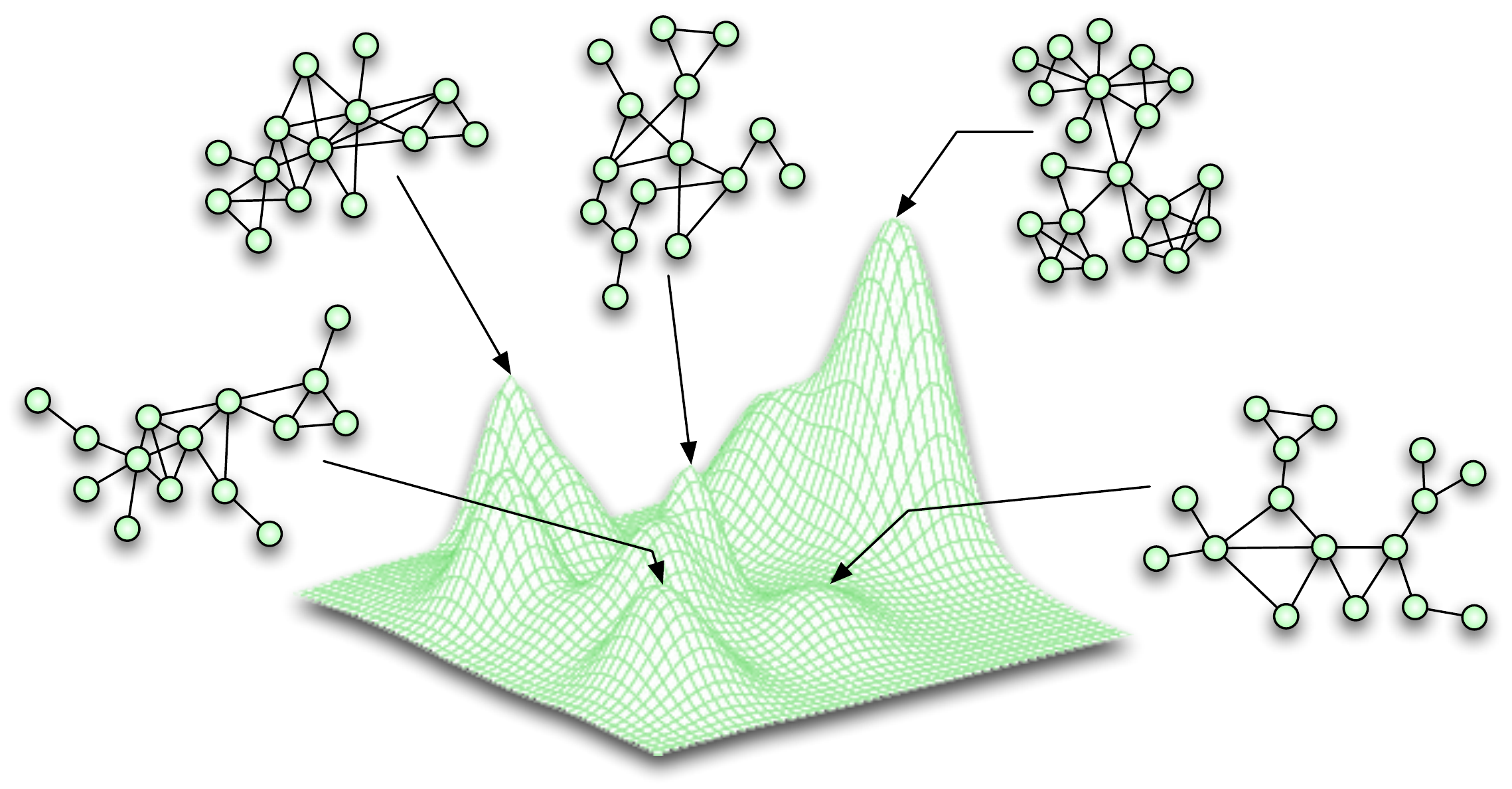}{graffle}{Fitness Landscape}{(modified from \cite{fitland}): We can represent software development as a walk through the landscape, towards the peaks which correspond to the optimal applications. Each point represents a unique combination of software services, and the roughness of the landscape indicates how difficult it is to reach an optimal software design \cite{fitland}. In this example, there is a global optimum, and several lower local optima.}{-2mm}{!b}{}{}

In biological ecosystems, fitness landscapes are virtually impossible to identify. This is both because there are large numbers of possible traits that can influence individual fitness, and because the environment changes over time and space \cite{begon96}. Within genetic algorithms, exact specification of a fitness landscape or function is common practice \cite{goldberg}. The designer of a Digital Ecosystem therefore faces a double challenge: first, to specify rules that govern the shape of the fitness function/landscape in a way that meaningfully maps landscape dynamics to user requests, and second, to evolve within this space, solution populations that are diverse enough to solve disparate problems, complex enough to meet user needs, and efficient enough to be preferable to those generated by other means.

The agents within a Digital Ecosystem will need to be like biological individuals in the sense that they reproduce, vary, interact, move, and die \cite{begon96}. Each of these properties contributes to the dynamics of the ecosystem. However, the way in which these individual properties are encoded may vary substantially depending on the intended purpose of the system \cite{chambers2001phg}.

\subsubsection{Networks and Spatial Dynamics}

\tfigure{scale=1.0}{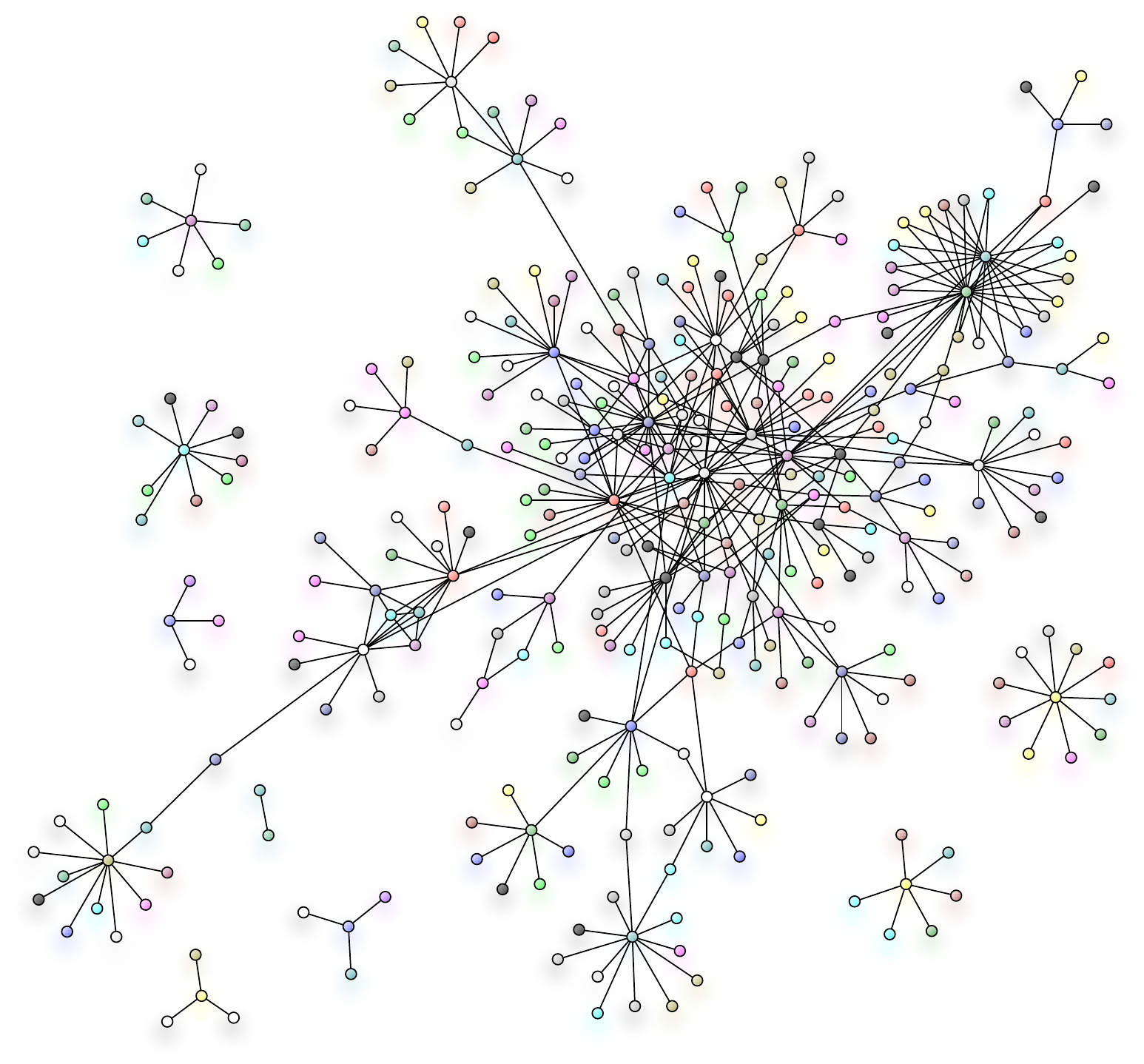}{graffle}{Abstract View of An Ecosystem}{Showing different populations (by the different colours) in different spatial areas, and their connection to one another by lines. Included are communities of populations that have become geographically separated and so are not connected to the main network of the ecosystem, and which could potentially give rise to allopatric (geographic) speciation \cite{lawrence1989hsd}.}{-2mm}{!h}{}{}

A key factor in the maintenance of diversity in biological ecosystems is spatial interactions, and several modelling systems have been used to represent these spatial interactions, including metapopulations\footnote{A metapopulation is a collection of relatively isolated, spatially distributed, local populations bound together by occasional dispersal between populations \cite{levins1969sda, hanski1999me, hanski2003mtf}.}, diffusion models, cellular automata and agent-based models (termed individual-based models in ecology) \cite{Greenetal2006}. The broad predictions of these diverse models are in good agreement. At local scales, spatial interactions favour relatively abundant species disproportionately. However, at a wider scale, this effect can preserve diversity, because different species will be locally abundant in different places. The result is that even in homogeneous environments, population distributions tend to form discrete, long-lasting patches that can resist an invasion by superior competitors \cite{Greenetal2006}. Population distributions can also be influenced by environmental variations such as barriers, gradients, and patches. The possible behaviour of spatially distributed ecosystems is so diverse that scenario-specific modelling is necessary to understand any real system \cite{suzie}. Nonetheless, certain robust patterns are observed. These include the relative abundance of species, which consistently follows a roughly log-normal relationship \cite{Bell}, and the relationship between geographic area and the number of species present, which follows a power law \cite{sizling2004pls}. The reasons for these patterns are disputed, because they can be generated by both spatial extensions of simple Lotka-Volterra competition models \cite{Hubbell}, and more complex ecosystem models \cite{Sole}. 

Landscape connectivity plays an important part in ecosystems. When the density of habitats within an environment falls below a critical threshold, widespread species may fragment into isolated populations. Fragmentation can have several consequences. Within populations, these effects include loss of genetic diversity and detrimental inbreeding \cite{GreenKirley}. At a broader scale, isolated populations may diverge genetically, leading to speciation, as shown in Figure \ref{figure2new}. 

From an information theory perspective, this phase change in landscape connectivity can mediate global and local search strategies \cite{Greenetal2000}. In a well-connected landscape, selection favours the globally superior, and pursuit of different evolutionary paths is discouraged, potentially leading to premature convergence. When the landscape is fragmented, populations may diverge, solving the same problems in different ways. Recently, it has been suggested that the evolution of complexity in nature involves repeated landscape phase changes, allowing selection to alternate between local and global search \cite{Greenetalinpress}. 

In a digital context, we can have spatial interactions by using a distributed system that consists of a set of interconnected locations, with agents that can migrate between these connected locations. In such systems the spatial dynamics are relatively simple compared with those seen in real ecosystems, which incorporate barriers, gradients, and patchy environments at multiple scales in continuous space \cite{begon96}. Nevertheless, depending on how the connections between locations are organised, such Digital Ecosystems might have dynamics closely parallel to spatially explicit models, diffusion models, or metapopulations \cite{suzie}. We will discuss later the use of a dynamic non-geometric spatial network, and the reasons for using this approach.

\subsubsection{Stability and Diversity in Complex Adaptive Systems}

Ecosystems are often described as \ac{CAS}: that is, systems made from diverse, locally interacting components that are subject to selection. Other \ac{CAS} include brains, individuals, economies, and the biosphere. All are characterised by hierarchical organisation, continual adaptation and novelty, and non-equilibrium dynamics. These properties appear to produce behaviour that is non-linear, historically contingent, subject to thresholds, and contains multiple basins of attraction \cite{Levin}.

In the previous subsections, we have advocated Digital Ecosystems that include agent populations evolving by natural selection in distributed environments. Like real ecosystems, digital systems designed in this way fit the definition of \ac{CAS}. The features of these systems, especially non-linearity and non-equilibrium dynamics, offer both advantages and hazards for adaptive problem-solving. The major hazard is that the dynamics of \ac{CAS} are intrinsically hard to predict because of the non-linear emergent self-organisation \cite{levin1999fdc}. This observation implies that designing a useful Digital Ecosystem will be partly a matter of trial and error. The occurrence of multiple basins of attraction in \ac{CAS} suggests that even a system that functions well for a long period may suddenly at some point transition to a less desirable state \cite{folke}. For example, in some types of system self-organising mass extinctions might result from interactions among populations, leading to temporary unavailability of diverse solutions \cite{newman1997mme}. This concern may be addressed by incorporating negative feedback or other mechanisms at the global scale. The challenges in designing an effective Digital Ecosystem are mirrored by the system's potential strengths. Non-linear behaviour provides the opportunity for scalable organisation and the evolution of complex hierarchical solutions, while rapid state transitions potentially allow the system to adapt to sudden environmental changes with minimal loss of functionality \cite{Levin}.

A key question for designers of Digital Ecosystems is how the stability and diversity properties of biological ecosystems map to performance measures in digital systems. For a Digital Ecosystem the ultimate performance measure is user satisfaction, a system-specific property. However, assuming the motivation for engineering a Digital Ecosystem is the development of scalable, adaptive solutions to complex dynamic problems, certain generalisations can be made. Sustained diversity \cite{folke}, is a key requirement for dynamic adaptation. In Digital Ecosystems, diversity must be balanced against adaptive efficiency because maintaining large numbers of poorly-adapted solutions is costly. The exact form of this trade-off will be guided by the specific requirements of the system in question. Stability \cite{Levin}, is likewise, a trade-off: we want the system to respond to environmental change with rapid adaptation, but not to be so responsive that mass extinctions deplete diversity or sudden state changes prevent control.

\subsection{Computing of Digital Ecosystems}
\label{compOfDE}

The value of creating analogies varies substantially depending on the behaviours or constructs being compared. For example, both organisms and computers have mechanisms to ensure data integrity. In computer systems, that integrity is absolute: data replication which introduces even the most minor change is considered to have failed, and is supported by mechanisms such as the \acl*{MD5} \cite{rivest1992rmm}. Similarly, in biological systems the genetic code is transcribed with a remarkable degree of fidelity due to elaborate evolved proof-reading and correction systems; there is, approximately, only one unforced error per one hundred bases copied \cite{Kunkel2004}. Despite their similar function, the operational control mechanisms for error-reduction in biological versus computing systems are radically different, making any parallel between the two misleading. Thus considerable finesse is required when inventing and applying such analogies. 

\changed{In the following paragraphs, we explore the analogies implicit \added{between biological ecosystems and potential Digital Ecosystems \cite{epi, acmMedes}. We will consider \aclp{MAS} to explore the references to \emph{agents} and \emph{migration}; followed by evolutionary computing and \aclp{SOA} for the references to \emph{evolution} and \emph{self-organisation}.}}

\subsubsection{Multi-Agent Systems}

A \emph{software agent} is a piece of software that acts, for a user in a relationship of agency, autonomously in an environment to meet its designed objectives \cite{wooldridge}. A \ac{MAS} is a system composed of several software agents, collectively capable of reaching goals that are difficult to achieve by an individual agent or monolithic system \cite{wooldridge}. Conceptually, there is a strong parallel between the software agents of a \ac{MAS} and the agent-based models of a biological ecosystem \cite{Greenetal2006}, despite the lack of evolution and migration in a \ac{MAS}. There is an even stronger parallel to a variant of \acp{MAS}, called \emph{mobile agent systems}, in which the mobility also mirrors the migration in biological ecosystems \cite{moaspaper}.

\tfigure{scale=1.0}{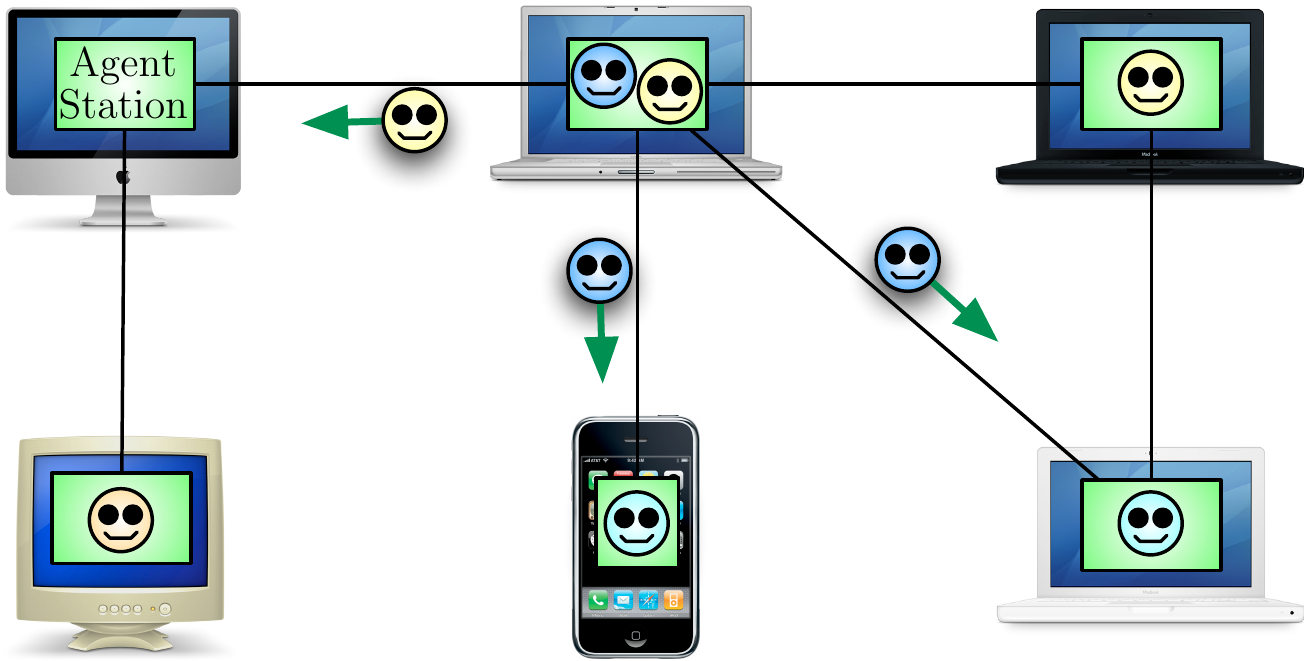}{graffle}{Mobile Agent System}{Visualisation that shows mobile agents as programmes that can migrate from one host to another in a network of heterogeneous computer systems and perform a task specified by its owner. On each host they visit, mobile agents need special software called an agent station, which is responsible for executing the agents and providing a safe execution environment \cite{agentStation}.}{-2mm}{!h}{}{}

The term \emph{mobile agent} contains two separate and distinct concepts: mobility and agency \cite{rothermel1998ma}. Hence, mobile agents are software agents capable of movement within a network \cite{moaspaper}. The mobile agent paradigm treats networks as multiple agent-friendly environments and the agents as programmatic entities that move from location to location, performing tasks for users. On each host they visit, mobile agents need software which is responsible for their execution in a safe  environment \cite{moaspaper}.

Generally, there are three types of design for mobile agent systems \cite{moaspaper}: (1) using a specialised operating system, (2) as operating system services or extensions, or (3) as application software. The first approach has the operating system providing the requirements of mobile agent systems directly \cite{svahnberg}. The second approach implements the mobile agent system requirements as operating system extensions, taking advantage of existing features of the operating system \cite{johansen1995oss}. Lastly, the third approach builds mobile agent systems as specialised application software that runs on top of an operating system, to provide for the mobile agent functionality, with such software being called an \emph{agent station} \cite{agentStation}. In this last approach, each agent station hides the vendor-specific aspects of its host platform, and offers standardised services to visiting agents. Services include access to local resources and applications; for example, web servers or web services, the local exchange of information between agents via message passing, basic security services, and the creation of new agents \cite{agentStation}. Also, the third approach is the most platform-agnostic, and is visualised in Figure \ref{mobileAgents}.

\subsubsection{Evolutionary Computing}

For evolving software in Digital Ecosystems evolutionary computing is the logical field from which to start. In \ac{BIC}, one of the primary sources of inspiration from nature has been evolution \cite{ecpaper}. Evolution has been clearly identified as the source of many diverse and creative solutions to problems in nature \cite{ec15, ec16}. However, it can also be useful as a problem-solving tool in artificial systems. Computer scientists and other theoreticians realised that the selection and mutation mechanisms so effective in biological evolution could be abstracted for implementation in a computational algorithm \cite{ecpaper}. Evolutionary computing is now recognised as a sub-field of artificial intelligence (more particularly computational intelligence) that involves combinatorial\added{, continuous optimisation, and mixed-integer} optimisation problems \cite{ec17}.

Evolutionary algorithms are based upon several fundamental principles from biological evolution, including reproduction, mutation, recombination (crossover), natural selection, and survival of the fittest. As in biological systems, evolution occurs by the repeated application of the above operators \cite{back1996eat}. An evolutionary algorithm operates on a set of individuals, called a population. An individual, in the natural world, is an organism with an associated fitness \cite{lawrence1989hsd}. Candidate solutions to an optimisation problem play the role of individuals in a population, and a cost function determines the environment within which the solutions live, analogous to the way the environment selects for the fittest individuals. Candidate solutions to an optimisation problem play the role of individuals in a population, and a cost function determines the environment by selecting for the fittest individuals. The number of individuals varies between different implementations and may also vary through time during the use of the algorithm. Each individual possesses some characteristics that are defined through its genotype, its genetic composition. These characteristics may be passed on to descendants of that individual \cite{back1996eat}. Processes of mutation (small random changes) and crossover (generation of a new genotype by the combination of components from two individuals) may occur, resulting in new individuals with genotypes different from the ancestors they will come to replace. These processes iterate, modifying the characteristics of the population \cite{back1996eat}. Which members of the population are kept, or are used as parents for offspring, will often depend upon some external characteristic, called the fitness (cost) function of the population. It is this that enables improvement to occur \cite{back1996eat}, and corresponds to the fitness of an organism in the natural world \cite{lawrence1989hsd}. Recombination and mutation create the necessary diversity and thereby facilitate novelty, while selection acts as a force increasing quality. Changed pieces of information resulting from recombination and mutation are randomly chosen. However, selection operators can be either deterministic, or stochastic. In the latter case, individuals with a higher fitness have a higher chance to be selected than individuals with a lower fitness \cite{back1996eat}.

There are different strands of what has become called evolutionary computing \cite{back1996eat}. The first is genetic algorithms. A second strand, evolution strategies, focuses \changed{on} \added{continuous and mixed-integer types of problems}. A third strand, evolutionary programming, originally developed from machine intelligence motivations. These areas developed separately for about fifteen years, but from the early nineties they are seen as different representatives (dialects) of one technology, called evolutionary computing \cite{eiben2003iec}. In the early nineties, another fourth stream following the general ideas had emerged, called genetic programming \cite{eiben2003iec}. \changed{Genetic programming \cite{ec25} can be considered as a variant of genetic algorithms where individual genotypes are represented by executable programmes. Specifically, solutions are represented as trees of expressions in an appropriate programming language, with the aim of evolving the most effective programme for solving a particular problem \cite{banzhaf1998gpi}.}

Many important questions remain to be answered in understanding the performance of evolutionary algorithms. For example, current evolutionary algorithms for evolving programmes (genetic programming) suffer from some weaknesses. First, while being moderately successful at evolving simple programmes, it is very difficult to scale them to evolve high-level software components \cite{mantere2005ese}. Second, the estimated fitness of a programme is normally given by a measure of how accurately it computes a given function, as represented by a set of input and output pairs, and therefore there is only a limited guarantee that the evolved programme actually does the intended computation \cite{mantere2005ese}. These issues are particularly important when evolving high-level, complex, structured software.

\changed{To evolve high-level software components in Digital Ecosystems, we will take advantage of the native method of software advancement, human developers, by using evolutionary computing for \emph{combinatorial optimisation} \cite{papadimitriou1998coa} of available software services.} \added{This would involve treating developer-produced software services as the functional building blocks, as the base unit in a genetic-algorithms-based process.} \changed{Such an approach will require a modular reusable paradigm to software development, such as \acp{SOA}, which are discussed in the following subsection.}

\subsubsection{Service-Oriented Architectures}

\label{soasection}

Our approach to evolving high-level software applications requires a modular reusable paradigm to software development. \acp{SOA} are the current state-of-the-art approach, being the current iteration of interface/component-based design from the 1990s, which was itself an iteration of event-oriented design from the 1980s, and before then modular programming from the 1970s \cite{histOfProg, krafzig2004ess}. Service-oriented computing promotes assembling application components into a loosely coupled network of services, to create flexible, dynamic business processes and agile applications that span organisations and computing platforms \cite{papazoglou2003soc}. This is achieved through a \ac{SOA}, an architectural style that guides all aspects of creating and using business processes throughout their life-cycle, packaged as services. This includes defining and provisioning infrastructure that allows different applications to exchange data and participate in business processes, loosely coupled from the operating systems and programming languages underlying the applications \cite{soa1w}. Hence, a \ac{SOA} represents a model in which functionality is decomposed into distinct units (services), which can be distributed over a network, and can be combined and reused to create business applications \cite{papazoglou2003soc}.

\tfigure{scale=0.25}{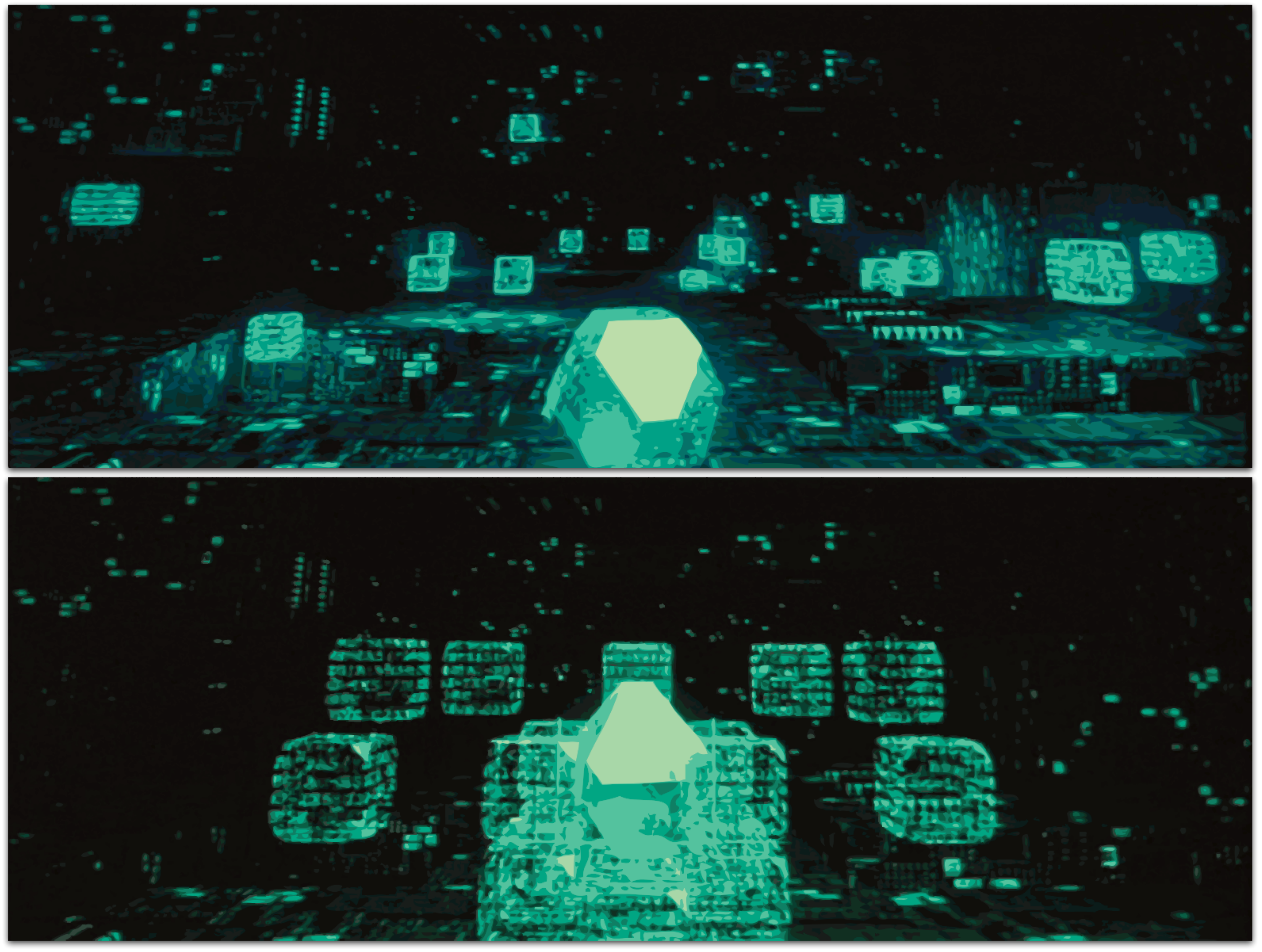}{graffle}{\added{Service-Oriented Architectures}}{\added{Abstract visualisations, with the first image showing the loosely joined services as cuboids, and the service orchestration as a polyhedron; and the second image showing their high interoperability and re-usability in forming applications, from the use of standardised interfaces and external service orchestration.}}{-1mm}{!b}{}{}

A \ac{SOA} depends upon service-orientation as its fundamental design principle. In a \ac{SOA} environment, independent services can be accessed without knowledge of their underlying platform implementation \cite{soa1w}. Services reflect a service-oriented approach to programming that is based on composing applications by discovering and invoking network-available services to accomplish some task. This approach is independent of specific programming languages or operating systems, because the services communicate with each other by passing data from one service to another, or by co-ordinating an activity between two or more services \cite{papazoglou2003soc}. So, the concepts of \acp{SOA} are often seen as built upon, and the development of, the concepts of modular programming and distributed computing \cite{krafzig2004ess}.

\acp{SOA} allow for an information system architecture that enables the creation of applications that are built by combining loosely coupled and interoperable services \cite{soa1w}. They typically implement functionality most people would recognise as a service, such as filling out an online application for an account, or viewing an online bank statement \cite{krafzig2004ess}. Services are intrinsically unassociated units of functionality, without calls to each other embedded in them. Instead of services embedding calls to each other in their source code, protocols are defined which describe how services can talk to each other, in a process known as orchestration, to meet new or existing business system requirements \cite{singh2005soc}. This is allowing an increasing number of third-party software companies to offer software services, such that \ac{SOA} systems will come to consist of such third-party services combined with others created in-house, which has the potential to spread costs over many users and uses, and promote standardisation both in and across industries \cite{chhatpar2008}. \added{For example, the travel industry now has a well-defined, and documented, set of both services and data, sufficient to allow any competent software engineer to create travel agency software using entirely off-the-shelf software services \cite{kotok2001eng, cardoso2005isw}. Other industries, such as the finance industry, are also making significant progress in this direction \cite{zimmermann2004sgw}.}

The vision of \acp{SOA} assembling application components from a loosely coupled network of services, that can create dynamic business processes and agile applications that span organisations and computing platforms\added{, as shown in Figure \ref{SOAvecFinal}}. It will be made possible by creating compound solutions that use internal organisational software assets, including enterprise information and legacy systems, and combining these solutions with external components residing in remote networks \cite{SOApaper0}. The great promise of \acp{SOA} is that the \emph{marginal cost} of creating the n-th application is virtually zero, as all the software required already exists to satisfy the requirements of other applications. Only their combination and orchestration are required to produce a new application \cite{tang2004ews, modi2008}. The key is that the interactions between the chunks are not specified within the chunks themselves. Instead, the interaction of services (all of whom are hosted by unassociated peers) is specified by users in an ad-hoc way, with the intent driven by newly emergent business requirements \cite{leymann2002wsa}. 

The pinnacle of \ac{SOA} interoperability, is the exposing of services on the internet as web services \cite{soa1w}. A web service is a specific type of service that is identified by a \acl*{URI}, whose service description and transport utilise open Internet standards. Interactions between web services typically occur as \acl*{SOAP} calls carrying \acl*{XML} data content. The interface descriptions of web services are expressed using the \ac{WSDL} \cite{SOApaper2}. The \ac{UDDI} standard defines a protocol for directory services that contain web service descriptions. \ac{UDDI} enables web service clients to locate candidate services and discover their details. Service clients and service providers utilise these standards to perform the basic operations of \acp{SOA} \cite{SOApaper2}. Service aggregators can then use the \ac{BPEL} to create new web services by defining corresponding compositions of the interfaces and internal processes of existing services \cite{SOApaper2}.

\ac{SOA} services inter-operate based on a formal definition (or contract, e.g. \ac{WSDL}) that is independent of the underlying platform and programming language. Service descriptions are used to advertise the service capabilities, interface, behaviour, and quality \cite{SOApaper2}. The publication of such information about available services provides the necessary means for discovery, selection, binding, and composition of services \cite{SOApaper2}. The expected behaviour of a service during its execution is described by its behavioural description (for example, as a workflow process). Also, included is a \acl*{QoS} description, which publishes important functional and non-functional service quality attributes, such as service metering and cost, performance metrics (response time, for instance), security attributes, integrity (transactional), reliability, scalability, and availability \cite{SOApaper2}. Service clients (end-user organisations that use some service) and service aggregators (organisations that consolidate multiple services into a new, single service offering) utilise service descriptions to achieve their objectives \cite{SOApaper2}. \changed{One of the most important and continuing developments in \acp{SOA} is \ac{SWS}, which make use of semantic descriptions for service discovery, so that a client can discover the services semantically \cite{SOAsemantic, cabral2004asw},} \added{i.e. using semantics to search for available web services}.

There are multiple standards available and still being developed for \acp{SOA} \cite{SOAstandards}, most notably of recent being \ac{REST} \cite{singh2005soc}. The software industry now widely implements a thin SOAP/WSDL/UDDI veneer atop existing applications or components that implement the web services paradigm \cite{SOApaper0}, but the choice of technologies will change with time. Therefore, the fundamentals of \acp{SOA} and its services are best defined generically, because \acp{SOA} are technology agnostic and need not be tied to a specific technology \cite{papazoglou2003soc}. Within the current and future scope of the fundamentals of \acp{SOA}, there is clearly potential to evolve complex high-level software applications from the modular services of \acp{SOA}, instead of the instruction level evolution currently prevalent in genetic programming \cite{overviewGP}.

\subsubsection{Distributed Evolutionary Computing}

Having previously introduced evolutionary computing, and the possibility of it occurring within a distributed environment, not unlike those found in mobile agent systems, leads us to consider a specialised form known as \acl*{DEC}. \changed{The motivation for using parallel or distributed evolutionary algorithms is twofold: first, to improve the speed of evolutionary processes by conducting concurrent evaluations of individuals in a population; second, to improve the problem-solving process, overcoming difficulties that face traditional evolutionary algorithms, such as maintaining diversity to avoid premature convergence \cite{muhlenbein1991eta, stender1993pga}.} The fact that evolutionary computing manipulates a population of independent solutions actually makes it well suited for parallel and distributed computation architectures \cite{cantupaz1998spg}. There are several variants of distributed evolutionary computing, leading some to propose a taxonomy for their classification \cite{nowostawski1999pga}, with there being two main forms \cite{cantupaz1998spg, stender1993pga}:
\begin{itemize}
\item multiple-population/coarse-grained migration/island models
\item single-population/fine-grained diffusion/neighbourhood models
\end{itemize}

In the coarse-grained \emph{island} models \cite{lin1994cgp, cantupaz1998spg}, evolution occurs in multiple parallel sub-populations (islands), each running a local evolutionary algorithm, evolving independently with occasional migrations of highly fit individuals among sub-populations. The core parameters for the evolutionary algorithm of the island-models are as follows \cite{lin1994cgp}:
\begin{itemize}
\item number of the sub-populations: 2, 3, 4, more
\item sub-population homogeneity
\begin{itemize}
\item size, crossover rate, mutation rate, migration interval
\end{itemize}
\item topology of connectivity: ring, star, fully-connected, random
\item static or dynamic connectivity
\item migration mechanisms: 
\begin{itemize}
\item isolated/synchronous/asynchronous
\item how often migrations occur 
\item which individuals migrate 
\end{itemize}
\end{itemize}

Fine-grained \emph{diffusion} models \cite{manderick1989fgp, stender1993pga} assign one individual per processor. A local neighbourhood topology is assumed, and individuals are allowed to mate only within their neighbourhood, called a deme\footnote{In biology a deme is a term for a local population of organisms of one species that actively interbreed with one another and share a distinct gene pool \cite{devisser2007ese}.}. The demes overlap by an amount that depends on their shape and size, and in this way create an implicit migration mechanism. Each processor runs an identical evolutionary algorithm which selects parents from the local neighbourhood, produces an offspring, and decides whether to replace the current individual with an offspring. However, even with the advent of multi-processor computers, and more recently multi-core processors, which provide the ability to execute multiple threads simultaneously \cite{newsArticle3}, this approach would still prove impractical in supporting the number of agents necessary to create a Digital Ecosystem. Therefore, we shall further consider the \emph{island} models.

\tfigure{scale=0.8}{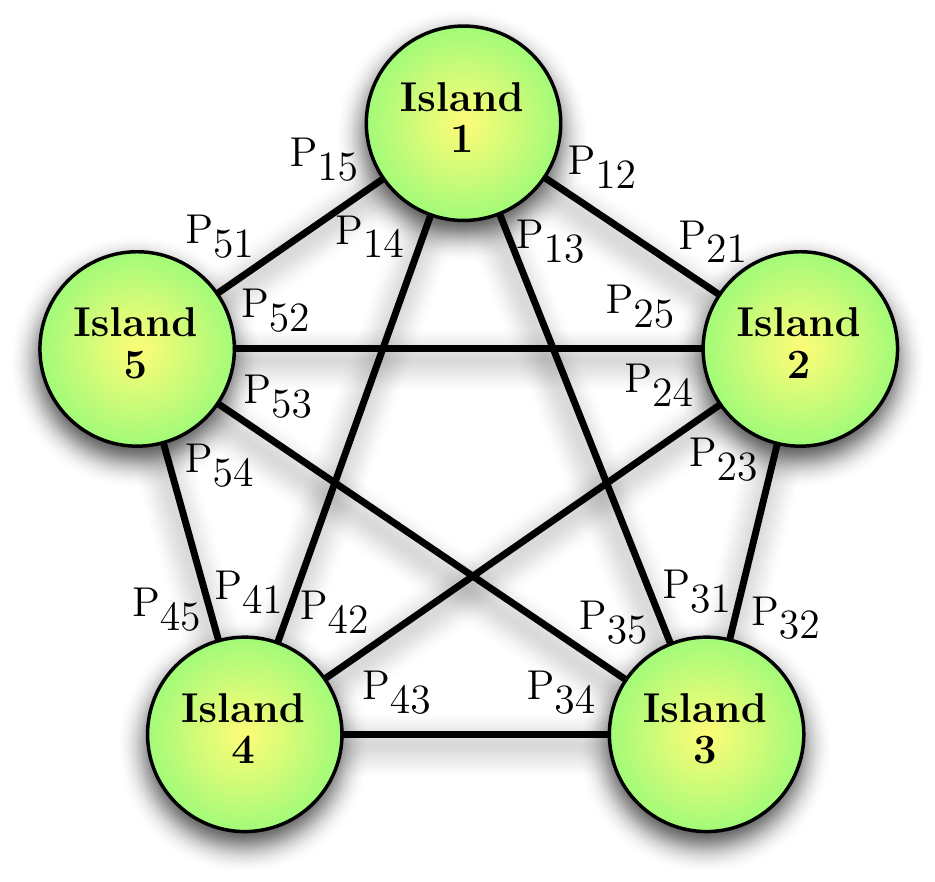}{graffle}{Island-Model of Distributed Evolutionary Computing}{\cite{lin1994cgp, cantupaz1998spg}: There \getCap{im1} This \getCap{im2}}{-4mm}{!h}{}{}

An example island-model \cite{lin1994cgp, cantupaz1998spg} is visualised in Figure \ref{islandModel}, in which there \setCap{are different probabilities of going from island \Circ{1} to island \Circ{2}, as there is of going from island \Circ{2} to island \Circ{1}.}{im1} This allows maximum flexibility for the migration process, and \setCap{mirrors the naturally inspired quality that although two populations have the same physical separation, it may be easier to migrate in one direction than the other, i.e. fish migration is easier downstream than upstream.}{im2} The migration of the \emph{island} models is like the notion of migration in nature, being similar to the metapopulation models of theoretical ecology \cite{levins1969sda}. \changed{However, the islands in this approach all work on the same exact problem, making it less analogous to biological ecosystems in which different locations can be environmentally different \cite{begon96}}. We will take advantage of this property later when defining \acp*{EOA} for Digital Ecosystems.

\subsection{Digital Business Ecosystems}

The questions we have raised are wide-ranging, and have motivated several interdisciplinary research teams, including those involved in an EU Framework VI project called \acp{DBE}. The \ac{DBE} is a proposed methodology for economic and technological innovation. Specifically, the \ac{DBE} is a software infrastructure for supporting large numbers of interacting business users and services \cite{dbebkintro}. The \ac{DBE} aims to be a next generation \ac{ICT} that will extend the \ac{SOA} concept with the automatic combining of available and applicable services in a scalable architecture, to meet business user requests for applications that facilitate business processes. In essence, the \ac{DBE} will be an internet-based environment in which businesses will be able to interact with each other in very effective and efficient ways \cite{dbebook}. 

The synthesis of the concept of Digital Business Ecosystems emerged by adding \cite{nachira} \emph{digital} in front of \emph{business ecosystem} \cite{moore1996}. The term Digital Business Ecosystem was used earlier, but with a focus exclusively on developing countries \cite{moore2003}. The generalisation of the term to refer to a new interpretation of what socio-economic development catalysed by \ac{ICT} means was new, emphasising the co-evolution between the business ecosystem and its partial digital representation: the digital ecosystem. The term Digital Business Ecosystem came to represent the combination of the two ecosystems \cite{dbebkintro}.

\tfigure{scale=0.35}{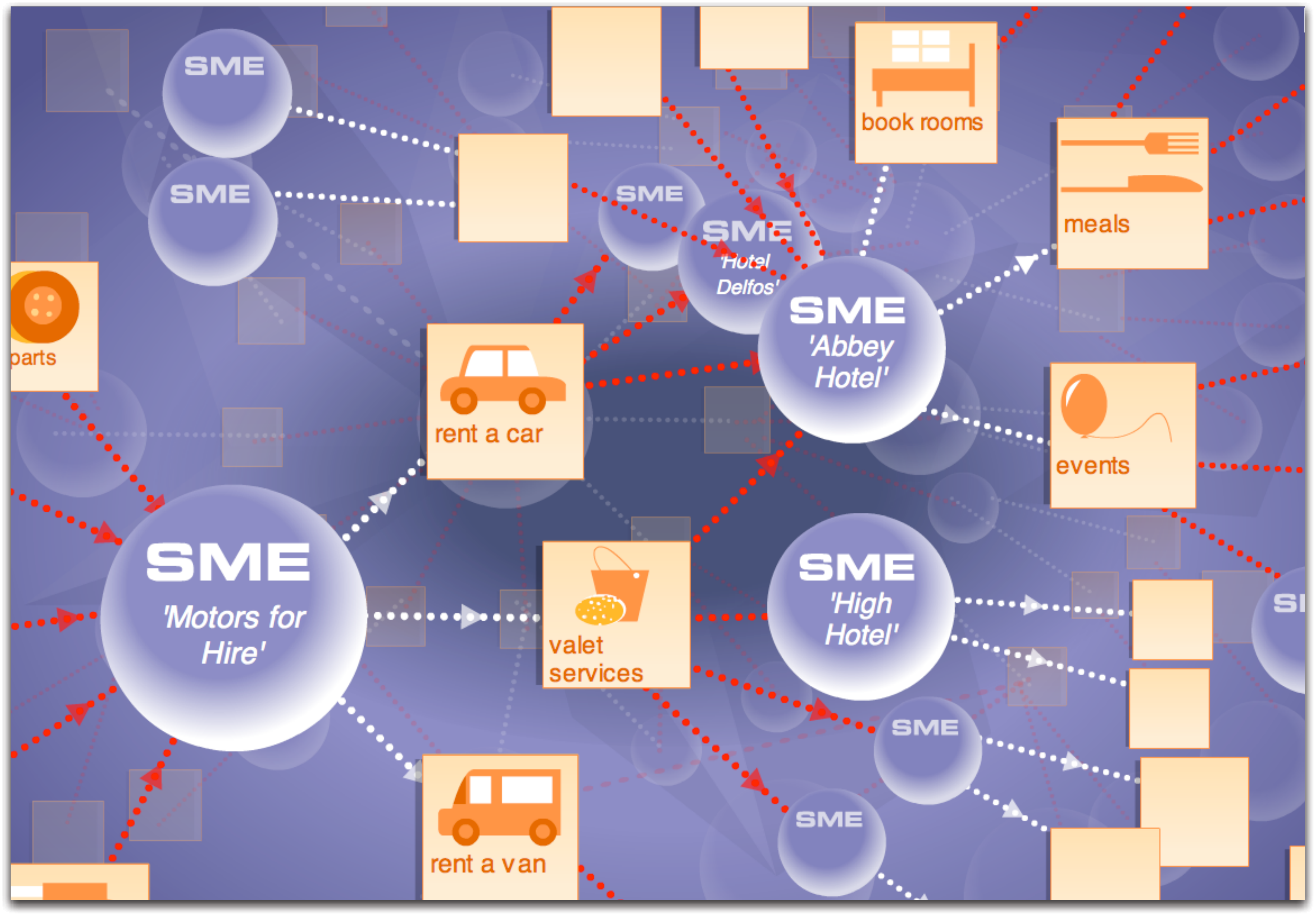}{graffle}{Business Ecosystem}{\cite{dbebkintro}: Conceptual visualisation \cite{dbevis} showing a Business Ecosystem of interacting \ac{SME} users, via the services they provide and consume. Creating a network of business ecosystems distributed over different geographical regions, business domains, and industry sectors.}{0mm}{}{}{}

The \emph{business ecosystem} is an economic community supported by a foundation of interacting organisations and individuals; i.e. the organisms of the business world. This economic community produces goods and services of value to customers, who are themselves members of the ecosystem \cite{moore1996}\tred{, as shown in Figure \ref{DBEprojectFinal}}. \setCap{A wealthy ecosystem sees a balance between co-operation and competition in a dynamic free market.}{DBEdescription} Regarding a particular \emph{business ecosystem}, two main different interpretations of its structure have been discussed in the literature. The keystone model has a structure in which a \emph{business ecosystem} is dominated by a large firm that is surrounded by many small suppliers \cite{iansiti2004kan}. This model works well when the central firm is healthy, but represents a significant weakness for the economy of the region when the dominant economic actor experiences difficulties \cite{moore1996}. This model also matches the economic structure of the USA where there is a predominant number of large enterprises at the centre of large value networks of suppliers \cite{iansiti2004kan}. However, the model for a \emph{business ecosystem} developed in Europe is less structured and more dynamic; it is composed mainly of \acp{SME}, but can accommodate large firms \cite{eurostat2006}. All actors complement one another, leading to a more dynamic division of labour, organised along one-dimensional value chains and two-dimensional value networks \cite{corallo2007}. This model is particularly well-adapted for the service and knowledge industries, where it is easier for small firms to reinvent themselves than, for instance, in the automotive industry which is dominated by large enterprises \cite{dbebkintro}.

In the DBE, the \emph{digital ecosystem} is the technical infrastructure, based on a \acl*{P2P} distributed software technology that transports, finds, and connects services and information over Internet links enabling networked transactions, and the distribution of all the digital objects present within the infrastructure \cite{dbebkintro}. Such organisms of the digital world encompass any useful digital representations expressed by languages (formal or natural) that can be interpreted and processed (by computer software and/or humans), e.g. software applications, services, knowledge, taxonomies, folksonomies, ontologies, descriptions of skills, reputation and trust relationships, training modules, contractual frameworks, laws \cite{dbebkintro}. So, the Digital Business Ecosystem is a biological metaphor that highlights the interdependence of all actors in the business environment, who co-evolve their capabilities and roles \cite{moore1996}, and which has attempted to develop an isomorphic model between biological behaviour and the behaviour of the digital ecosystem, leading to an evolutionary, self-organising, and self-optimising environment built upon an underlying \ac{SOA} \cite{dbebook}. 

The \ac{DBE} aims to help local economic actors become active players in globalisation \cite{dini2008bid}, valorising their local culture and vocations, enabling them to interact and create value networks at the global level. Increasingly this approach, dubbed glocalisation, is being considered a successful strategy of globalisation that preserves regional growth and identity \cite{robertson1994gog, swyngedouw1992mqg, khondker2004}, and has been embraced by the mayors and decision-makers of thousands of municipalities \cite{glocalforum2004}, because of the possible tension between globalisation and localisation when adopting \acp{ICT} \cite{castells2000}. 

The \ac{DBE} represents a \ac{B2B} interaction concept supported by a software platform (digital ecosystem) that is intended to have the desirable properties of biological ecosystems \cite{dbebokpaolo}, and its researchers also recognise the importance of \acp{SOA} in creating Digital Ecosystems \cite{razavi2007cmd, dbebook}. So, we will consider using the concept of a \emph{business ecosystem} as a potential user base for Digital Ecosystems.

\section{Ecosystem-Oriented Architectures}

We will now define the architectural principles of Digital Ecosystems. We will use our understanding of theoretical biology from section \ref{bioOfDE}, mimicking the processes and structures of life, evolution, and ecology of biological ecosystems. We will achieve this by combining elements from mobile agents systems, \acl*{DEC}, and \acp{SOA} from section \ref{compOfDE}, to create a hybrid architecture which is the digital counterpart of biological ecosystems.

We will refer to the agents of Digital Ecosystems as \emph{Agents}, populations as \emph{Populations}, and the habitats as \emph{Habitats}, to distinguish their new hybrid definitions from their original biological and computing definitions.

\subsection{Agents}

The Agents of the Digital Ecosystem are functionally analogous to the organisms of biological ecosystems, including the behaviour of migration and the ability to be evolved \cite{begon96}, and will be achieved through using a hybrid of different technologies. The ability to migrate is provided by using the paradigm of \emph{agent mobility} from mobile agent systems \cite{moaspaper}, with the Habitats of the Digital Ecosystem provided by the facilities of \emph{agent stations} from mobile agent systems \cite{agentStation}, i.e. a distributed network of locations to migrate to and from. The Habitats, and the Habitat network will be discussed later. The ability of the Agents to be evolved is in two parts: first, by using the interoperability of services from \acp{SOA} \cite{soa1w} to aggregate Agents; and second, the use of evolutionary computing \cite{eiben2003iec} for \emph{combinatorial optimisation} \cite{papadimitriou1998coa} at the Habitats to evolve \changed{aggregations of Agents, that are optimal} \added{compared to user requests for applications}. The Agents will take advantage of the interoperability of \acp{SOA} \cite{soa1w}, by acting in a relationship of agency \cite{wooldridge} to the user supplied \acp{SWS}, which will be \ac{SOA} compliant \cite{SOApaper2}. We can then evolve high-level software applications by using evolutionary computing for combinatorial optimisation of the available Agents, or rather the services they represent, in a genetic-algorithms-based \cite{goldberg} process. This makes an Agent, of the Digital Ecosystem, a \setCap{lightweight entity consisting primarily of a pointer to the \ac{SWS} it represents,}{serviceCap} including the service's \setCap{executable component and semantic description. A software service can be a software only service, e.g. data encryption, or provide a front-end to a real-world service, e.g. selling books}{service2cap}, as shown in Figure \ref{service}.

\tfigure{scale=0.6}{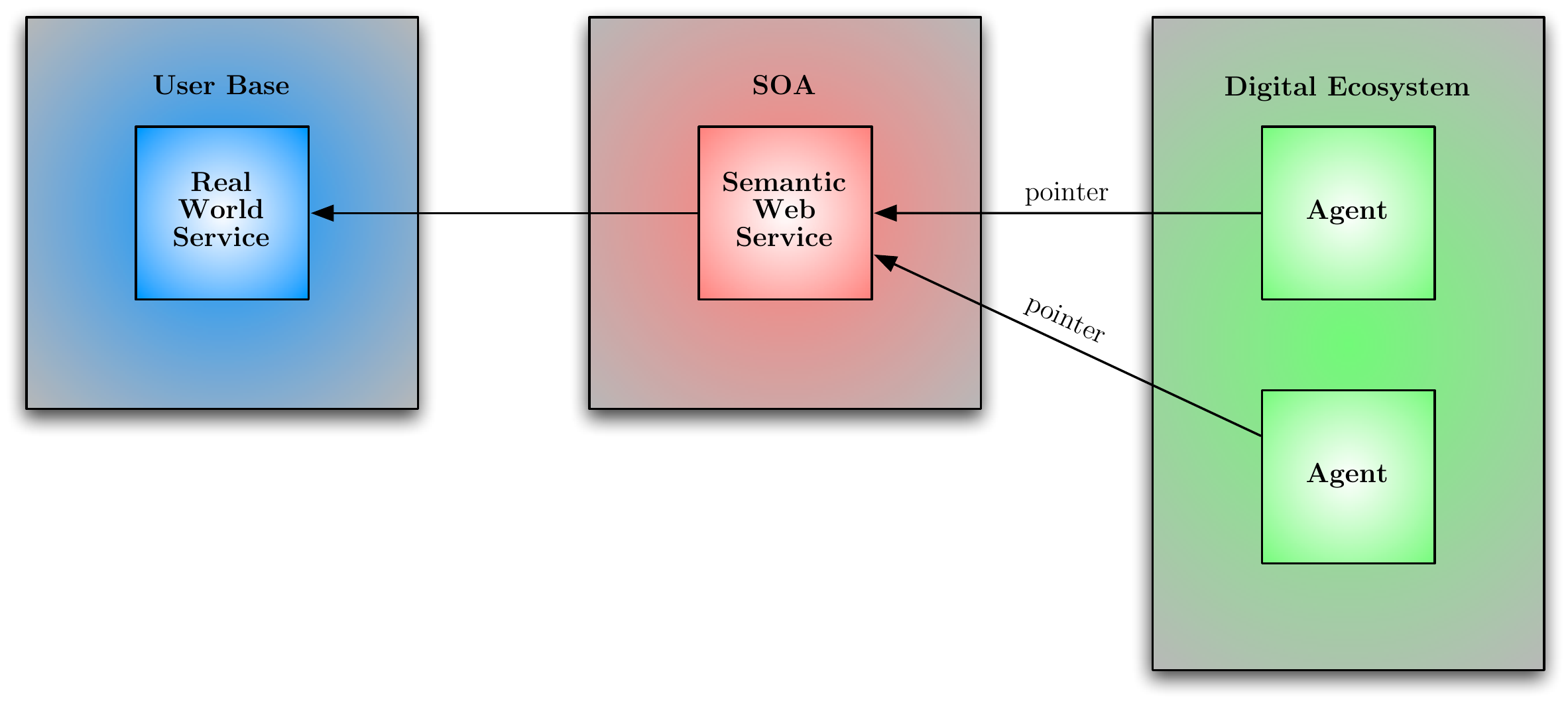}{graffle}{Agent of the Digital Ecosystem}{A \getCap{serviceCap} which is \acp{SOA} compliant and therefore includes an \getCap{service2cap}.}{-2mm}{}{}{}

An organism within Digital Ecosystems is an Agent, or an Agent aggregation created using evolutionary optimisation in response to a user request for an application. These Agents will migrate through the Habitat network of the Digital Ecosystem, adapting to find niches where they are useful in fulfilling other user requests for applications. The Agents will interact, evolve and adapt over time to the environment, thereby serving the ever-changing requirements of the user base.

The executable component, of a \ac{SWS} that an Agent represents, is equivalent to the DNA of an organism, whose sequence encodes the genetic information of living organisms and has two primary functions \cite{lawrence1989hsd}: the holder of virtually all information in inheritance, and the controller of protein synthesis for the construction and operation of its organism. Equivalently, the executable component is also the inheritable component from one generation to the next, and defines the objects and behaviour of its service's run-time instantiation.

The genotype of an individual describes the genetic constitution (DNA) of an individual, independent of its physical existence (the phenotype) \cite{lawrence1989hsd}. Equivalently, the semantic description, of a \ac{SWS} that an Agent represents, describes the functionality of the executable component. The phenotype of an individual arises from the combination of an organism's DNA and the environment \cite{lawrence1989hsd}. Equivalently, the run-time instantiation, of a service that an Agent represents, results from instantiating the executable component in the run-time environment. This differentiation between genotype and phenotype is fundamental for escaping local optima, and is often lacking in artificial evolutionary systems \cite{shackleton2000irg}, having instead a one-to-one genotype-phenotype mapping, in which the phenotype is directly encoded in the genotype with no differentiation provided by instantiation (development) \cite{shackleton2000irg}. Neutral genotype-phenotype mappings have this differentiation between the genotype and phenotype \cite{ec39}, which more strongly parallels biological evolution \cite{banzhaf1994gpm}. We therefore expect the use of a neutral genotype-phenotype mapping to help Digital Ecosystems demonstrate behaviour more akin to biological ecosystems.

\subsubsection{Agent Aggregation}

\tfigure{scale=0.8}{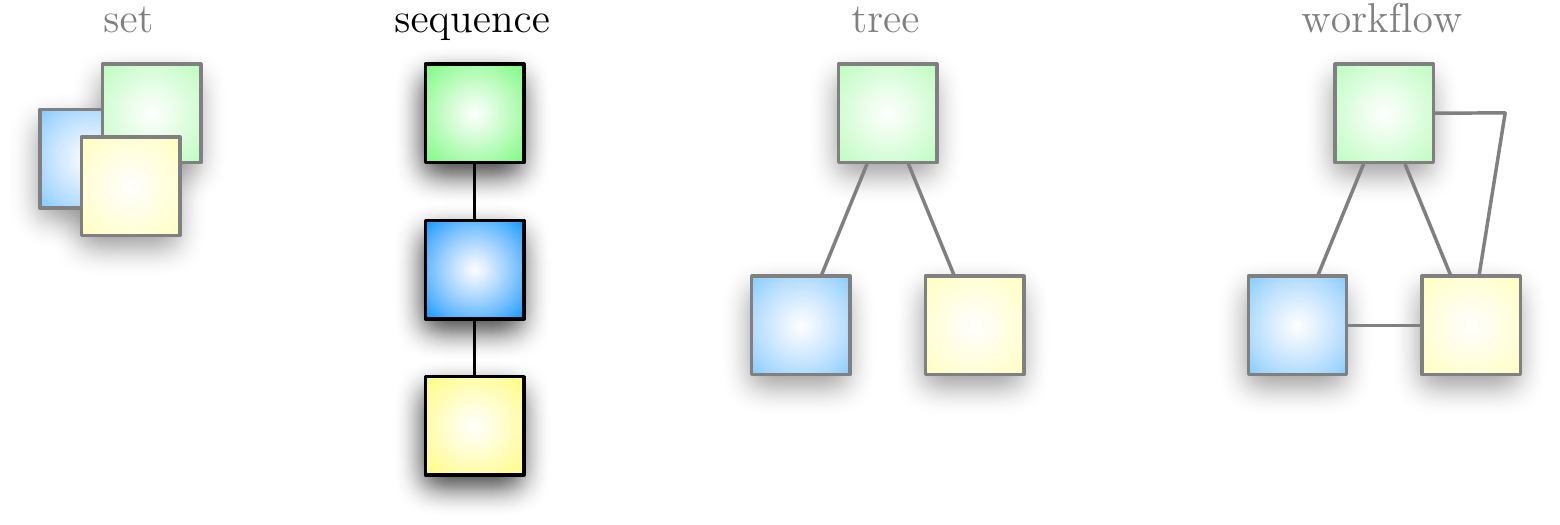}{graffle}{Structure of Aggregated Agents}{\getCap{structureCap}. Instead of \getCap{structure2}.}{-3mm}{!b}{-5mm}{}

\setCap{The executable component of a \ac{SWS}, that an Agent represents \added{via a pointer to the \ac{SWS}}, is equivalent to an organism's DNA and is the gene (functional unit) in the evolutionary process \cite{lawrence1989hsd}. So, the Agents should be aggregated as a sequence, like the sequencing of genes in DNA \cite{lawrence1989hsd}\added{, such that an aggregation of Agents is a sequence of Agents}.}{structureCap} It could be argued that the Agents should be aggregated as \setCap{an unordered set, or, based on service orchestration, into a tree or workflow}{structure2}, as shown in Figure \ref{agentStructures}. However, the aggregated structure of the Agents should not be the orchestration structure of the collection of software services that the Agents represent, not only because the service orchestration of the run-time instantiation is application domain-specific (e.g. trees in supply chain management \cite{lambert2000isc}, workflows in the travel industry \cite{benatallah2005frd}), but because it would also move it undesirably towards a one-to-one genotype-phenotype mapping \cite{shackleton2000irg}.

\subsection{Habitats}

The Habitats are the nodes of the Digital Ecosystem, and are functionally analogous to the habitats of a biological ecosystem \cite{lawrence1989hsd}. Their functionality is provided by using \setCap{the \emph{agent stations} from mobile agent systems \cite{agentStation} (to provide a distributed environment in which Agent migration can occur), with evolutionary computing \cite{eiben2003iec} for the Agent interaction (instead of traditional agent interaction mechanisms \cite{wooldridge}), and the island-model of \acl*{DEC} \cite{lin1994cgp} for the connectivity between Habitats}{habnet}\tred{, as shown in Figure \ref{architecture2}.} There will be a Habitat for each user, which the users will typically run locally, and through which they will submit requests for applications. Supporting this functionality, Habitats have the following core functions:

\begin{itemize}
\item Provide a subset of the Agents and Agent-sequences available globally, relevant to the user that the Habitat represents, and stored in what we will call an Agent-pool (for reasons that will be explained later). 
\item Accelerate, via the Agent-pool, the Populations instantiated to evolve optimal Agent-sequences in response to user requests for applications.
\item Manage the inter-Habitat connections for Agent migration.
\item For service providers; manage the distribution of Agents (which represent their services) to other users of the Digital Ecosystem, via the network of interconnected Habitats.
\end{itemize}

The collection of Agents at each Habitat (peer) will change over time, as the more successful Agents spread throughout the Digital Ecosystem, and as the less successful Agents are deleted. Successive user requests over time to their dedicated Habitats will make this process possible, because the continuous and varying user requests for applications provide a dynamic evolutionary pressure on the Agents, which have to evolve to better satisfy those requests. So, the Agents will recombine and evolve over time, constantly seeking to increase their effectiveness for the user base. The Agent is the base unit of the evolutionary process in Digital Ecosystems, in the same way that the gene is the base unit for evolution in biological ecosystems \cite{begon96}. So, the collection of Agents at each Habitat provides an Agent-pool, similar to a gene-pool, which is all the genes in a population \cite{lawrence1989hsd}. Additionally, it also stores Agent-sequences evolved from the Habitat's Populations, and Agent-sequences that migrate to the Habitat from other users' Habitats, because they can potentially accelerate future Populations instantiated to respond to user requests.

\tfigure{scale=0.8}{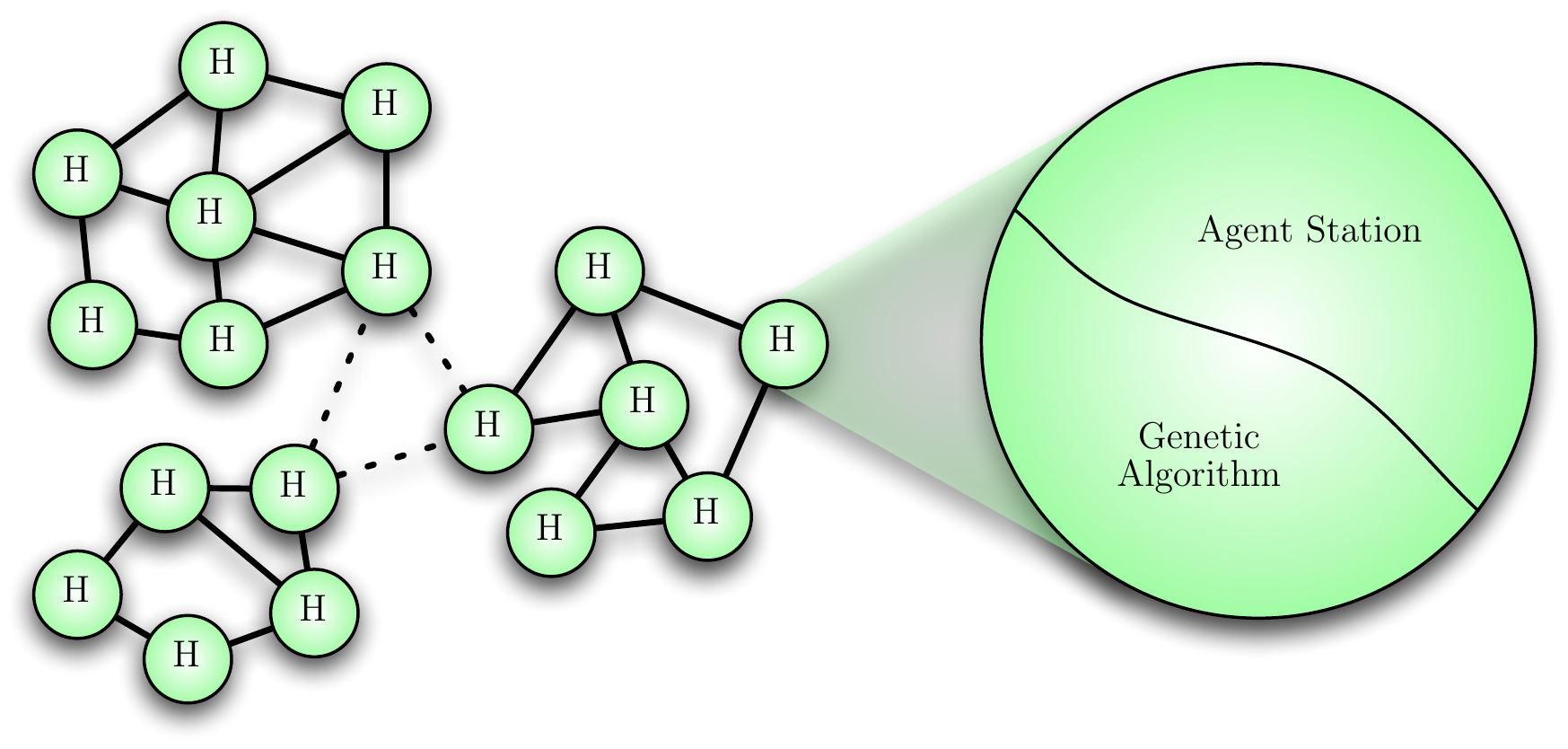}{graffle}{Habitat Network}{Uses \getCap{habnet}.}{-3mm}{!h}{}{}

The landscape, in energy-centric biological ecosystems, defines the connectivity between habitats \cite{begon96}. Connectivity of nodes in the digital world is generally not defined by geography or spatial proximity, but by information or semantic proximity. For example, connectivity in a peer-to-peer network is based primarily on bandwidth and information content, and not geography. The island-models of \acl*{DEC} use an information-centric model for the connectivity of nodes (islands) \cite{lin1994cgp}. However, because it is generally defined for one-time use (to evolve a solution to one problem and then stop) it usually has a fixed connectivity between the nodes, and therefore a fixed topology \cite{cantupaz1998spg}. \changed{So, supporting evolution in the Digital Ecosystem that has a dynamic multi-objective \emph{selection pressure} (fitness landscape \cite{wright1932} with many peaks) requires a re-configurable network topology, such that Habitat connectivity can be dynamically adapted based on the observed migration paths of the Agents between the users within the Habitat network. Therefore}, based on the island-models of \acl*{DEC} \cite{lin1994cgp}, each connection between the Habitats is bi-directional and there is a probability associated with moving in either direction across the connection, with the connection probabilities affecting the rate of migration of the Agents. \changed{Additionally, the connection probabilities will be updated by the success or failure of Agent migration using the concept of Hebbian learning \cite{hebb}: the Habitats which do not successfully exchange Agents will become less strongly connected, and the Habitats which do successfully exchange Agents will achieve stronger connections.} This leads to a topology that adapts over time, resulting in a network that supports and resembles the connectivity of the user base. When we later consider an example user base, we will further discuss a resulting topology. 

\changed{When a new user joins the Digital Ecosystem, a Habitat will be created for them, and most importantly connected to the correct cluster(s) in the Habitat network.} A new user's Habitat can be connected randomly to the Habitat network, as it will dynamically reconnect based upon the user's behaviour. \changed{User profiling can also be used to help connect a new user's Habitat to the optimal part of the network, finding a similar user or asking the user to identify a similar user and then cloning their Habitat's connections.} Also, when a new Habitat is created, its Agent-pool should be created by merging the Agent-pools of the Habitats to which it is initially connected.

\subsubsection{Agent Migration}
\label{migrationHistory}

The Agents will migrate through the interconnected Habitats combining with one another in Populations to meet user requests for applications. The migration path from the current Habitat is dependent on the \emph{migration probabilities} between the Habitats. The migration of an Agent within the Digital Ecosystem is initially triggered by deployment to its user's Habitat, for distribution to other users who will potentially make use of the service the Agent represents. When a user deploys a service, its representative Agent must be generated and deployed to their Habitat. It is then copied to the Agent-pool of the user's Habitat, and from there the migration of the Agent occurs, which involves migrating (copying) the agent probabilistically to all the connected Habitats. The Agent is copied rather than moved, because the Agent may also be of use to the providing user. The copying of an Agent to a connected Habitat depends on the associated migration probability. If the probability were one, then it would definitely be sent. When migration occurs, depending on the probabilities associated with the Habitat connections, an exact copy of the Agent is made at a connected Habitat. The copy of the Agent is identical until the new Agent's \emph{migration history} is updated, which differentiates it from the original. The successful use of the migrated Agent, in response to user requests for applications, will lead to further migration (distribution) and therefore availability of the Agent to other users.

The connections joining the Habitats are reinforced by successful Agent and Agent-sequence migration. The success of the migration, the \emph{migration feedback}, leads to the reinforcing and creation of migration links between the Habitats, just as the failure of migration leads to the weakening and negating of migration links between the Habitats. The success of migration is determined by the usage of Agents at the Habitats to which they migrate. When an Agent-sequence is found and used in responding to a user request, then the individual Agent \emph{migration histories} can be used to determine where they have come from and update the appropriate connection probabilities. If the Agent-sequence was fully or partly evolved elsewhere, then where the sequence or sub-sequences were created needs to be passed on to the connection probabilities, because the value in an Agent-sequence is the unique ordering and combination it provides of the individual Agents contained within. So, it is necessary to manage the feedback to the connection probabilities for migrating Agent-sequences, and not just the individual Agents contained within the sequence, including the partial use of an Agent-sequence in a newly evolved one. Specifically, the mechanism for \emph{migration feedback} needs to know the Habitats where migrating Agent-sequences were created, to create new connections or reinforce existing connections to these Habitats. The global effect of the Agent migration and \emph{migration feedback} on the Habitat network is the clustering of Habitats around the communities present within the user base, and will be discussed later in more detail.

The \emph{escape range} is the number of escape migrations available to an Agent upon the risk of death (deletion). If an Agent migrates to a Habitat and is not used after several user requests, then it will have the opportunity to migrate (move not copy) randomly to another connected Habitat. After this happens several times the Agent will be deleted (die). The \emph{escape range} will be dynamically responsive to the size of the Habitat cluster that the Agent exists within. This creates a dynamic time-to-live \cite{comer1988iti} for the Agents, in which Agents that are used more will live longer and distribute farther than those that are used less.

\subsection{Populations}

\tfigure{width=\textwidth}{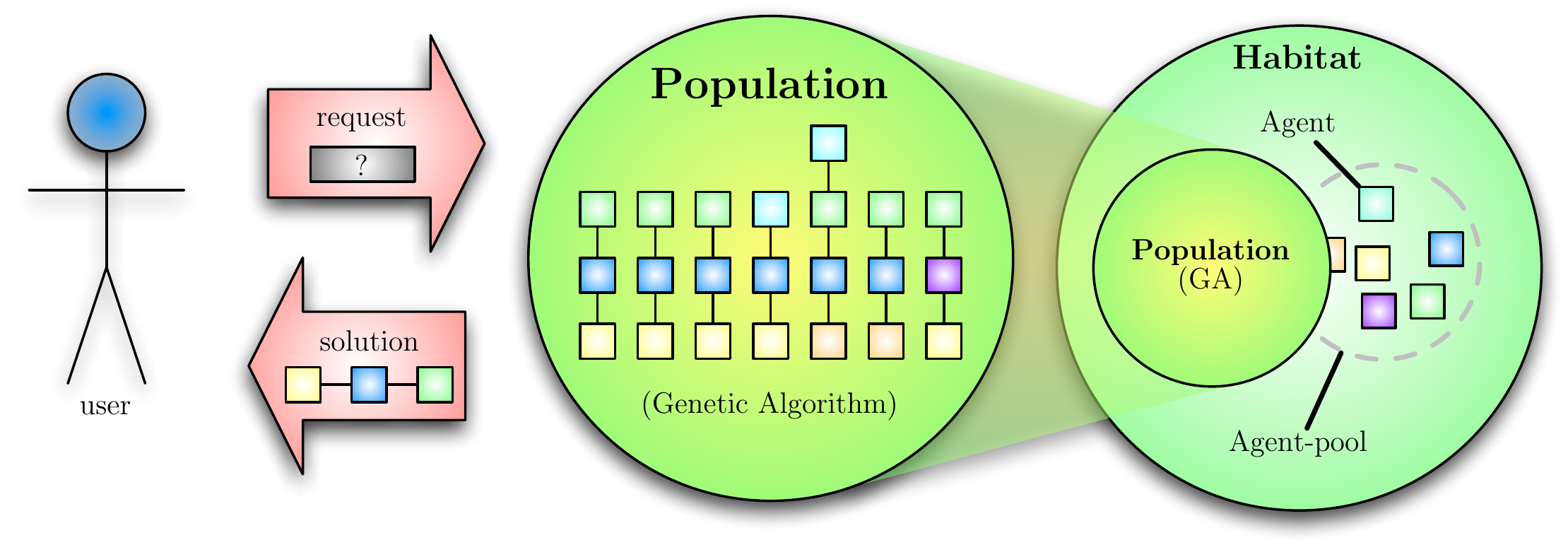}{graffle}{User Request to the Digital Ecosystem}{(modified from \cite{kpic}): A user \getCap{picUser}. \getCap{picUserReq} (Agent-pool).}{-5mm}{!h}{}{}

The Populations of the Digital Ecosystem are functionally equivalent to the evolving, self-organising populations of a biological ecosystem, and are achieved through using evolutionary computing. A population in biological ecosystems is all the members of a species that occupy a particular area at a given time \cite{lawrence1989hsd}. Our Population is also \emph{all the members of a species that occupy a particular area at a given time}, like an island from the island-models of \acl*{DEC} \cite{lin1994cgp}. The use of \acl*{DEC} to accelerate the Populations will be explained later.

\changed{The users \setCap{will formulate queries to the Digital Ecosystem by creating a request as a \emph{semantic description}, like those being used and developed in \acp{SOA} \cite{SOAsemantic}, specifying an application they desire and submitting it to their Habitat}{picUser}}\added{, instead of using the semantic descriptions to search directly for the web services.} This description enables the definition of a metric for evaluating the \emph{fitness} of a composition of Agents, as a distance function between the \emph{semantic description} of the request and the Agents' \emph{semantic descriptions}. \setCap{A Population is then instantiated in the user's Habitat in response to the user's request, seeded from the Agents available at their Habitat}{picUserReq} (i.e. its Agent-pool). \changed{This allows the evolutionary optimisation to be accelerated in the following three ways: first, the Habitat network provides a subset of the Agents available globally, localised to the specific user it represents; second, making use of Agent-sequences previously evolved in response to the user's earlier requests; and third, taking advantage of relevant Agent-sequences evolved elsewhere in response to similar requests by other users.} The Population then proceeds to evolve the optimal Agent-sequence(s) that fulfils the user request, and as the Agents are the base unit for evolution, it searches the available Agent-sequence combination space. For an evolved Agent-sequence that is executed (instantiated) by the user, it then migrates to other peers (Habitats) becoming hosted where it is useful, to combine with other Agents in other Populations to assist in responding to other user requests for applications

\subsubsection{Evolution}

Evolution in biological ecosystems leads to both great \emph{diversity} and high \emph{specialisation} of its organisms \cite{begon96}. In Digital Ecosystems the \emph{diversity} of evolution will provide for the wide range of user needs and allow for quick responses to the changing of these user needs, while the \emph{specialisation} will simultaneously provide solutions which are tailored to fulfil specific user requests. We will consider the issue of \emph{diversity} in a later subsection, because it is achieved through evolution in a distributed environment, which will be discussed later. In biological ecosystems, evolutionary \emph{specialisation} is localised to a population within its micro-habitat, which allows for the creation of niches (high specialisation) \cite{lawrence1989hsd}. So, a Population is instantiated in the user's own Habitat, where the collection of Agents is chosen for the user, and the micro-Habitat is provided by the user request. There is nothing to preclude more than one Population being instantiated in a user's Habitat at any one time, provided there are computational resources sufficiently available.

A \emph{selection pressure} is the sum aggregate of the forces acting upon a population, resulting in genetic change through natural selection \cite{lawrence1989hsd}. Those organisms best fit to survive the selection pressures operating upon them will pass on their biological \emph{fitness} to their progeny through the inheritance process \cite{lawrence1989hsd}. The \emph{fitness} of an individual Agent-sequence within a Population is determined by a \emph{selection pressure}, applied as a \emph{fitness function} \cite{eiben2003iec} instantiated from the user request, and works primarily on comparing the \emph{semantic descriptions} of the Agents with the \emph{semantic description} of the user request. The \emph{selection pressure} selects for those Agent-sequences that are \emph{fit} and capable of surviving the environment to reproduce, and against those that do not have sufficient fitness and therefore die before passing on their genes, thereby providing the direction for genetic change. In biology fitness is a measure of an organism's success in its environment \cite{lawrence1989hsd}, and its definition here will be further explained in the next subsection.

Genes are the functional unit in biological evolution \cite{lawrence1989hsd}; whereas here the functional unit is the Agent. Therefore, the evolutionary process of a Population provides a combinatorial optimisation \cite{papadimitriou1998coa} of the Agents available, when responding to a user request. So, it does not change or mutate the Agents themselves. In biology a mutation is a permanent transmissible change (over the generations) in the genetic material (DNA) of an individual, and recombination (e.g. crossover) is the formation within the offspring of alleles (gene combinations), which are not present in the parents \cite{lawrence1989hsd}. As in genetic algorithms \cite{goldberg}, mutations will occur by switching Agents in and out of the Agent-sequence structure, and recombination (crossover) will occur by performing a crossing of two Agent-sequences.

As the Digital Ecosystem receives more and more sophisticated requests, so more and more complex applications are evolved and become available for use by the users. To achieve this evolution, specifically the Agent-sequence recombination and optimisation, is a very significant challenge, because of the range of services that must be catered for and the potentially huge number of factors that must be considered for creating an applicable \emph{fitness function}. First, to construct ever more complex software solutions, requires modularity, which is provided by the paradigm of service interoperability from \acp{SOA} \cite{soa1w}. Second, two of the most important issues are that of defining \emph{fitness} and managing \emph{bloat}, which we will discuss next. Finally, there is a huge body of work and continuing research regarding theoretical approaches to evolutionary computing \cite{eiben2003iec}, including the extensive use of genetic algorithms for practical real-world problem solving \cite{ducheyne2003fiu}. In defining Digital Ecosystems we should make use of the current state-of-the-art, and future developments, in the areas of evolutionary computing \cite{jin2005csf} and service interoperability \cite{soa1w}.

\subsubsection{Fitness}

In biology \emph{fitness} is a measure of how successful an organism is in its environment, i.e. its phenotype \cite{lawrence1989hsd}. The fitness of an Agent-sequence within a Population would also, ideally, be based upon its phenotype, the run-time instantiation, and nothing else. However, such an approach would be impractical, because it is currently infeasible to execute all the Agent-sequences of a Population at every generation, and not least because of the computational resources that would be required. The other concern is one of practicality, by which we mean that it may not even be possible to perform a live execution for the executable components of an Agent-sequence; for example, if they are for buying an item from an online retailer. These are well known issues in evolutionary computing, which is why \emph{fitness functions} are often defined as simulated input/output pairs to test functionality \cite{mantere2005ese}. In Digital Ecosystems we can use historical usage information, but this would be insufficient initially, because such information would not be available at the time of an Agent's deployment. However, because each Agent also carries a \emph{semantic description}, a specification of what it does, the \emph{fitness function} can measure a complete Agent-sequence's collective \emph{semantic descriptions} relative to the \emph{semantic description} of a user request. So, initially the \emph{fitness function} should be based primarily on comparing the \emph{semantic descriptions} of the Agent-sequences to the \emph{semantic description} of the user request, ever increasingly augmented with the growing usage information available for the Agents. In biological terms the genotype will be used as the phenotype, combined with any available past fitness of the phenotype; with the Agent's \emph{semantic description} (genotype) therefore acting as a guarantee of its expected behaviour. So, for any newly deployed Agent a one-to-one genotype-phenotype mapping \cite{shackleton2000irg} will initially exist, until sufficient usage information is available. While the use of such a mapping is undesirable, it is temporary, and necessary to allow Digital Ecosystems to operate effectively.

We have already suggested that the primary driver of the evolutionary process should initially be the extent by which an Agent-sequence can verifiably satisfy the specified requirements. This could be measured probabilistically, or using theorem-proving to validate the system, though automatic theorem proving is notoriously slow \cite{slagle1974atp, schumann2001atp}. However, there will also be other pressures on the fitness. For example, one may seek the most parsimonious solution to a problem (one that provides exactly the specified features and no more), or the cheapest solution, or one with a good reputation. Some aspects of fitness will be implicit in the evolutionary process (e.g. Agents which are often used will gain more fitness) while others will require explicit measures (e.g. price, or user satisfaction). One way to handle this multiplicity of fitness values (some qualitative) is to explicitly recognise the multi-objective nature of the optimisation problem. In this way, we are seeking not the single best solution, but a range of possible compromises that can be made most optimally. The set of solutions for which there are no better compromises is called the Pareto-set, and evolutionary techniques have been adapted to solve such problems with considerable success \cite{veldhuizen2000mea}. The main point is that selection has to be driven not by an absolute value of fitness, but rather by a notion of what it means for one solution to be better than another. We say one solution dominates another if it is better in at least one respect, and no worse in any of the others \cite{fonseca1995oea}.

\subsubsection{Bloat}

\label{bloat}

If the repetition of Agents is allowed within evolving Agent-sequences, then the search space can become countably infinite, because the nature of the problem to be solved may not allow us to determine what the length of a solution is beforehand. Therefore, a variable length approach must be adopted, which is common in genetic programming \cite{ec25}. When variable length representations of solutions are used, a well-known phenomenon arises, called \emph{bloat}, in which the individuals of an evolving population tend to grow in size without gaining any additional advantage \cite{langdon1997fcb}. The bloat phenomenon can cause early termination of an evolutionary process due to the exhaustion of the available memory, and can also significantly reduce performance, because typically longer sequences have higher fitness computation costs \cite{ratle2001abp}. Bloat is not specific to genetic programming, and is inherent in search techniques with discrete variable length representations \cite{langdon1998esv}. It is a fundamental area of research within search-based approaches such as \aclp*{GA}, \acl*{GP} and other approaches not based on populations such as simulated annealing\added{whenever non-crossover based recombination occurs (i.e. mutation), which is further explained in} \cite{langdon1998esv}. However, considerable work on bloat has been done in connection with \acl*{GP} \cite{langdon1998gpa, banzhaf1998gpi}, and we believe that the \aclp*{GA} community generally, and the genetic-algorithms-based approach of our Digital Ecosystems specifically, can benefit directly from this research. While bloat is a phenomenon which was first observed in practice \cite{ec25}, theoretical analyses have been attempted \cite{banzhaf2002scr}. One should take care with these approaches as implementations will always deal with finite populations, while theoretical approaches often deal with infinite populations \cite{ec25}, and this difference can be important. Yet, both theoretical and empirical approaches are required to understand bloat. There are many factors contributing to bloat, and while the phenomenon may appear simple, the reasons are not. There are several theories to explain why it occurs, and, as we shall discuss, some measures that can be taken for its prevention.

There are several different qualitative theories which attempt to explain bloat, and they can be considered in two groups. First, protection against crossover and bias removal (which can be considered jointly) and second, the nature of programme search spaces \cite{banzhaf2002scr}. First, near the end of a run a Population consists of mostly fit individuals, and any crossover is likely to be detrimental to the fitness of the offspring. In any sequence of Agents there may be Agents that do not contribute semantically to the complete functionality of the sequence if, for example, their functionality was not requested by the user or if it is duplicated in the sequence; analogously to \acl*{GP} \cite{banzhaf2002scr}, we can call these redundant Agents bloat. The genotype can then be grown further without affecting the phenotype if Agents with similar functionality are added; but, as the genotype grows larger, crossover is more likely to transfer redundant Agents to the new off-springs (assuming uniform crossover). Second, above a certain threshold size, the distribution of functionality does not vary with the size of the search space \cite{banzhaf2002scr}. Thus, if we randomly sample long and short Agent-sequences above a length threshold, they will likely have the same functionality and fitness. So, as a search process progresses we are more likely to sample longer Agent-sequences, as mutation results in more of them (all other things being equal) and this will give rise to the bloating phenomenon.

Each of the stages of construction of a \acl*{GA} (i.e. choice of fitness function, selection method and genetic operator) can affect bloat. It has been shown that even small differences in the fitness function can cause a difference: a single programme glitch in an otherwise \emph{flat fitness landscape} (from the neutral theory of molecular evolution \cite{kimura:ntm}) is sufficient to significantly increase the average programme size of an infinite population \cite{mcphee2001EuroGP}. If a fitness-proportional selection method is used, individuals with zero fitness will be discontinued as they have zero probability of being selected as parents \cite{blickle1996css}. However, if tournament selection method is used, then there is a finite chance that individuals with zero fitness will be selected to be parents \cite{blickle1996css}. Finally, the choice of genetic operator affects the size of the programmes which are sampled; standard crossover on a flat landscape heavily over samples the shorter programmes \cite{poli01exact}. There are other factors that may affect bloat, for example, how the population is initialised, or the choice of representation used, such as a neutral genotype-phenotype mapping, which can actually alleviate bloat \cite{miller2001wbcgpbp}.

Bloat is a fact, whatever the reasons, happening in this type of optimisation and needs to be controlled if the space is to be searched effectively. One solution is to apply a hard limit to the size of the sets that can be sampled \cite{langdon1997fcb}: this enables the search algorithm to keep running without having out-of-memory run-time errors, but poses questions on how to set this hard limit. An alternative but similar method is to apply a \emph{parsimony pressure}, where a term is added to the fitness function which chastises big sets in preference for smaller sets \cite{soule1998ecg}. In this approach, individuals larger than the average size are evaluated with a reduced probability, biasing the search to smaller sets, while providing a dynamic limit which adapts to the average size of individuals in a changing population \cite{soule1998ecg}.

\subsection{The Digital Ecosystem}

\tfigure{scale=0.8}{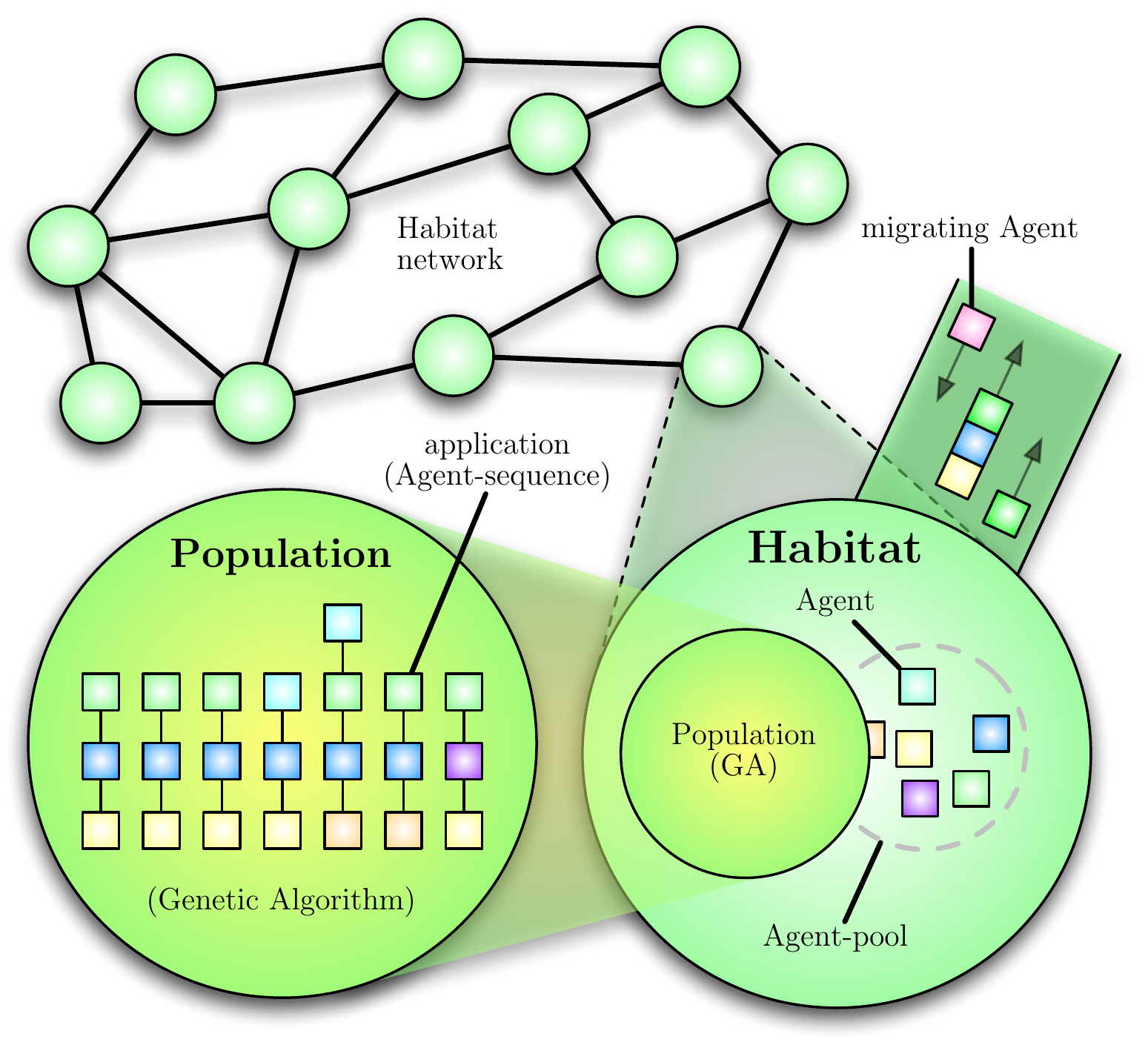}{graffle}{Digital Ecosystem}{A network of interconnected Habitats, combined \getCap{digEco}. Agents travel along the peer-to-peer connections; in every node (Habitat) local optimisation is performed through an evolutionary algorithm, where the search space is determined by the Agents present at the node.}{-3mm}{}{}{}

The Digital Ecosystem supports the automatic combining of numerous Agents (which represent services), by their interaction in evolving Populations to meet user requests for applications, in a scalable architecture of distributed interconnected Habitats. The sharing of Agents between Habitats ensures the system is scalable, while maintaining a high evolutionary specialisation for each user. The network of interconnected Habitats is equivalent to the abiotic environment of biological ecosystems \cite{begon96}; combined \setCap{with the Agents, the Populations, the Agent migration for \acl*{DEC}, and the environmental selection pressures provided by the user base, then the union of the Habitats creates the Digital Ecosystem}{digEco}, which is summarised in Figure \ref{DE}. The continuous and varying user requests for applications provide a dynamic evolutionary pressure on the Agent sequences, which have to evolve to better fulfil those user requests, and without which there would be no driving force to the evolutionary self-organisation of the Digital Ecosystem.

In the Digital Ecosystem, local and global optimisations concurrently operate to determine solutions to satisfy different optimisation problems. The global optimisation here is not a decentralised super-peer based control mechanism \cite{risson2006srt}, but the completely distributed peer-to-peer network of the interconnected Habitats, and therefore not susceptible to super-peer failure. \added{This is because we are seeking the digital counterparts of biological ecosystems, which are not centralised, but distributed.} It provides a novel optimisation technique inspired by biological ecosystems, working at two levels: a first optimisation, migration of Agents which are distributed in a peer-to-peer network, operating continuously in time; this process feeds a second optimisation, based on evolutionary combinatorial optimisation, operating locally on single peers and is aimed at finding solutions that satisfy locally relevant constraints. So, the local search is improved through this twofold process to yield better local optima faster, as the distributed optimisation provides prior sampling of the search space through computations already performed in other peers with similar constraints. \added{We will further elucidate this two stage process in the remainder of this section, starting with the practical example described in the following paragraph.} This novel form of distributed evolutionary computing will be discussed further below, once we have discussed a topology resulting from an example user base.

If we consider an example user base for the Digital Ecosystem, the use of \acp{SOA} in its definition means that \acf{B2B} interaction scenarios \cite{krafzig2004ess} lend themselves to being a potential user base for Digital Ecosystems. So, we can consider the \emph{business ecosystem} of \ac{SME} networks from \acp{DBE} \cite{dbebkintro}, as a specific class of examples for \ac{B2B} interaction scenarios; and in which the \ac{SME} users are requesting and providing software services, represented as Agents in the Digital Ecosystem, to fulfil the needs of their business processes. \acp{SOA} promise to provide potentially huge numbers of services that programmers can combine, via the standardised interfaces, to create increasingly more sophisticated and distributed applications \cite{SOApaper2}. The Digital Ecosystem extends this concept with the automatic combining of available and applicable services, represented by Agents, in a scalable architecture, to meet user requests for applications. These Agents will recombine and evolve over time, constantly seeking to improve their effectiveness for the user base. From the SME users' point of view the Digital Ecosystem provides a network infrastructure where connected enterprises can advertise and search for services (real-world or software only), putting a particular emphasis on the composability of loosely coupled services and their optimisation to local and regional, needs and conditions. To support these SME users the Digital Ecosystem is satisfying the companies' business requirements by finding the most suitable services or combination of services (applications) available in the network. A composition of services is an Agent-sequence in the Habitat network that can move from one peer (company) to another, being hosted only in those where it is most useful in satisfying the SME users' business needs.

\added{To define the computing model of our ecosystem-oriented distributed evolutionary computing, we provide a reference \ac{UML} class diagram in Figure \ref{classDiagram} to complement the detailed explanations of the functionality of each class of objects. This package represents our ecosystem-oriented distributed evolutionary computing, with \emph{Agents} that can move between \emph{Habitats}, and for which the optimal \emph{Agent}-sequence (or combination) is determined through evolutionary computing at the \emph{Populations} of the \emph{Habitats}.}

\tfigure{scale=0.85}{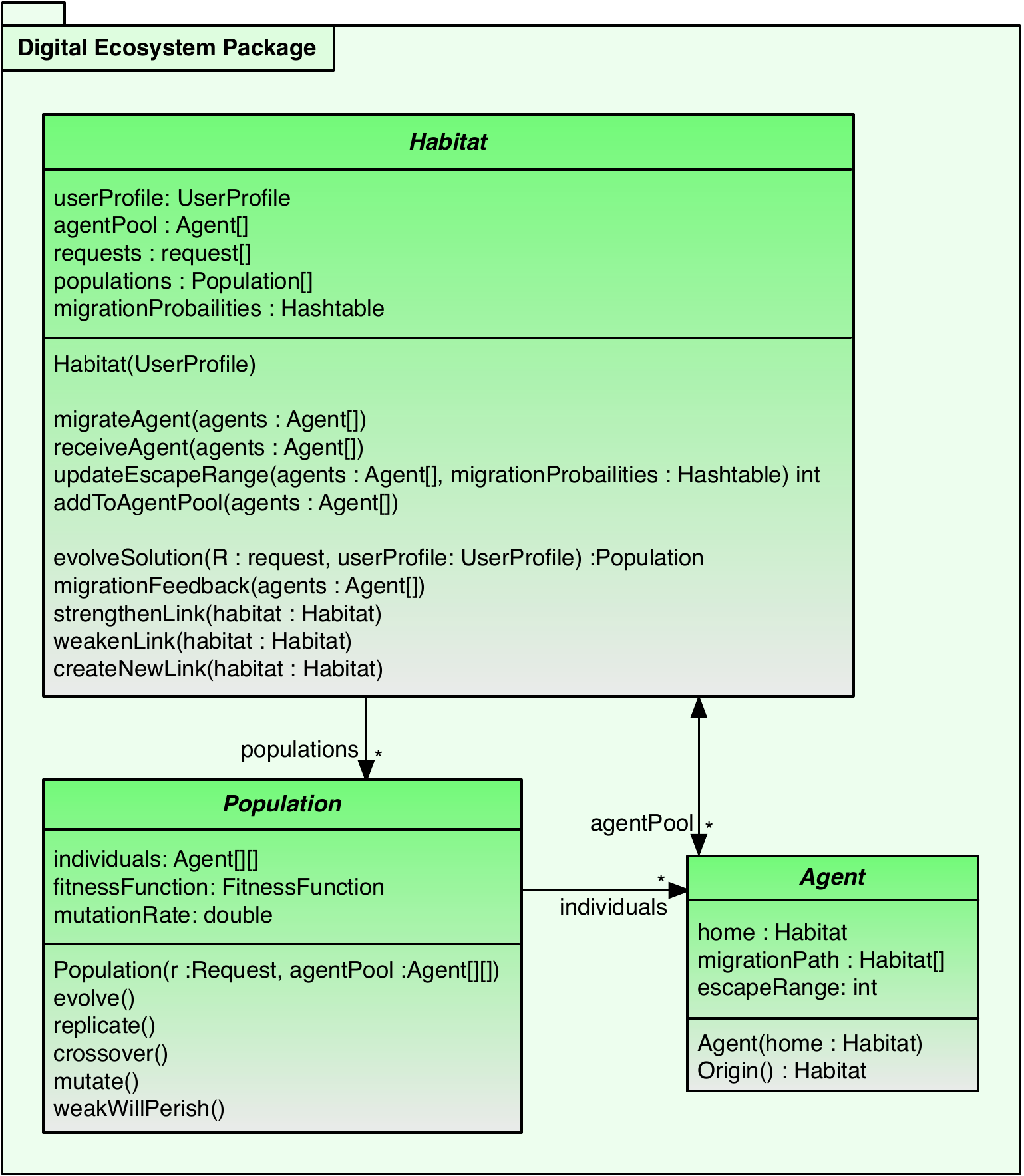}{graffle}{\added{Reference UML Class Diagram}}{\added{This package represents the computing model of our ecosystem-oriented distributed evolutionary computing, with \emph{Agents} (which represent simple algorithms) that can move between \emph{Habitats}, for which the optimal \emph{Agent-sequence} is determined through evolutionary computing at the \emph{Populations} in a scalable architecture of distributed interconnected \emph{Habitats}.}}{0mm}{!b}{}{}

\subsubsection{Topology}

The Digital Ecosystem allows for the connectivity in the Habitats to adapt to the connectivity within the user base, with a cluster of Habitats representing a community within the user base. If a user is a member of more than one community, the user's Habitat will be in more than one cluster. This leads to a network topology that will be discovered with time, and which reflects the connectivity within the user base. Similarities in requests by different users will reinforce behavioural patterns, and lead to clustering of the Habitats within the ecosystem, which can occur over geography, language, etc. This will form communities for more effective information sharing, the creation of niches, and will improve the responsiveness of the system. The connections between the Habitats will be self-managed, through the mechanism of Agent migration defined earlier. Essentially, successful Agent migration will reinforce Habitat connections, thereby increasing the probability of future Agent migration along these connections. If a successful multi-hop migration occurs, then a new link between the start and end Habitats can be formed. Unsuccessful migrations will lead to connections (migration probabilities) decreasing, until finally the connection is closed.

\tfigure{scale=0.8}{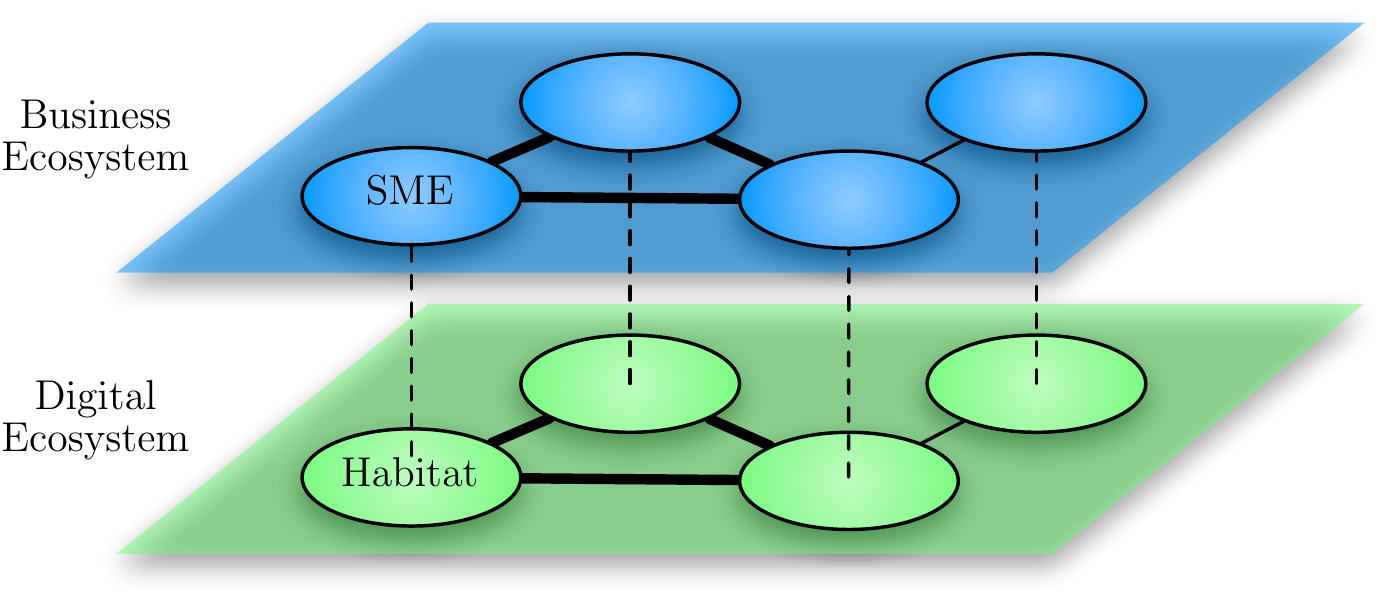}{graffle}{Digital Business Ecosystem}{Business ecosystem, network of \acp{SME} \cite{dbebkintro}, using the Digital Ecosystem. \getCap{bizEcoCap}.}{-3mm}{!h}{}{}

If we consider the \emph{business ecosystem} - a network of \acp{SME} from \acp{DBE} \cite{dbebkintro} - as an example user base, such business networks are typically small-world networks \cite{white2002nst, antionella}. They have \setCap{many strongly connected clusters (communities), called sub-networks (quasi-complete graphs), with a few connections between these clusters (communities) \cite{swn1}. Graphs with this topology have a very high clustering coefficient and small characteristic path lengths \cite{swn1}.}{archComTop} \setCap{As the connections between Habitats are reconfigured depending on the connectivity of the user base, the Habitat clustering will therefore be parallel to the business sector communities}{bizEcoCap}, as shown in Figure \ref{DBE}. The communities will cluster over language, nationality, geography, etc. -- all depending on the user base. So, the Digital Ecosystem will take on a topology similar to that of the user base.

\tfigure{scale=0.65}{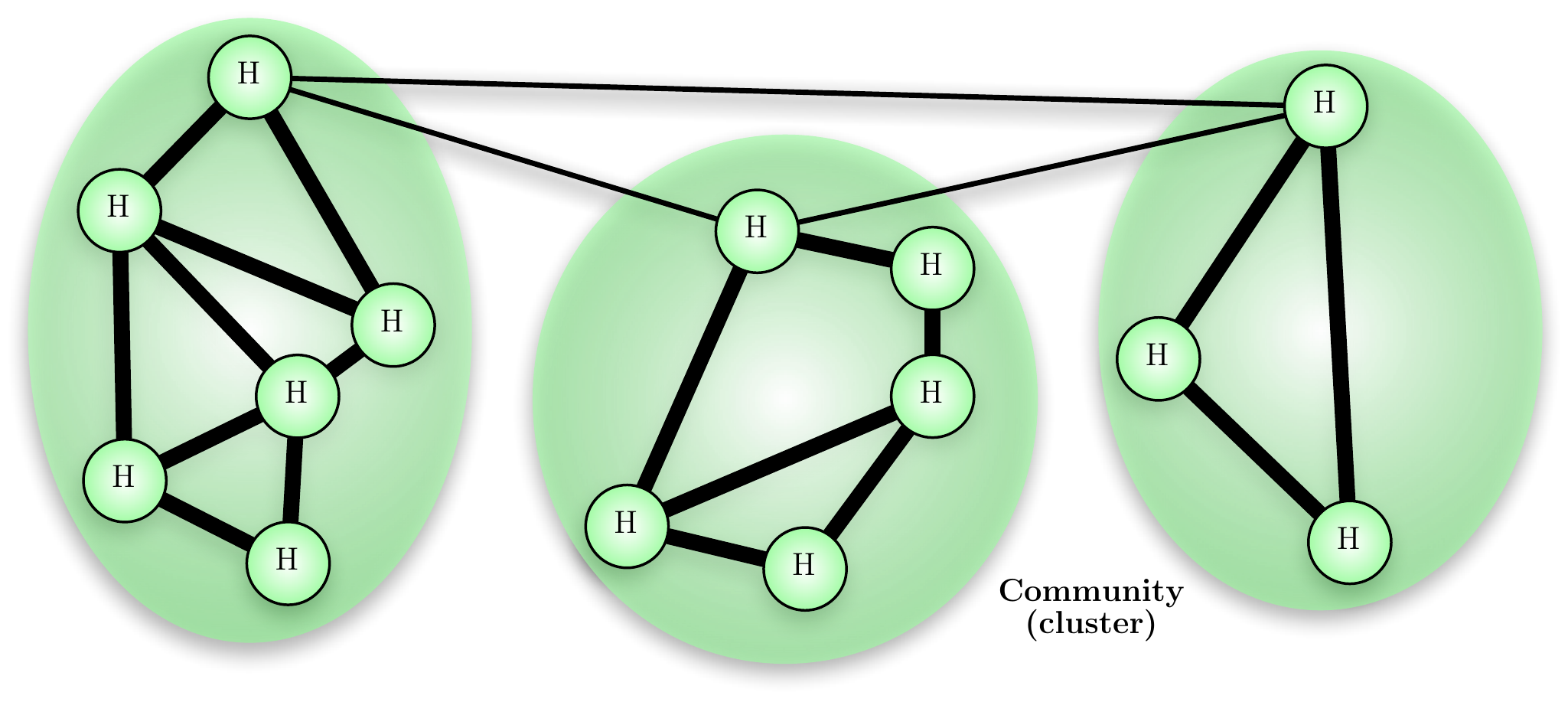}{graffle}{Habitat Clustering}{Topology adapted to the small-world network of a business ecosystem of \acp{SME} from \acp{DBE} \cite{dbebkintro}, having \getCap{archComTop}}{-3mm}{!b}{}{}

Fragmentation of the Habitat network can occur, but only if dictated by the structure of the user base. The issue of greater concern is when individual Habitats become totally disconnected, which could only occur under certain conditions. One condition is that the Agents within the Agent-pool consistently fail to satisfy user requests. Another condition is when the Agents and Agent-sequences they share are undesirable to the users that are within the migration range of these Agents and Agent-sequences. These scenarios can arise because the Habitat is located within the wrong cluster, in which case the user can be asked to join another cluster within the Habitat network, assuming the user base is of sufficient size to provide a viable alternative.

\subsubsection{Distributed Evolution}
\label{novel}

The Digital Ecosystem is a hybrid of \acp{MAS}, more specifically of \emph{mobile} agent systems, \acp{SOA}, and \emph{distributed} evolutionary computing, which leads to a novel form of evolutionary computation. \changed{The novelty arrises from the creation of multiple evolving Populations responding to \emph{similar} requests, whereas in the island-models of \acl*{DEC} there are multiple evolving populations responding to only one request \cite{lin1994cgp}.} So, in our Digital Ecosystem different \setCap{requests are evaluated on separate \emph{islands} (Populations), with their evolution accelerated by the sharing of solutions between the evolving Populations (islands), because they are working to solve similar requests (problems).}{similarCap} This is shown in Figure \ref{similar}, where the dashed \setCap{yellow lines connecting the evolving Populations indicate similarity in the requests being managed.}{similar2}

\tfigure{scale=0.55}{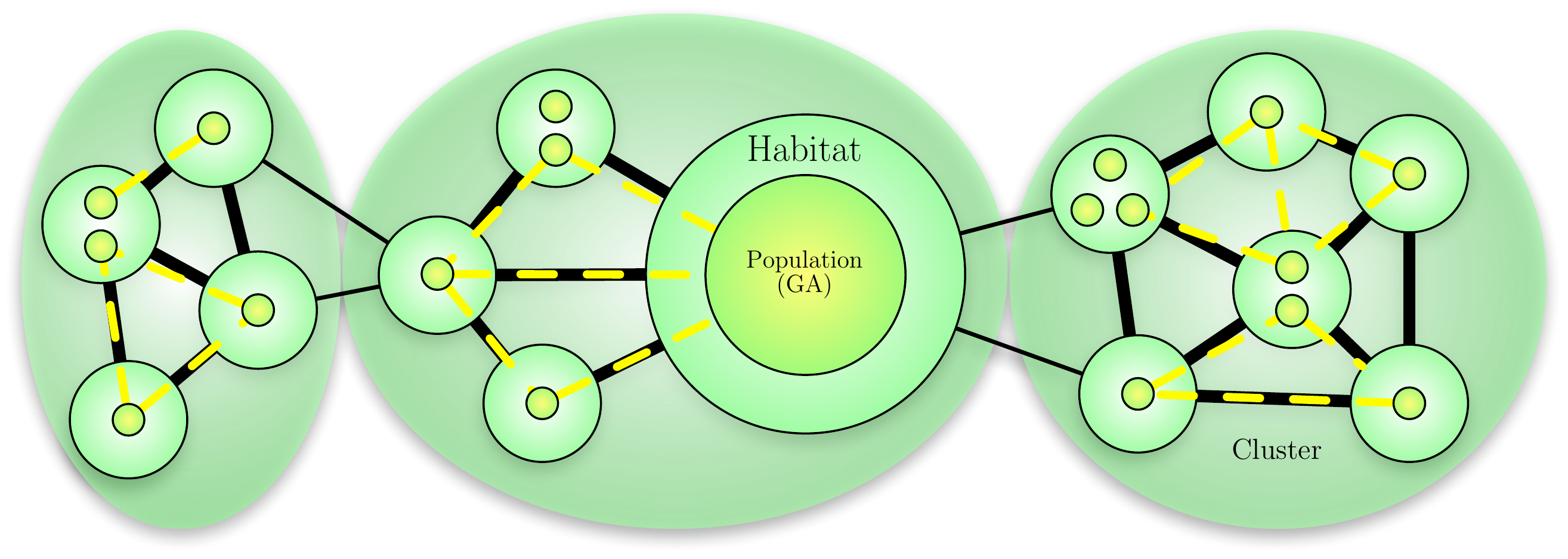}{graffle}{Distributed Evolution in the Digital Ecosystem}{Different \getCap{similarCap} The \getCap{similar2}}{-1mm}{!h}{}{}

If we again consider the \emph{business ecosystem} of \acp{SME} from \acp{DBE} \cite{dbebkintro} as an example user base, then in Figure \ref{similar} the four Habitats, in the left cluster, could be travel agencies, and the three with linked evolving Populations are looking for similar package holidays. So, an optimal solution found and used in one Habitat will be migrated to the other connected Habitats and integrate into any evolving Populations via the local Agent-pools. This will help to optimise the search for similar package holidays at the Habitats of the other travel agencies. This also works in a time-shifted manner, because an optimal solution is stored in the Agent-pool of the Habitats to which it is migrated, being available to optimise a similar request placed later.

The distributed architecture of Digital Ecosystems favours the use of Pareto-sets for fitness determination, because Pareto optimisation for multi-objective problems is usually most effective with spatial distribution of the populations, as partial solutions (solutions to different niches) evolve in different parts of a \emph{distributed population} \cite{detoro2002ppg} (i.e. different Populations in different Habitats). By contrast, in a single population, individuals are always interacting with each other, via crossover, which does not allow for this type of specialisation \cite{back1996eat}.

This approach requires the Digital Ecosystem to have a sufficiently large user base, so that there can be communities within the user base, and therefore allow for similarity in the user requests. Assuming a user base of hundreds of users, then there would be hundreds of Habitats, in which there will be potentially three or more times the number of Populations at any one time. Then there will be thousands of Agents and Agent-sequences (applications) available to meet the requests for applications from the users. In such a scenario, there would be a sufficient number of users for the Digital Ecosystem to find similarity within their requests, and therefore apply our novel form of \acl*{DEC}.

\subsubsection{Agent Life-Cycle}
\label{agentLifeCycle}

An Agent is created to represent a user's service in the Digital Ecosystem, and its life-cycle begins \setCap{with deployment to its owner's Habitat for distribution within the Habitat network.}{lifeCycleCap} The Agent is then migrated to any Habitats connected to the owner's Habitat, to make it available in other Habitats where it could potentially be useful. The Agent is then available to the local evolutionary optimisation, to be \setCap{used in evolving the optimal Agent-sequence in response to a user request. The optimal Agent-sequence is then registered at the Habitat}{lifeCycle2}, being stored in the Habitat's Agent-pool. \setCap{If an Agent-sequence solution is then executed, an attempt is made to migrate (copy) it to every other connected Habitat, success depending on the probability associated with the connection.}{lifeCycle3} The Agent life-cycle is shown in Figure \ref{lifeCycle}. 

\tfigure{scale=0.8}{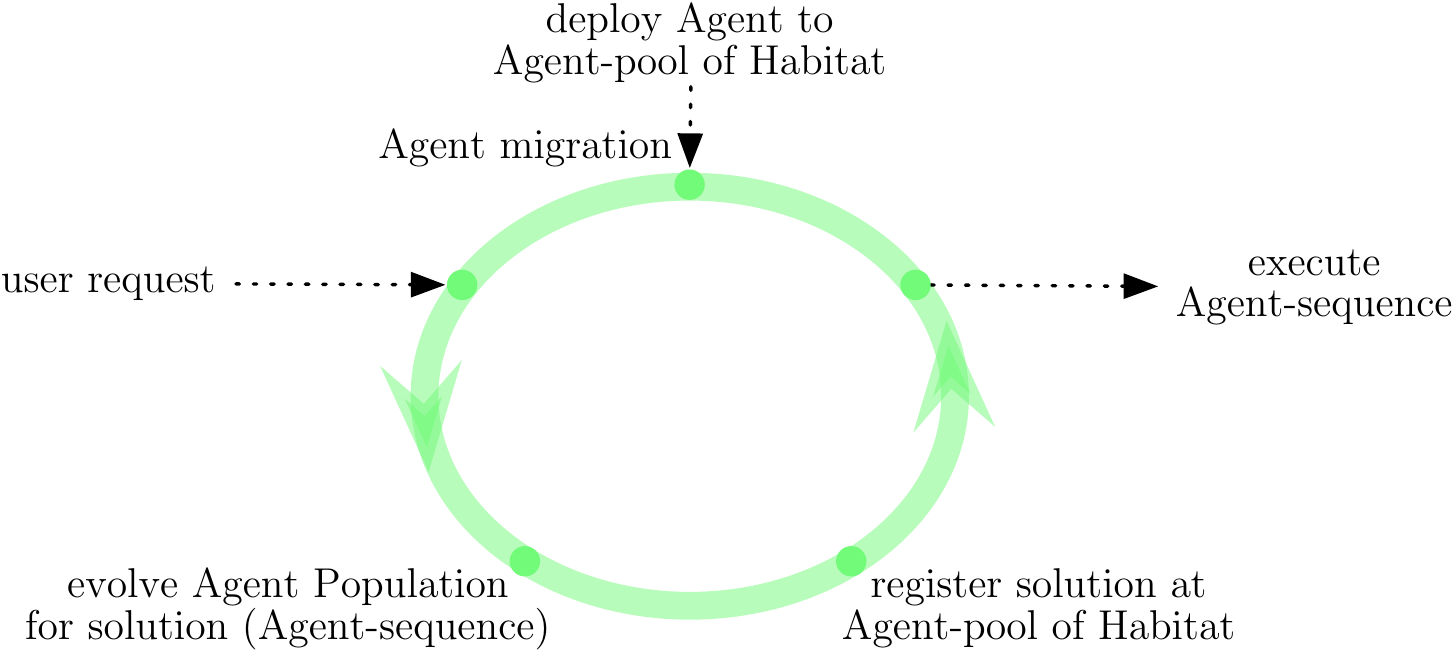}{graffle}{Agent Life-Cycle}{Begins \getCap{lifeCycleCap} It can then be \getCap{lifeCycle2}. \getCap{lifeCycle3}}{0mm}{!h}{}{}

An Agent can also be deleted if after several successive user requests at a Habitat it remains unused; it will have a small number of \emph{escape migrations}, in which it is not copied, but is randomly moved to another connected Habitat. If the Agent fails to find a niche before running out of \emph{escape migrations}, then it will be deleted.

\section{Simulation and Results}
\label{simRes}

We simulated the Digital Ecosystem, based upon our \acl*{EOA}, and recorded key variables to determine whether it displayed behaviour typical of biological ecosystems\added{, and so provide proof of concept}. We followed the \acl*{EOA} from the previous section, and used the \emph{business ecosystem} of \acp{SME} from \acp{DBE} \cite{dbebkintro} as an example user base.

\subsection{Agents: Semantic Descriptions}
\label{descriptions}

An Agent represents a user's service, including the \emph{semantic description} of the business process involved, and is based on existing and emerging technologies for \emph{semantically capable} \acp{SOA} \cite{SOAsemantic}, such as the \acl*{OWL-S} semantic markup for web services \cite{martin2004bsw}. \added{So, we can then make use of the ability to search for web services semantically.} \changed{Therefore}, we simulated a service's \emph{semantic description} \setCap{with an abstract representation consisting of a set of}{as3} numeric tuples, to simulate the properties of a \emph{semantic description}. Each \setCap{tuple representing an \emph{attribute} of the \emph{semantic description}, one integer for the \emph{attribute identifier} and one for the \emph{attribute value}, with both ranging between one and a hundred.}{agentSemantic2} \setCap{Each simulated Agent had a semantic description}{as4}, with between three and six tuples, an example of which is shown in Figure \ref{BMLprocess}. 

\mfigure{\begin{center}\bold{A = \{(1,25), (2,35), (3,55), (4,6), (5,37), (6,12)\}}\end{center}}{Agent Semantic Descriptions}{\getCap{as4} \getCap{as3} between three and six numeric tuples; each \getCap{agentSemantic2}}{BMLprocess}{0mm}{-3mm}{!t}

\subsection{User Base}

Throughout the simulations we assumed a hundred users, which meant that at any time the number of users joining the network equalled those leaving. The Habitats of the users were randomly connected at the start, to simulate the users going online for the first time. The users then produced Agents (services) and requests for business applications. Initially, the users each deployed five Agents to their Habitats, for migration (distribution) to any Habitats connected to theirs (i.e. their community within the \emph{business ecosystem}). Users were simulated to deploy a new Agent after the submission of three requests for business applications, and were chosen at random to submit their requests. \setCap{A simulated user request consisted of an abstract \emph{semantic description}, as a list of sets of numeric tuples to represent the properties of a desired business application}{semanticRequest}. The use of the numeric tuples made it comparable to the \emph{semantic descriptions} of the services represented by the Agents; while the list of sets (two level hierarchy) and a much longer length provided sufficient complexity to support the sophistication of business applications. An example is shown in Figure \ref{userRequest}.

\mfigure{\begin{center}\bold{R = [\{(1,23),(2,45),(3,33),(4,6),(5,8),(6,16)\}, \{(1,84),(2,48),(3,53),(4,11),(5,16)\}]}\end{center}}{User Request}{\getCap{semanticRequest}; each \getCap{agentSemantic2}}{userRequest}{0mm}{-3mm}{!h}

The user requests were handled by the Habitats instantiating evolving Populations, which used evolutionary computing to find the optimal solution(s), Agent-sequence(s). It was assumed that the users made their requests for business applications \emph{accurately}, and always used the response (Agent-sequence) provided.

\subsection{Populations: Evolution}
\label{fitnessFunction}

Populations of Agents, $[A_1, A_1, A_2, ...]$, were evolved to solve user requests, seeded with Agents and Agent-sequences from the Agent-pool of the Habitats in which they were instantiated. A dynamic population size was used to ensure exploration of the available combinatorial search space, which increased with the average length of the Population's Agent-sequences. The optimal combination of Agents (Agent-sequence) was evolved to the user request $R$, by an artificial \emph{selection pressure} created by a \emph{fitness function} generated from the user request $R$. An individual (Agent-sequence) of the Population consisted of a set of attributes, ${a_1, a_2, ...}$, and a user request essentially consisted of a set of required attributes, ${r_1, r_2, ...}$. So, the \emph{fitness function} for evaluating an individual Agent-sequence $A$, relative to a user request $R$, was
\begin{equation}
fitness(A,R) = \frac{1}{1 + \sum_{r \in R}{|r-a|}},
\label{ff}
\end{equation}
where $a$ is the member of $A$ such that the difference to the required attribute $r$ was minimised. Equation \ref{ff} was used to assign fitness values between 0.0 and 1.0 to each individual of the current generation of the population, directly affecting their ability to replicate into the next generation. The evolutionary computing process was encoded with a low mutation rate, a fixed selection pressure and a non-trapping fitness function (i.e. did not get trapped at local optima). The type of selection used was fitness-proportional and non-elitist, fitness-proportional meant that the fitter the individual the higher its probability of surviving to the next generation \cite{blickle1996css}. Non-elitist meant that the best individual from one generation was not guaranteed to survive to the next generation; it had a high probability of surviving into the next generation, but it was not guaranteed as it might have been mutated \cite{eiben2003iec}. \added{So, individuals replicated proportionally to their fitness, i.e. the higher the fitness the more they replicated, with their fitness values used as probabilities of their inclusion in future populations.} Crossover (recombination) was then applied to a randomly chosen 10\% of the surviving population, a one-point crossover, by aligning two parent individuals and picking a random point along their length, and at that point exchanging their tails to create two offspring \cite{eiben2003iec}. Mutations were then applied to a randomly chosen 10\% of the surviving population; \added{one} point \changed{mutation} \added{per chosen individual}, which were randomly located \added{(through use of uniformly random function to select a point (Agent) along the Agent-sequence)}\changed{.} \added{These point mutations} \changed{consisted} of insertions, \added{in which} \changed{an Agent was inserted into an Agent-sequence}, replacements, \added{in which} \changed{an Agent was replaced in an Agent-sequence}, and deletions, \added{in which} \changed{an Agent was deleted from an Agent-sequence} \cite{lawrence1989hsd}. \added{So, the remainder of the Population is not subject to any kind of recombination, crossover or mutation}. The issue of bloat was controlled by augmenting the \emph{fitness function} with a \emph{parsimony pressure} \cite{soule1998ecg} which biased the search to shorter Agent-sequences, evaluating longer than average length Agent-sequences with a reduced fitness, and thereby providing a dynamic control limit which adapted to the average length of the ever-changing evolving Agent Populations.

\subsection{Semantic Filter}
\label{semanticFilter}

The simulation of the Digital Ecosystem complied with the \acl*{EOA} defined in the previous section, but there was the possibility of model error in the \emph{business ecosystems} of the user base (\acp{SME} from \acp{DBE} \cite{dbebkintro}), because while the abstract numerical definition for the simulated \emph{semantic descriptions}, of the services and requests the users provide, makes it widely applicable, it was unclear that it could accurately represent \emph{business services}. So we created a \emph{semantic filter} to show \setCap{the numerical semantic descriptions, of the simulated services (Agents) and user requests, in a human readable form.}{bmlcap1} The basic properties of any business process are cost, quality, and time \cite{davenport1990nie}; so this was followed in the semantic filter. \tred{It} \setCap{translated numerical semantic descriptions for one community within the user base, showing it in the context of the travel industry}{capbml2}, as shown in Figure \ref{BMLreal}. \setCap{The simulation still operated on the numerical representation for operational efficiency, but the semantic filter essentially assigned meaning to the numbers.}{capbml3} The output from the semantic filter, in Figure \ref{BMLreal}, shows that the numerical semantic descriptions were a reasonable modelling assumption, abstracting sufficiently rich textual descriptions of business services.

\mfigure{\begin{center}Agent's semantic description:\end{center}
\vspace{-6mm}
\begin{quote}\begin{center}\bold{\{\red{(1,25)}, \blue{(2,35)}, \yellow{(3,55)}, \green{(4,6)}, \purple{(5,37)}, \turquoise{(6,12)}\}}\end{center}
\end{quote}
\begin{center}
(with semantic filter):
\end{center}
\vspace{-6mm}
\begin{quote}\begin{center}
\bold{\{\red{(Business, Airline)}, \blue{(Company, British Midland)}, \yellow{(Quality, Economy)}, \green{(Cost, 60)}, \purple{(Depart, Edinburgh)}, \turquoise{(Arrive, London)}\}}\end{center}\end{quote}
\begin{center}
user request:
\end{center}
\vspace{-6mm}
\begin{quote}\bold{[\{\red{(1,23)}, \blue{(2,45)}, \yellow{(3,33)}, \green{(4,6)}, \purple{(5,8)}, \turquoise{(6,16)}\}, \{\red{(1,84)}, \blue{(2,48)}, \yellow{(3,53)}, \green{(4,11)}, \grey{(7,16)}, \brown{(8,34)}\}, \{\red{(1,23)}, \blue{(2,45)}, \yellow{(3,53)}, \green{(4,6)}, \purple{(5,16)}\turquoise{(6,53)}\}, \{\red{(1,86)}, \blue{(2,48)}, \yellow{(3,33)}, \green{(4,25)}, \grey{(7,55)}\brown{(8,23)}\}, \{\red{(1,25)}, \blue{(2,52)}, \yellow{(3,53)}, \green{(4,5)}, \purple{(5,55)}, \turquoise{(6,37)}\}, \{\red{(1,86)}, \blue{(2,48)}, \yellow{(3,43)}, \green{(4,25)}, \grey{(7,37)}, \brown{(8,40)}\}, \{\red{(1,22)}, \blue{(2,77)}, \yellow{(3,82)}, \green{(4,9)}, \purple{(5,35)}, \turquoise{(6,8)}\}]}
\end{quote}
\begin{center}
(with semantic filter):
\end{center}
\vspace{-6mm}
\begin{quote}
\bold{[\{\red{(Business, Airline)}, \blue{(Company, Air France)}, \yellow{(Quality, Economy)}, \green{(Cost, 60)}, \purple{(Depart, Edinburgh)}, \turquoise{(Arrive, Paris)}\}, \{\red{(Business, Hotel)}, \blue{(Company, Continental)}, \yellow{(Quality, 3*)}, \green{(Cost, 110)}, \grey{(Location, Paris)}, \brown{(Nights, 3)}\}, \{\red{(Business, Airline)}, \blue{(Company, Air France)}, \yellow{(Quality, Economy)},\green{(Cost,60)},\purple{(Depart, Paris)}, \turquoise{(Arrive, Monte Carlo)}\}, \{\red{(Business, Hotel)}, \blue{(Company, Continental)}, \yellow{(Quality, 2*)}, \green{(Cost, 250)}, \grey{(Location, Monte Carlo)}, \brown{(Nights, 2)}\}, \{\red{(Business, Airline)}, \blue{(Company, KLM)}, \yellow{(Quality, Economy)}, \green{(Cost, 50)}, \purple{(Depart, Monte Carlo)}, \turquoise{(Arrive, London)}\}, \{\red{(Business, Hotel)}, \blue{(Company, Continental)}, \yellow{(Quality, 3*)}, \green{(Cost, 250)}, \grey{(Location, London)}, \brown{(Nights, 4)}\}, \{\red{(Business, Airline)}, \blue{(Company, Air Espana)}, \yellow{(Quality, First)}, \green{(Cost, 90)}, \purple{(Depart, London)}, \turquoise{(Arrive, Edinburgh)}\}]}
\end{quote}}{Semantic Filter}{Shows \getCap{bmlcap1} The semantic filter \getCap{capbml2}. \getCap{capbml3}}{BMLreal}{0mm}{-3mm}{!h}

\subsection{Ecological Succession}
\label{secSuccession}

We then compared some of the Digital Ecosystem's dynamics with those of biological ecosystems, to determine if it had been \tred{imbued} with the properties of biological ecosystems. A biological ecosystem develops from a simpler to a more mature state, by a process of succession, where the genetic variation of the populations changes with time \cite{begon96}. \setCap{So, it becomes increasingly more complex through this process of succession, driven by the evolution of the populations within the ecosystem \cite{connell111msn}.}{succession} Equivalently, the Digital Ecosystem's increasing complexity comes from the Agent Populations being evolved to meet the dynamic selection pressures created by the user requests. 

\tfigure{scale=0.8}{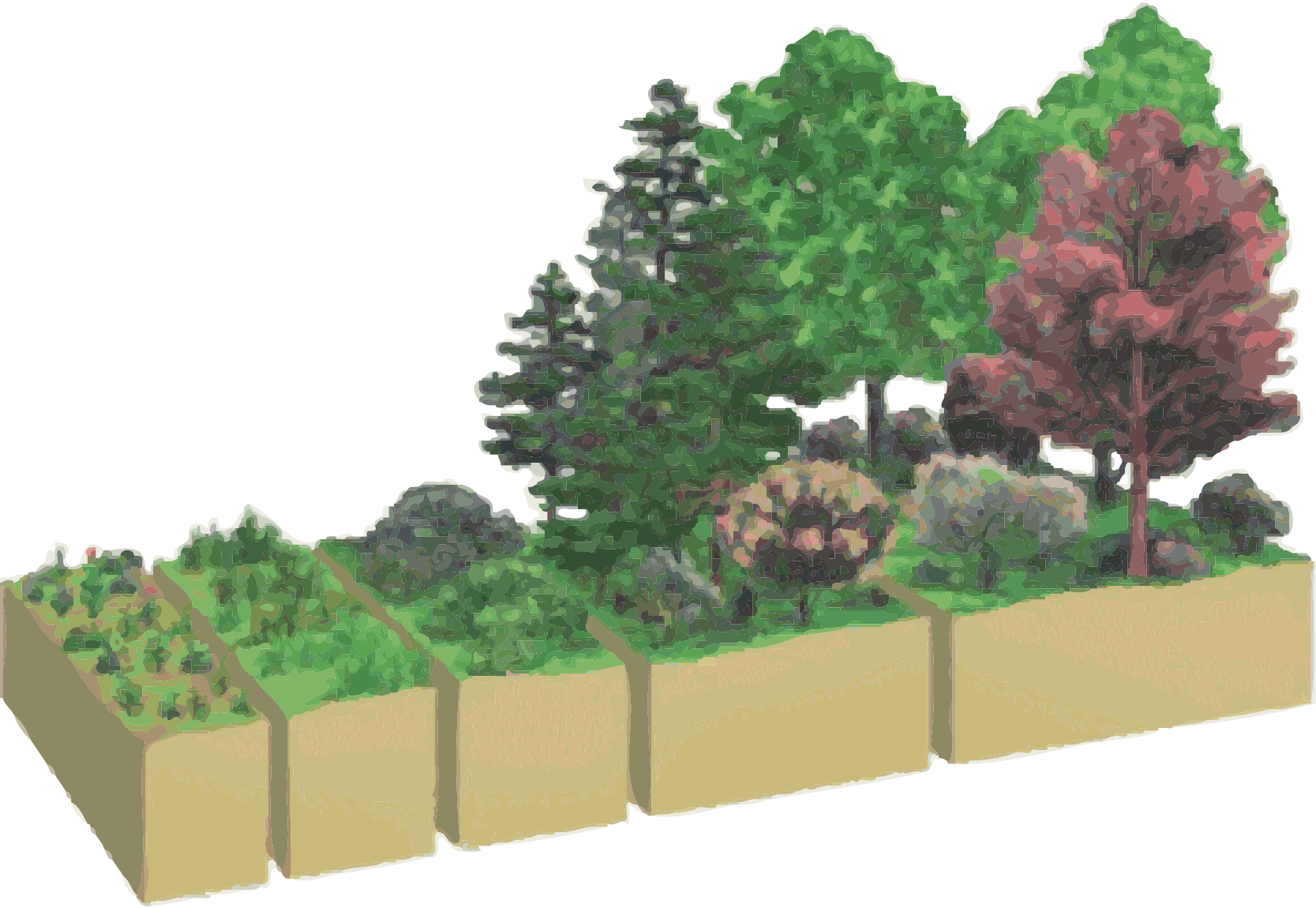}{pdf}{Ecological Succession}{(modified from \cite{davis2008}): \getCap{succession2} \getCap{succession3}. \getCap{succession}}{-1mm}{}{}{}

\setCap{The formation of a mature ecosystem}{succession2}, ecological succession, \setCap{is the slow, predictable, and orderly changes in the composition and structure of an ecological community, for which there are defined stages in the increasing complexity \cite{begon96}, as shown}{succession3} in Figure \ref{successionNew}. Succession may be initiated either by the formation of a new, unoccupied habitat (e.g., a lava flow or a severe landslide) or by some form of disturbance (e.g. fire, logging) of an existing community. The former case is often called \emph{primary succession}, and the latter \emph{secondary succession} \cite{begon96}. The trajectory of ecological change can be influenced by site conditions, by the interactions of the species present, and by more stochastic factors such as availability of colonists or seeds, or weather conditions at the time of disturbance. Some of these factors contribute to predictability of successional dynamics; others add more probabilistic elements \cite{gotelli1995pe}. Trends in ecosystem and community properties of succession have been suggested, but few appear to be general. For example, species diversity almost necessarily increases during early succession upon the arrival of new species, but may decline in later succession as competition eliminates opportunistic species and leads to dominance by locally superior competitors \cite{connell111msn}. Net Primary Productivity\footnote{Net Primary Productivity is the net flux of carbon from the atmosphere into green plants per unit time \cite{lawrence1989hsd}.}, biomass, and trophic level properties all show variable patterns over succession, depending on the particular system and site \cite{gotelli1995pe}. Generally, communities in early succession will be dominated by fast-growing, well-dispersed species, but as the succession proceeds these species will tend to be replaced by more competitive species \cite{begon96}.

We then considered existing theories of complexity for ecological succession and how it would apply to Digital Ecosystems, seeking a high-level understanding that would apply equally to both biological and digital ecosystems. As succession leads communities, of an ecosystem, to states of \emph{dynamic equilibrium}\footnote{Dynamic Equilibrium is when opposing forces of a system are proceeding at the same rate, such that its state is unchanging with time \cite{begon96}.} within the environment \cite{begon96}, the complexity has to increase initially or there would be no ecosystem, and presumably this increase eventually stops, because there must be a limit to how many species can be supported. The period in between is more complicated. If we consider the neutral biodiversity theory \cite{Hubbell}, which basically states network aspects of ecosystems are negligible, we would probably get a relatively smooth progression, because although you would get occasional extinctions, they would be randomly isolated events whose frequency would eventually balance arrivals, not self-organised crashes like in systems theory. In systems theory \cite{systemsTheory}, when a new species arrives in an ecological network, it can create a positive feedback loop that destabilises part of the network and drives some species to extinction. Ecosystems are constantly being perturbed, so it is reasonable to assume that a species that persists will probably be involved in a stabilising interaction with other species. So, the whole ecological network evolves to resist invasion. That would lead to a spiky succession process, perhaps getting less spiky over time. 

So, which theory is more applicable to the Digital Ecosystem depends on the extent that a species in the ecosystem acts independently, competing entities (smooth succession) \cite{Hubbell} versus tightly co-adapted ecological partners (spiky succession) \cite{systemsTheory}. Our Digital Ecosystem despite its relative complexity is quite simple compared to biological ecosystems. It has the essential and fundamental processes, but no sophisticated social mechanisms. Therefore, the smooth succession of the neutral biodiversity theory \cite{Hubbell} was more probable.

\tfigure{scale=0.8}{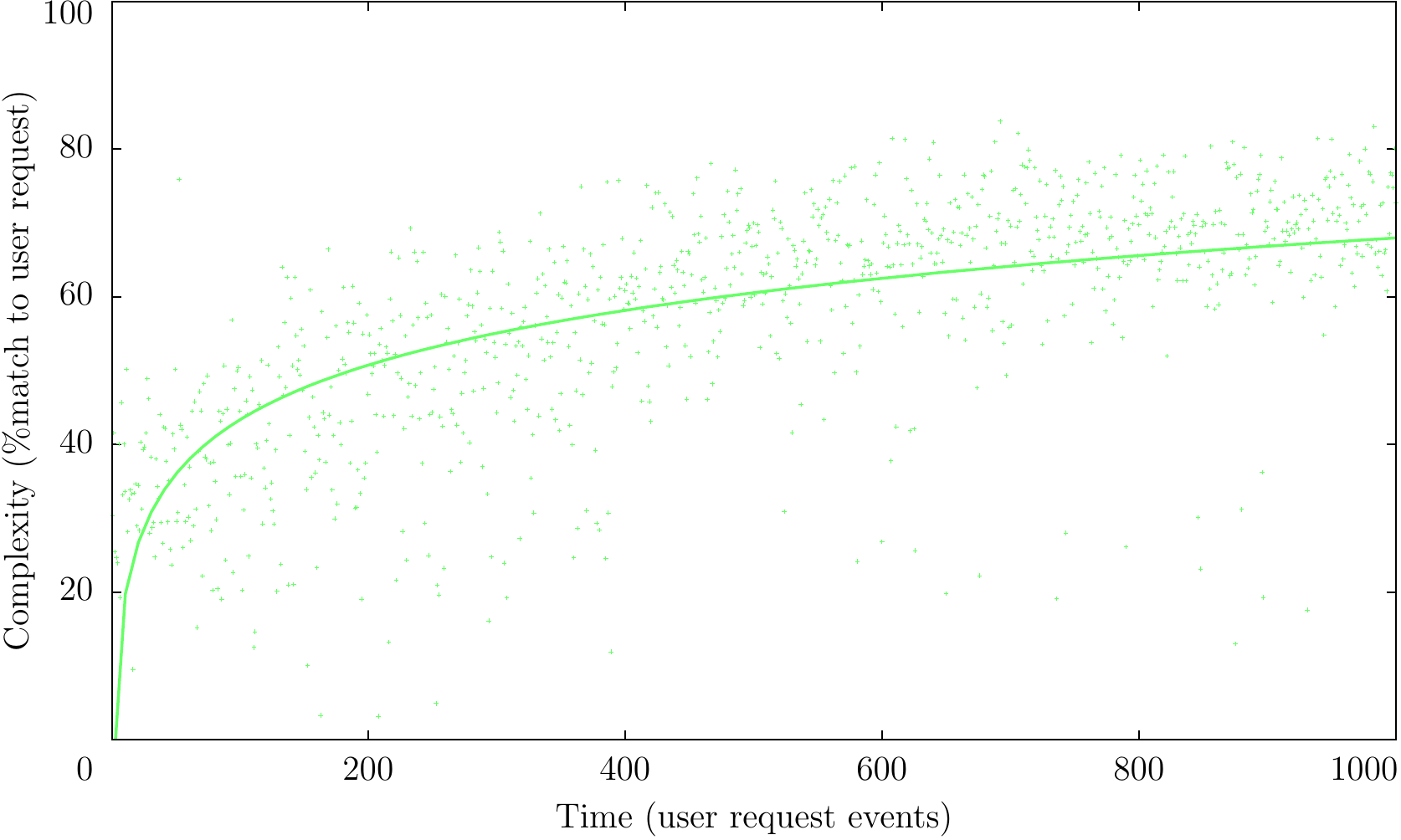}{graph}{Graph of Succession in the Digital Ecosystem}{\getCap{DigEcoSucCap} Still, at \getCap{DigEcoSuc2Cap}}{-2mm}{}{}{}

\label{ecosucexp}

As the increasing complexity of the Digital Ecosystem comes from its evolving Agent Populations responding to user requests, the effectiveness of the evolved Agent-sequences (responses) can provide a measure of its complexity over time. So, in simulation we measured the effectiveness of its responses over a thousand user requests, i.e. until it had reached a mature state like a biological ecosystem \cite{begon96}, and graphed a typical run in Figure \ref{DigEcoSuc}. The range and diversity of Agents at initial deployment were such that 70\% fulfilment of user requests was possible, increasing to 100\% fulfilment as more Agents were deployed. The Digital Ecosystem performed as expected, adapting and improving over time, reaching a mature state as seen in the graph of Figure \ref{DigEcoSuc}. The succession of the Digital Ecosystem followed the smooth succession of the neutral biodiversity theory \cite{Hubbell}, shown by the \emph{tight distribution} and \emph{equal density} of the points around the best fit curve of the graph in Figure \ref{DigEcoSuc}. The variation in the percentage responsiveness, over the successive user request events, came from the differential rates of adaption at the Habitats. Still, by \setCap{the end of the simulation run, the Agent-sequences had evolved and migrated over an average of only ten user requests per Habitat, and collectively had already reached near 70\% effectiveness for the user base.}{DigEcoSuc2Cap} \setCap{The formation of a mature biological ecosystem, ecological succession, is a relatively slow process \cite{begon96}, and the simulated Digital Ecosystem acted similarly in reaching a mature state.}{DigEcoSucCap}

\subsection{Species Abundance}

\tfigure{scale=0.8}{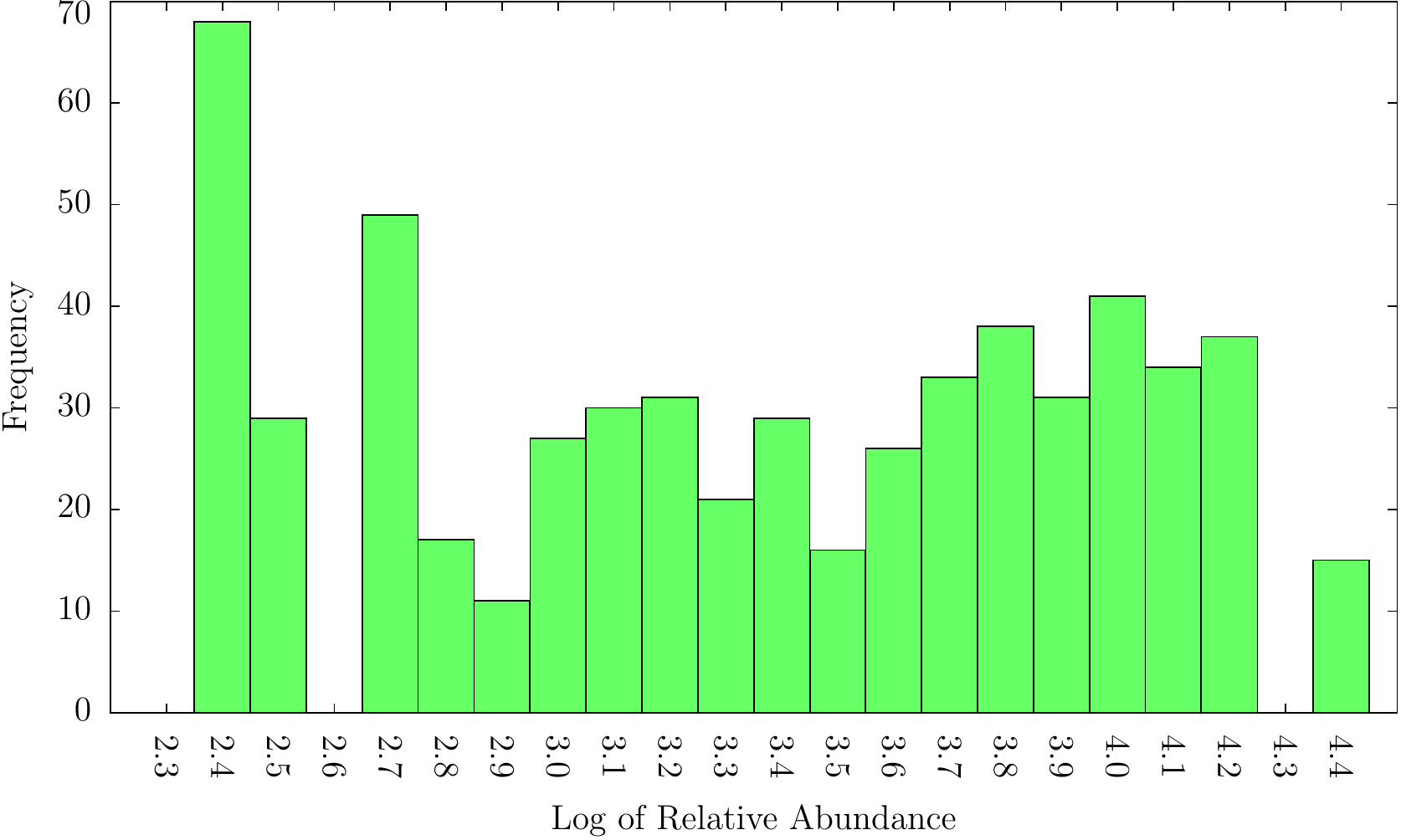}{graph}{Graph of Relative Abundance in the Digital Ecosystem}{Relative abundance \getCap{speciesAbundance} However, \getCap{specAbund2}.}{-2mm}{}{}{}

In ecology, \emph{relative abundance} \setCap{is a measure of the proportion of all organisms in a community belonging to a particular species \cite{Bell}. A relative abundance distribution provides the inequalities in population size within an ecosystem and therefore an indicator of biodiversity, with the distribution of most biological ecosystems taking a log-normal form \cite{Bell}.}{speciesAbundance} So, for Digital Ecosystems this measures globally the abundance of different solutions relative to one another.

A snapshot of the Agents (organisms) within the Digital Ecosystem, for a typical simulation run, was taken after a thousand user requests, i.e. once it had reached a mature state. In biology a species is a series of populations within which significant gene flow can and does occur, so groups of organisms showing a very similar genetic makeup \cite{lawrence1989hsd}. We therefore chose to define species within Digital Ecosystems similarly, as a grouping of genetically similar digital organisms (based on their semantic descriptions), with no more than 10\% variation within the species group. Relative abundance was calculated for each species and grouped by frequency in Figure \ref{SpeciesHistogram}. In contrast to expectations from biological ecosystems, relative abundance in \setCap{the Digital Ecosystem did not conform to the expected log-normal}{specAbund2} \cite{Bell}. We suggest that the high frequency for the lowest relative abundance was caused by the dynamically re-configurable topology of the Habitat network, which allowed species of small abundance to survive as their respective Habitats were clustered by the Digital Ecosystem. Therefore, it also most likely skewed the other frequencies of the relative abundance measure.

\subsection{Species-Area Relationship}

\tfigure{scale=0.8}{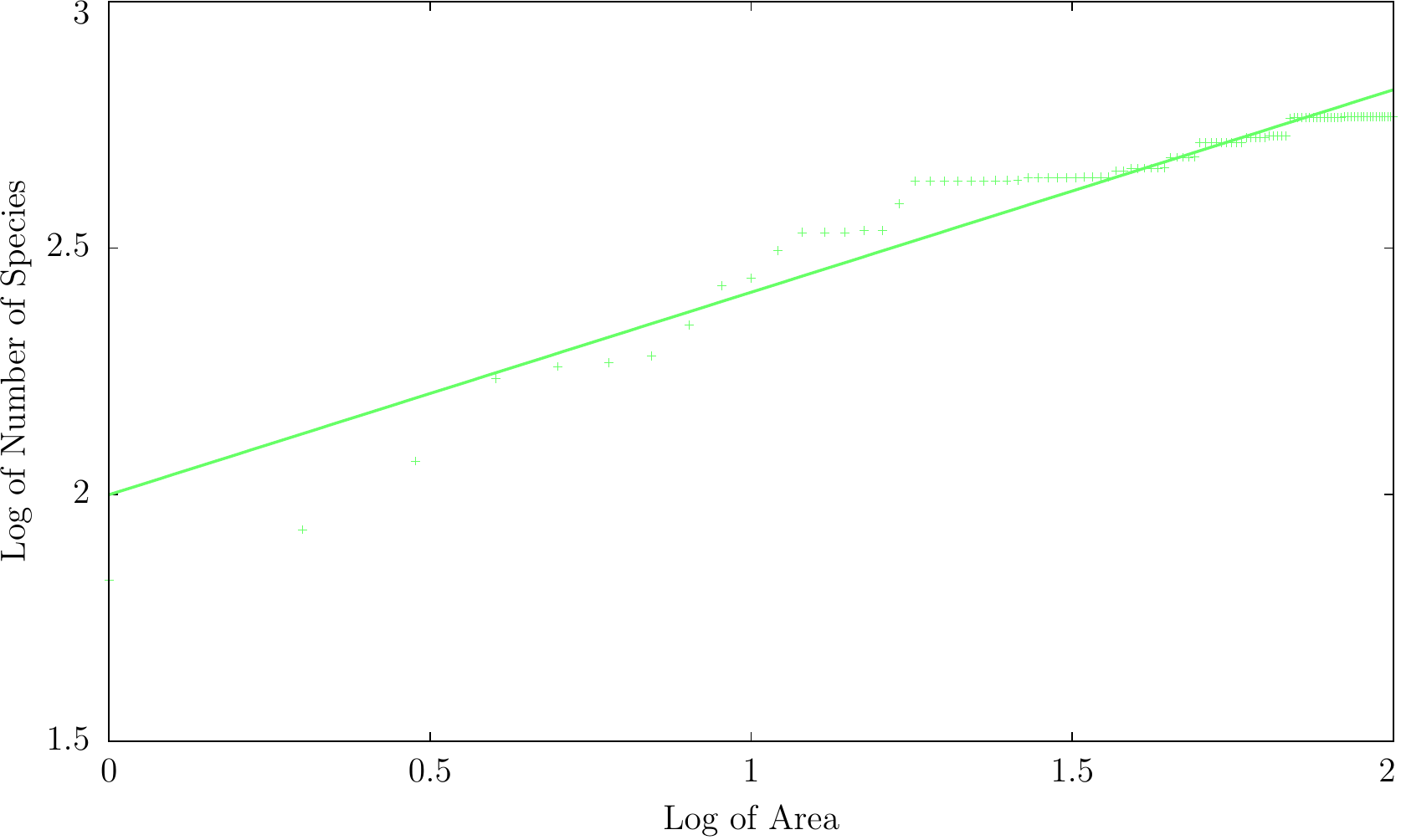}{graph}{Graph of Species-Area in the Digital Ecosystem}{\getCap{speciesArea}, which the Digital Ecosystem also demonstrates.}{-2mm}{!b}{}{}

\setCap{In ecology the \emph{species-area} relationship measures diversity relative to the spatial scale, showing the number of species found in a defined area of a particular habitat or habitats of different areas \cite{sizling2004pls}, and is commonly found to follow a power law in biological ecosystems}{speciesArea} \cite{sizling2004pls}. For Digital Ecosystems this relationship represents how similar solutions are to one another at different Habitat scales. 

Again, a snapshot of the Agents (organisms) within the Digital Ecosystem, for a typical simulation run, was taken once it had reached a mature state, after a thousand user requests. For this experiment, we assumed each Habitat to have an area of one unit. Then, the number of species, at $n$ randomly chosen Habitats, was measured, where $n$ ranged between one and a hundred. For each $n$, ten sets of measurements were taken at different random sets of Habitats to calculate averaged results, and the $log_{10}$ values of these results are depicted in the graph of Figure \ref{speciesArea}. The distribution of species diversity over a spatial scale in the Digital Ecosystem demonstrates behaviour similar to biological ecosystems, also following a power law \cite{sizling2004pls}. However, diversity at fine spatial scales appears to be lower than predicted by the line of best fit. This may be explained by higher specialisation at some Habitats, making them more like micro-habitats in terms of a reduced species diversity \cite{lawrence1989hsd}.

The majority of the experimental results indicate that Digital Ecosystems behave like their biological counterparts, and suggest that incorporating ideas from theoretical ecology can contribute to useful self-organising properties in Digital Ecosystems, which can assist in generating scalable solutions to complex dynamic problems. 

\section{Discussion}
\label{endChap2}

Creating the digital counterpart of biological ecosystems was not without apparent compromises; the temporary one-to-one genotype-phenotype mapping for Agents, the information-centric dynamically re-configurable network topology, and the \emph{species abundance} result are inconsistent with biological ecosystems. 

\tfigure{scale=0.8}{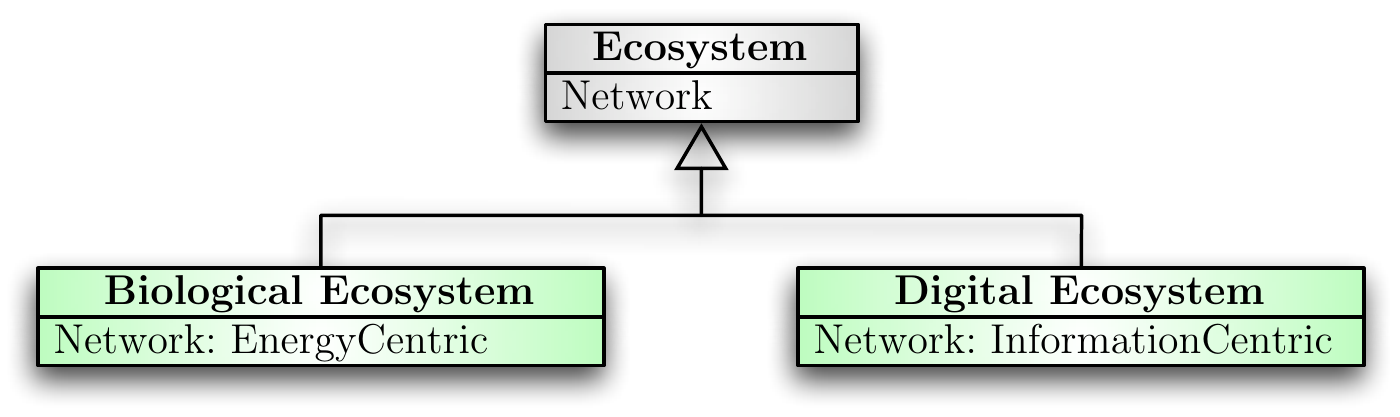}{graffle}{Hypothetical Abstract Ecosystem Definition}{If there were an abstract ecosystem class in the \acs{UML}, \getCap{ecoCapClass}. \getCap{ecoCap2Class}}{-4mm}{!h}{}{}

We would argue that these differences are not compromises, but features unique to Digital Ecosystems. As we discussed earlier, biomimicry, when done well, is not slavish imitation; it is inspiration using the principles which nature has demonstrated to be successful design strategies \cite{biomimicry}. Hypothetically, if there were an abstract definition of an ecosystem, defined as an abstract ecosystem class, \setCap{then the Digital Ecosystem and biological ecosystem classes would both inherit from the abstract ecosystem class, but implement its attributes differently}{ecoCapClass}, as shown in the \acf{UML} class diagram of Figure \ref{ecoClass}. \setCap{So, we would argue that the apparent compromises in mimicking biological ecosystems are actually features unique to Digital Ecosystems.}{ecoCap2Class}

Service-oriented architectures promise to provide potentially huge numbers of services that programmers can combine via standardised interfaces, to create increasingly sophisticated and distributed applications \cite{SOApaper2}. The Digital Ecosystem extends this concept with the automatic combining of available and applicable services in a scalable architecture to meet user requests for applications. This is made possible by a fundamental paradigm shift, from a pull-oriented approach to a push-oriented approach. So, instead of the pull-oriented approach of generating applications only upon request in \acp{SOA} \cite{singh2005soc}, the Digital Ecosystem follows a push-oriented approach of distributing and composing applications pre-emptively, as well as upon request. Although the use of \acp{SOA} in the definition of Digital Ecosystems provides a predisposition to business \cite{krafzig2004ess}, it does not preclude other more general uses. The \acl*{EOA} definition of Digital Ecosystems is intended to be inclusive and interoperable with other technologies, in the same way that the definition of \acp{SOA} is with grid computing and other technologies \cite{singh2005soc}. For example, Habitats could be executed using a distributed processing arrangement, such as grid computing \cite{singh2005soc}, which would be possible because the Habitat network topology is information-centric (instead of location-centric).

\section{Conclusions}

We have determined the fundamentals for a new class of system, created through combining understanding from theoretical ecology \cite{levins1969sda}, evolutionary theory \cite{ec16}, \acp{MAS} \cite{moaspaper}, distributed evolutionary computing \cite{lin1994cgp}, and \acp{SOA} \cite{soa1w}. So, the digital counterpart of biological ecosystems, having their properties of self-organisation, scalability and sustainability \cite{Levin}: being similarly complex systems that show emergent behaviour, since they are more than the sum of their constituent parts. Therefore, we have created the first interpretation of Digital Ecosystems where the word ecosystem is more than just a metaphor, which we have confirmed experimentally.

The \acl*{EOA} of Digital Ecosystems includes a novel form of distributed evolutionary computing, an optimisation technique working at two levels: a first optimisation, migration of Agents which are distributed in a peer-to-peer network, operating continuously in time; this process feeds a second optimisation, based on evolutionary computing, operating locally on single peers and is aimed at finding solutions that satisfy locally relevant constraints. So, the local search is improved through this twofold process to yield better local optima faster, as the distributed optimisation provides prior sampling of the search space through computations already performed in other peers with similar constraints. We have also defined the interaction of Digital Ecosystems with \emph{business ecosystems} \cite{moore1996}, specifically in supporting and enabling them to create \acp{DBE}.

The ever-increasing challenge of software complexity in creating progressively more sophisticated and distributed applications, makes the design and maintenance of efficient and flexible systems a growing challenge \cite{newsArticle1, slashdot, newsArticle3}, for which current software development techniques have hit a \emph{complexity wall} \cite{lyytinen2001nwn}. In response we have created Digital Ecosystems, the digital counterparts of biological ecosystems, possessing their properties of self-organisation, scalability and sustainability \cite{Levin}; \aclp*{EOA} that overcome the challenge by automating the search for new algorithms in a scalable architecture, through the evolution of software services in a distributed network.

\section{Future Work}

In creating Digital Ecosystems, the digital counterpart of biological ecosystems, we naturally asked their likeness to the biological ecosystems from which they were inspired. Further to this, we could consider the applicability of other aspects of ecosystems theory in understanding and analysing the dynamics of Digital Ecosystems. For example, \emph{energy pyramids}\footnote{Energy pyramids show the dissipation of energy at trophic levels, positions that organisms occupy in a food chain, e.g. producers or consumers \cite{odum1968efe}.} of biological ecosystems, what is their equivalent in Digital Ecosystems? Given that Digital Ecosystems are information-centric, whereas biological ecosystems are energy-centric \cite{begon96}, they would undoubtedly be \emph{information pyramids}, but further definition would naturally require more research.

While evolutionary theory \cite{ec16} was well understood within computer science, under the auspices of evolutionary computing \cite{eiben2003iec}, ecosystems theory \cite{begon96}, until our efforts, was not. Similarly, while evolutionary theory is well understood within linguistics \cite{croft2000elc} and economics \cite{nelson1982ete}, equally ecosystems theory is not \cite{mufwene2001ele}. So, using our efforts as a \emph{case study}, we could follow the same process to create Language Ecosystems and Economic Ecosystems. For example, there are many separate efforts within linguistics using evolution to model language change \cite{christiansen2003le}, but there is no unifying framework, which has resulted from different linguists independently adopting elements of evolutionary theory \cite{christiansen2003le}. So, we could provide a wide-ranging and encompassing definition of Language Ecosystems, which would unify the many disparate efforts in linguistics aimed at understanding language evolution.

Conceptualising ecosystems has been an inherent part of this work, which presents us with an opportunity to formalise our current and future efforts to improve the cross-disciplinary knowledge transfer required \cite{caes}. In the creation of Digital Ecosystems we considered aspects of biological ecosystems, including \acl*{ABM} \cite{Greenetal2006} and \acf{CAS} \cite{Levin}, and then constructed their counterparts in Digital Ecosystems. After which we considered the possibility of a Generic Ecosystem definition, and without which some of the counterparts we constructed appeared to be compromised, when they were actually the realisation of generic abstract concepts in Digital Ecosystems. Most notably the network structure, which is energy-centric in biological ecosystems \cite{begon96}, while information-centric in Digital Ecosystems. So, there is potential to create a Generic Ecosystem definition, using a suitable modelling technique such as \ac{CAS} \cite{waldrop1992ces}, which would abstractly define the key properties of an ecosystem, and would theoretically be applicable to any domain where the modelling technique has been applied. Therefore, the Generic Ecosystem definition would provide a framework for the application of ideas, concepts, and models from biological ecosystems to other classes of ecosystems, including Digital Ecosystems, Language Ecosystems and Economic Ecosystems. A \emph{design pattern} is a general reusable solution to a commonly occurring problem in software design \cite{gamma1995dpe}. It is not a finished design that can be transformed directly into code, but a description or template for how to solve a problem that can be used in many different situations \cite{gamma1995dpe}. For example, object-oriented design patterns typically show relationships and interactions between classes or objects, without specifying the final application classes or objects that are involved \cite{gamma1995dpe}. \aclp*{BDP} would extend this concept to catalogue common interactions between biological structures using a pattern-oriented modelling approach \cite{grimm2005pom}, which when applied would endow software systems with the desirable properties of biological systems, such as self-organisation, self-management, scalability and sustainability.

An open-source simulation framework for Digital Ecosystems \cite{eveSim} was created by the \acf{DBE} project \cite{DBE}, and is currently supported by the \acl*{OPAALS} project \cite {OPAALS} to assist further research into Digital Ecosystems, including the wider implications of interacting with social systems, such as \emph{business ecosystems} of \acp{SME}. In an old market-based economy, made up of sellers and buyers, the parties exchange property \cite{delcloque2001dii}. While in a new network-based economy, made up of servers and clients in a \emph{business ecosystem} \cite{moore1996}, the parties share access to services and experiences \cite{delcloque2001dii}. Digital Ecosystems are a platform for the network-based economy of \emph{business ecosystems}, providing mechanisms for the creation of \acp{DBE}. One such mechanism the Digital Ecosystem could provide to the network-based economy of \emph{business ecosystems} \cite{moore1996}, would be a \emph{futures market}\footnote{An auction market in which participants buy and sell commodities for an agreed price, that the sellers have yet to produce \cite{hull2005ffa}.} for services. As each service (Agent) consists of an executable component and a semantic description, the later acting as a guarantee of behaviour, and the evolving Agent Populations only requiring the guarantees (semantic descriptions) to operate, the actual executable component of a service (Agent) is only required once an application (Agent-sequence) has been assembled. Therefore, service (Agent) evolution could operate entirely on the semantic descriptions, with business users only needing to supply the executable component of a service (Agent) once there is a demand, i.e. when the semantic description of one of their services has been used in the construction of an application which meets the request of another business user. Therefore, creating a \emph{futures market} for evolving services within Digital Business Ecosystems.

A partial reference implementation \cite{eveNet} for our Digital Ecosystem, was created by the \acp{DBE} project \cite{DBE}, and we expect that once completed will be deployed as part of the software platform intended for the regional deployment of their \emph{Digital Ecosystems} \cite{den4dek, OPAALS}. These \emph{Digital Ecosystems} (distributed adaptive open socio-technical systems, with properties of self-organisation, scalability and sustainability, inspired by natural ecosystems \cite{OPAALS}) are emerging as a novel approach to the catalysis of sustainable regional development driven by \acp{SME} \cite{abcdbe, OPAALS}. The community focused on the deployment of \emph{Digital Ecosystems}, REgions for Digital Ecosystems Network (REDEN) \cite{reden}, is supported by projects such as the Digital Ecosystems Network of regions for (4) DissEmination and Knowledge Deployment (DEN4DEK) \cite{den4dek}, a thematic network that aims to share experiences and disseminate knowledge that will allow regions to plan an effective deployment of \emph{Digital Ecosystems} at all levels (economic, social, technical and political) to produce real impacts in the economic activities of European regions through the improvement of \ac{SME} business environments. So, the next major step in our research will be to collect real world data, confirming that Digital Ecosystems operate effectively with \emph{business ecosystems} in creating Digital Business Ecosystems.

\begin{acknowledgements}

The authors would like to thank for their encouragement and suggestions: Paolo Dini of the \acl*{LSE}, Thomas Heistracher and his group of the \acl*{STU}, Jonathan Rowe of the \acl*{UBHAM}, and Miguel Vidal of \acl*{SUN}. The Digital Ecosystem model was constructed through interacting with these people and others. This work was supported by the EU-funded \acl*{OPAALS} Network of Excellence (NoE), Contract No. FP6/IST-034824.

\end{acknowledgements}

\bibliographystyle{spbasic}      
\bibliography{references}   

\end{document}